\documentclass{aa}  

\usepackage{graphicx}
\usepackage{txfonts}
\usepackage{hyperref}
\usepackage{adjustbox}
\usepackage{url}
\begin{document} 

\def\teff{${T}_{\rm eff}$}
\def\kms{{km\,s}$^{-1}$}
\def\logg{$\log g$}
\def\micro{$\xi_{\rm t}$}
\def\macro{$\zeta_{\rm RT}$}
\def\rad{$v_{\rm r}$}
\def\vsini{$v\sin i$}
\def\ebv{$E(B-V)$}
\def\kepler{\textit{Kepler}}

   \title{New mercury-manganese stars and candidates from LAMOST DR4}

\author{E.~Paunzen\inst{1}
        \and S.~H{\"u}mmerich\inst{2,3}
        \and K.~Bernhard\inst{2,3}
        }
\institute{Department of Theoretical Physics and Astrophysics, Masaryk University, Kotl\'a\v{r}sk\'a 2, 611\,37 Brno, Czech Republic, \email{epaunzen@physics.muni.cz}
\and Bundesdeutsche Arbeitsgemeinschaft f{\"u}r Ver{\"a}nderliche Sterne e.V. (BAV), D-12169 Berlin, Germany, 
\and American Association of Variable Star Observers (AAVSO), 49 Bay State Rd, Cambridge, MA 02138, USA}

\date{}

 
  \abstract
   {}
   {The present work presents our efforts at identifying new mercury-manganese (HgMn/CP3) stars using spectra obtained with the Large Sky Area Multi-Object Fiber Spectroscopic Telescope (LAMOST).}
   {Suitable candidates were searched for among pre-selected early-type spectra from LAMOST DR4 using a modified version of the MKCLASS code that probes several \ion{Hg}{ii} and \ion{Mn}{ii} features. The spectra of the resulting 332 candidates were visually inspected. Using parallax data and photometry from Gaia DR2, we investigated magnitudes, distances from the Sun, and the evolutionary status of our sample stars. We also searched for variable stars using diverse photometric survey sources.}
   {We present 99 bona fide CP3 stars, 19 good CP3 star candidates, and seven candidates. Our sample consists of mostly new discoveries and contains, on average, the faintest CP3 stars known (peak distribution 9.5\,$\le$\,$G$\,$\le$\,13.5\,mag). All stars are contained within the narrow spectral temperature-type range from B6 to B9.5, in excellent agreement with the expectations and the derived mass estimates (2.4\,$\le$\,M$_\odot$\,$\le$\,4 for most objects). Our sample stars are between 100\,Myr and 500\,Myr old and cover the whole age range from zero-age to terminal-age main sequence. They are almost homogeneously distributed at fractional ages on the main sequence $\le$\,80\,\%, with an apparent accumulation of objects between fractional ages of 50\,\% to \,80\,\%. We find a significant impact of binarity on the mass and age estimates. Eight photometric variables were discovered, most of which show monoperiodic variability in agreement with rotational modulation.}
	 {Together with the recently published catalogue of APOGEE CP3 stars, our work significantly increases the sample size of known Galactic CP3 stars, paving the way for future in-depth statistical studies.}


   \maketitle

%
\section{Introduction} \label{introduction}

A significant fraction of early B- to early F-type main-sequence stars exhibits peculiar atmospheric compositions that indicate abundance stratification due to atomic diffusion in their outer layers (e.g. \citealt{michaud70}, \citealt{richer00}). These objects, which are generally referred to as chemically peculiar (CP) stars, have been divided into several groups according to their peculiarities. The four main groups are the CP1 stars (the metallic-line or Am/Fm stars), the CP2 stars (the magnetic Bp/Ap stars), the CP3 stars (the mercury-manganese or HgMn stars), and the CP4 stars (the He-weak stars) \citep{preston74}. Further groups have been defined, such as the $\lambda$ Bootis stars, which show peculiarly low surface abundances of iron-peak elements \citep{murphy17}. In general, concerning the strength of the chemical peculiarities, there is a continuous transition from normal to peculiar stars \citep{loden87}.

Significant progress has been made in the understanding of the atomic diffusion process, for example concerning its interplay with parameters such as magnetic field strength and mass loss (\citealt{alecian17,alecian19}). Nevertheless, further refinement is necessary before the models will be able to reproduce the observed complex abundance patterns of CP stars in detail.

On average, CP3 stars rotate considerably slower than chemically normal main-sequence stars of similar temperatures \citep{mathys04}. They are encountered in the temperature range of 10\,500\,K\,$\leq$\,\teff\,$\leq\,$16\,000\,K and mainly characterised by the presence of strong atmospheric overabundances of Hg and Mn as compared to the solar values (up to 6 and 3 dex, respectively; \citealt{smith96}, \citealt{ghazaryan16}). However, their spectra show numerous other peculiarities and often indicate strikingly different abundance patterns from one star to another. As a general rule, heavy elements are overabundant, with the strength of the overabundances increasing with atomic number \citep{castelli04,ghazaryan16}. A good summary of the observed peculiarities in CP3 stars is provided by \citet{ghazaryan16}.

Several studies indicated a multiplicity incidence of more than 50\,\% among CP3 stars \citep{smith96}. \citet{hubrig95} and \citet{schoeller10}, for instance, estimated that values may be as high as 67\,\% and 91\,\%, respectively. The tidal interaction induced by the presence of a companion star is thought to significantly impact the observed chemical peculiarities.

An inhomogeneous surface element distribution with obvious signs of secular evolution was identified in several CP3 stars (e.g. \citealt{hubrig95}, \citealt{adelman02}, \citealt{hubrig06}, \citealt{kochukhov07}, \citealt{briquet10}, \citealt{korhonen13}). Several mechanisms were proposed to explain the observed abundance inhomogeneities, such as time-dependent atomic diffusion \citep{alecian09} and atomic diffusion in the presence of a weak, multi-polar magnetic field \citep{alecian12,alecian13}. CP3 stars have traditionally been grouped with the non-magnetic CP stars. While they certainly lack the strong organised magnetic fields of the CP2 and CP4 stars (the so-called magnetic CP stars or mCP stars), several studies indicated the presence of weak or tangled fields \citep{hubrig10,hubrig12}. These claims were disputed by other investigations (see the discussion and references in \citealt{kochukhov13}), and this remains a controversial issue. Recent evidence for the detection of variable, weak, and tangled magnetic fields in CP3 stars was presented by \citet{hubrig20}.

Although theoretical modelling predicted the occurrence of pulsational instability via the $\kappa$ mechanism \citep{turcotte03,alecian09} and CP3 stars share their position in the Hertzsprung-Russell (HR) diagram with pulsating variables such as the slowly pulsating B (SPB) stars, they are among the least photometrically variable CP stars \citep{adelman98}, which suggests that important parameters are missing from the pulsation models. Since the advent of high-precision space photometry, however, an increasing number of photometrically variable CP3 stars has been identified. Still, no consensus has yet been reached as to the cause of the observed photometric variations, and arguments have been put forth in favour of both rotational as well as pulsational variability, with the more recent studies favouring rotational modulation due to abundance spots (cf. e.g. the discussion and references in \citealt{huemmerich18}).

Recently, \citet{chojnowski20} reported their efforts at identifying CP3 stars using SDSS/APOGEE $H$-band spectra. The authors identified suitable targets via the presence of \ion{Mn}{ii} and \ion{Ce}{iii} lines and presented a catalogue of 260 newly identified objects, thereby more than doubling the sample size of known Galactic CP3 stars (from $\sim$194 to $\sim$454 known objects; cf. \citealt{chojnowski20}).

Here we present our efforts at identifying new CP3 stars using archival spectra from the fourth data release (DR4) of the Large Sky Area Multi-Object Fiber Spectroscopic Telescope (LAMOST) of the Chinese Academy of Science \citep{lamost1,lamost2}. We have identified 99 bona fide CP3 stars, 19 very good CP3 star candidates, and seven candidates. Most of these stars are new discoveries; only 15 stars of our sample are contained in the catalogue of \citet{chojnowski20}. Thus, including the very good candidates, we further increase the number of known CP3 stars to more than 550 objects. Furthermore, with a broad peak distribution between $G$ magnitudes 9.5 and 13.5, our sample contains on average the faintest CP3 stars known.

Together with the work of \citet{chojnowski20}, our study significantly increases the sample size of known Galactic CP3 stars, which will benefit future in-depth statistical studies and theoretical modelling attempts alike. Section \ref{dataanalysis} discusses spectroscopic data and sample selection process and details the spectral classification workflow. Section \ref{results} provides a discussion of our results, such as spectral classifications, evolutionary status, and photometric variability of our sample stars. We conclude in Section \ref{conclusion}.

\section{Spectroscopic data and sample selection} \label{dataanalysis}

This section provides descriptions of the LAMOST spectral archive, the MKCLASS spectral classification code, the sample selection process, and the spectral classification workflow.

\subsection{The Large Sky Area Multi-Object Fiber Spectroscopic Telescope (LAMOST)} \label{LAMOST}

The LAMOST telescope\footnote{The LAMOST telescope is also referred to as the Guo Shou Jing telescope, in dedication to the famous Chinese astronomer, hydraulic engineer, mathematician, and politician of the Yuan Dynasty.} \citep{lamost1,lamost2} is located at the Xinglong Observatory in Beijing, China. It is a Schmidt telescope with an effective aperture of 3.6$-$4.9\,m (field of view of about 5$\degr$), which was designed to collect 4000 spectra in a single exposure (spectral resolution R\,$\sim$\,1800, limiting magnitude $r$\,$\sim$\,19\,mag, wavelength coverage 3700 to 9000\,\AA). The LAMOST telescope is therefore well suited to carry out large-scale spectral surveys and is currently executing a survey of the entire available northern sky. LAMOST spectra are made available to the public in consecutive data releases via the LAMOST spectral archive.\footnote{\url{http://www.lamost.org}} Alreading containing a total of 14.48 million spectra as of DR7 (31 March 2020), the LAMOST archive is a prime resource for researchers, whose exploitation has only just begun.

Several investigations used LAMOST spectra to study CP stars. For instance, \citet{hou15} and \citet{qin19} searched for CP1 stars in the low-resolution spectra of DR1 and DR5, respectively. Our efforts in the identification and classification of mCP stars using spectra from DR4 led to the identification of a sample of 1002 mCP stars, comprising mostly new discoveries \citep{huemmerich20}.

\subsection{The MKCLASS code} \label{MKCLASS}

\subsubsection{General overview} \label{MKCLASS_1}

The MKCLASS code was written by Richard O. Gray to classify stellar spectra on the Morgan-Keenan-Kellman (MKK) system in a way similar to a human classifier \citep{gray14}. In its current version (v1.07), MKCLASS works with spectra in the violet-green region (3800-5600\,\AA) in either rectified or flux-calibrated format. Several studies employed MKCLASS for the classification of a multitude of stellar spectra (e.g. \citealt{gray14,gray16,huemmerich18,huemmerich20}) and demonstrated that its results compare favourably to manually derived classifications if input spectra of sufficient signal-to-noise (S/N) are provided. Typical temperature class and luminosity class uncertainties are of the order of 0.6 and 0.5 subclass, respectively \citep{gray14}.

\begin{table*}
\caption{Characteristics of the spectral libraries used as standard star libraries for the modified version of the MKCLASS code in the present study.}
\label{table_libraries}
\begin{tabular}{ll}
\hline
\hline
Library & Characteristics \\
\hline
\textbf{\textit{libr18}} & spectral range from 3800–4600\,\AA, resolution of 1.8\,\AA\ (R\,$\sim$\,2200), normalised and rectified spectra; \\
& all luminosity classes (Ia-V) \\
\hline
\textbf{\textit{libnor36}} & spectral range from 3800–5600\,\AA, resolution of 3.6\,\AA\ (R\,$\sim$\,1100), flux-calibrated and normalised spectra; \\
& all luminosity classes (Ia-V) \\
\hline
\textbf{\textit{libsynth}} & spectral range from 3800–4600\,\AA, smoothed to a resolution of 3.0\,\AA\ and an output spacing of 0.5\,\AA, flux-calibrated \\
& and normalised synthetic spectra; only dwarf spectra (luminosity class V); spectral types B5 to F5 \\
\hline
\textbf{\textit{liblamost}} & spectral range from 3800–5600\,\AA, resolution R\,$\sim$\,1800, flux-calibrated and normalised spectra; \\
& only dwarf and giant spectra (luminosity classes V and III); spectral types B3 to G0 \\
\hline
\hline
\end{tabular}                                                                                                                                       
\end{table*}

As a first step, MKCLASS assigns a rough spectral type, which is subsequently refined by direct comparison with spectra from MKK standard star libraries. Two libraries of standard spectra are recommended for use with MKCLASS, which were obtained with the Gray/Miller spectrograph on the 0.8\,m reflector of the Dark Sky Observatory (North Carolina, USA): the \textit{libr18} library, which consists of rectified spectra in the spectral range 3800–4600\,\AA\ (resolution of 1.8\,\AA), and the \textit{libnor36} library, which encompasses flux-calibrated and normalised spectra in the spectral range from 3800–5600\,\AA\ (resolution of 3.6\,\AA). However, it is also possible to employ additional spectral libraries specifically tailored to the needs of the researcher.

In its current implementation, MKCLASS is capable of identifying a set of spectral peculiarities that are relevant to the classification of objects such as CP1 and CP2 stars, barium stars, or carbon-rich giants. Unfortunately, the original code does not probe any lines relevant to the identification of CP3 stars. For more information on the MKCLASS code and its usage, we refer the reader to \citet{gray14} and the corresponding website.\footnote{\url{http://www.appstate.edu/~grayro/mkclass/}}

\subsubsection{Customising the MKCLASS code} \label{MKCLASS_2}

Our previous LAMOST-based study \citep{huemmerich20} was concerned with the identification and classification of mCP stars using low-resolution spectra from DR4. To this end, we produced a modified version of the MKCLASS code (termed MKCLASS\_mCP) capable of probing a number of spectral features related to Si, Cr, Sr, Eu, and He abundances, which are helpful in the identification and classification of these objects (cf. Table \ref{table_lines}). Furthermore, in addition to using the \textit{libr18} and \textit{libnor36} libraries that come with MKCLASS, two further standard star libraries were employed: the \textit{libsynth} library, which contains spectra synthesised with the SPECTRUM code\footnote{\url{http://www.appstate.edu/~grayro/spectrum/spectrum.html}} \citep{SPECTRUM} and ATLAS9 model atmospheres \citep{ATLAS9}, and the \textit{liblamost} library, which consists of a grid of suitable high S/N LAMOST DR4 spectra chosen from the list of \citet{gray16}. The characteristics of the different libraries, which have also been employed in the present study, are summarised in Table \ref{table_libraries}. For more information on the MKCLASS\_mCP code and the \textit{libsynth} and \textit{liblamost} libraries, the reader is referred to \citet{huemmerich20}.

%
%
%

In our search for mCP stars, suitable candidates were chosen from a colour-selected sample of early-type objects via the presence of the conspicuous 5200\,\AA\ flux depression, which is a characteristic of this subgroup of CP stars and relatively easy to identify even at the low resolution of the employed LAMOST spectra \citep{huemmerich20}. Unfortunately, no similarly convenient spectral feature is available for the identification of CP3 stars. Therefore, we altered the MKCLASS\_mCP code to additionally probe suitable \ion{Hg}{ii} and \ion{Mn}{ii} lines and blends in the blue-violet spectral region, which are relevant to the identification of CP3 stars. The choice of lines was influenced by the resolution of the LAMOST spectra and the availability of neighbouring continuum windows to probe a certain feature. After some experimentation, we came up with the lines and blends presented in Table \ref{table_lines} (bold font) as best choice. Although the \ion{Mn}{ii} 4252\,\AA\ and \ion{Mn}{ii} 4259\,\AA\ lines are partly resolved in LAMOST spectra, we chose to probe the blend of the region from 4248\,\AA\ to 4262\,\AA, which yielded more promising results. These features were searched for in the spectral type range from O6 to A3, in addition to the other lines and blends listed in Table \ref{table_lines}.

We note that at the resolution of the LAMOST low-resolution spectra, all features listed in Table \ref{table_lines} are, to some extent, blended with other absorption lines. Nevertheless, the specified elements generally constitute the main contributors to these blends in the investigated groups of CP stars. For example, the 3984\,\AA\ feature, which is particularly relevant to the identification of CP3 stars, is a complex blend, which, apart from the \ion{Hg}{ii} line, may include contributions from \ion{Cr}{i} 3983.896\,\AA, \ion{Mn}{i} 3983.683\,\AA, \ion{Fe}{i} 3983.835\,\AA, and \ion{Fe}{i} 3983.959\,\AA. However, the listed lines mostly play a part in the cooler CP2 stars \citep{preston73}. In the temperature regime of the CP3 stars, the density of lines is rather low and we generally do not expect significant blending of the \ion{Hg}{ii} 3984\,\AA\ line \citep{cowley75}. This is confirmed by data from the Vienna Atomic Line Database (VALD; \citealt{1999A&AS..138..119K}), which indicate no other strong lines in the corresponding wavelength region for the investigated effective temperature range.

Nevertheless, at LAMOST resolution, the 3984\,\AA\ feature also contains contributions from the \ion{Y}{ii} line at 3982.59\,\AA. Y is strongly overabundant in CP stars \citep{1971Ap&SS..10..156G,2019AJ....158..157M}, in particularly in CP3 stars. From the compilation of \citet{ghazaryan18}, we derive mean Y abundances of +0.86\,dex, +1.26\,dex, and +2.64\,dex as compared to the solar value for, respectively, CP1, CP2, and CP3 stars. However, considering the strength of the \ion{Y}{ii} spectrum, at the resolution of the LAMOST spectra, an extreme overabundance of Y would be needed for a significant detection of the \ion{Y}{ii} 3982.59\,\AA\ line alone. \citet{Monier_2020}, for example, illustrate the spectrum of the CP3 star HR 8937, which shows an Y overabundance of +4\,dex. Line depths in this spectrum reach 12\,\% for the \ion{Y}{ii} line, which should be retraceable from LAMOST spectra, and 25\,\% for the neighbouring \ion{Hg}{ii} line. At low resolution, the corresponding blend of both lines would therefore be clearly dominated by the \ion{Hg}{ii} line.

In summary, while contributions from the \ion{Y}{ii} 3982.59\,\AA\ line are expected, we are confident that the \ion{Hg}{ii} 3894\,\AA\ line is the main contributor to the 3984\,\AA\ blend observed in our sample stars. This is in line with the common practice to use the 3984\,\AA\ feature to identify CP3 stars in classification resolution spectra (e.g. \citealt{gray09}), in particular as the \ion{Hg}{ii} 3984\,\AA\ line is generally not prominent in the other groups of CP stars \citep{gray09}. For instance, none of the CP2 star LAMOST spectra investigated in \citet{huemmerich20} shows a feature of comparable strength in the 3984\,\AA\ region. Together with the obvious \ion{Mn}{ii} features observed in our sample stars, the strong 3984\,\AA\ blend seems to be a unique characteristic of the group of CP3 stars.

We stress that the altered version of the MKCLASS code, which was derived from the MKCLASS\_mCP code and is referred to hereafter as MKCLASS\_HgMn, was specifically created to work with LAMOST low-resolution spectra. Applying the code to spectra of other resolutions will require a corresponding update of the peculiarity classification routine and, perhaps, of the employed standard star libraries. We also caution against using the \textit{libsynth} and \textit{liblamost} libraries outside the context of this study, in particular as they do not contain stars of luminosity classes I and II (\textit{liblamost}) and I, II, and III (\textit{libsynth}).

\begin{table}
\caption{Absorption lines and blends identified by the modified versions of the MKCLASS code. Normal font denotes features identified by the MKCLASS\_mCP code \citep{huemmerich20}, bold font the additional lines used for the identification of CP3 stars by the MKCLASS\_HgMn code. The columns denote: (1) Blend/Line. (2) Wavelength (\AA). (3) Spectral range in which the corresponding feature was probed. At the resolution of the LAMOST low-resolution spectra, all features listed in Table \ref{table_lines} are, to some extent, blended with other absorption lines. Nevertheless, the specified elements generally constitute the main contributors to these blends in the investigated groups of CP stars.}
\label{table_lines}
\begin{center}
\begin{tabular}{cll}
\hline
\hline
(1) & (2) & (3) \\
\textbf{Blend/line} & \textbf{Wavelength (\AA)} & \textbf{SpT\_range} \\
\hline
(\ion{Si}{ii}/\ion{Cr}{ii}/\ion{Sr}{ii}) & 4076/77 & B7$-$F5 \\
\hline
(\ion{Si}{ii}/\ion{Eu}{ii}) & 4128/30 & B3$-$F2 \\
\hline
\ion{Si}{ii} & 3856 & B3$-$F2 \\
\ion{Si}{ii} & 4200 & B3$-$A2 \\
\ion{Si}{ii} & 5041 & B3$-$F2 \\
\ion{Si}{ii} & 5056 & B3$-$F2 \\
\ion{Si}{ii} & 6347 & B3$-$F2 \\
\ion{Si}{ii} & 6371 & B3$-$F2 \\
\hline
\ion{Cr}{ii} & 3856 & B7$-$F2 \\
\ion{Cr}{ii} & 4172 & B7$-$F2 \\
\hline
\ion{Sr}{ii} & 4216 & B7$-$F2 \\
\hline
\ion{Eu}{ii} & 4205 & B7$-$F2 \\
\hline
\ion{He}{i} & 4009 & B0$-$A0 \\
\ion{He}{i} & 4026 & B0$-$A0 \\
\ion{He}{i} & 4144 & B0$-$A0 \\
\ion{He}{i} & 4387 & B0$-$A0 \\
\hline
\textbf{\ion{Hg}{ii}} & \textbf{3984} & \textbf{O6$-$A3} \\
\hline
\textbf{\ion{Mn}{ii}} & \textbf{4136} & \textbf{O6$-$A3} \\
\textbf{\ion{Mn}{ii}} & \textbf{4206} & \textbf{O6$-$A3} \\
\textbf{\ion{Mn}{ii}} & \textbf{4252/9}$^{a}$ & \textbf{O6$-$A3} \\
\hline
\multicolumn{3}{l}{Notes: $^{a}$ Although these lines are partly resolved in low-} \\
\multicolumn{3}{l}{resolution LAMOST spectra, we chose to probe the blend} \\
\multicolumn{3}{l}{of the region from 4248 to 4262\,\AA.} \\
\end{tabular}
\end{center}
\end{table}

\subsection{Sample selection process} \label{target}

As a first step, we cross-matched the complete LAMOST DR4 archive with the $Gaia$ DR2 catalogue \citep{gaia1,gaia2,gaia3} and utilised the $G$ versus $(BP-RP)$ diagram to impose a cut on the investigated spectral type range, selecting only stars with $(BP-RP)$\,$<$\,0.45\,mag (i.e. hotter than spectral type mid F). The limit was extended down to this colour range in order not to exclude reddened early-type stars. Because this colour-based approach will still miss highly reddened early-type objects, such as stars in the Galactic disk, or objects with bad $Gaia$ photometry, additional B- and A-type stars were selected via the spectral types listed in the DR4 VizieR online catalogue \citep{DR4}.\footnote{\url{http://cdsarc.u-strasbg.fr/viz-bin/cat/V/153}} From the resulting list of stars, we excluded apparent supergiants and objects having spectra with a S/N of less than 50 in the Sloan $g$ band. About 45\,000 spectra were discarded in this manner. This cut was imposed because the detection of CP star features in objects with S/N\,$<$\,50 is difficult \citep{2011AN....332...77P}. In the case of objects having more than one spectrum in the DR4 catalogue, we considered only the spectrum with the highest Sloan $g$ band S/N.

In a second step, all remaining spectra were classified with the MKCLASS\_HgMn code using all four libraries (\textit{libr18}, \textit{libnor36}, \textit{libsynth}, \textit{liblamost}). To identify CP3 star candidates, we counted the number of detections $N$\textsubscript{det}($\lambda$) of a peculiarly strong \ion{Hg}{ii} and \ion{Mn}{ii} feature at or around the specified wavelength (\AA) with the different standard star libraries (cf. Table \ref{table_lines}). The resulting value (0\,$\le$\,$N$\textsubscript{det}($\lambda$)\,$\le$\,4) provides an estimation of significance: $N$\textsubscript{det}($\lambda$)\,=\,4 was considered to be a robust detection, while $N$\textsubscript{det}($\lambda$)\,<\,2 detections were viewed with caution. For convenience, the number of detections of the different \ion{Mn}{ii} features were added up to yield $N$\textsubscript{det}(Mn\_all). A sample output of MKCLASS\_HgMn is provided in column two of Table \ref{table_MKCLASS_HgMn_output}.

\begin{figure*}
        \includegraphics[width=\textwidth]{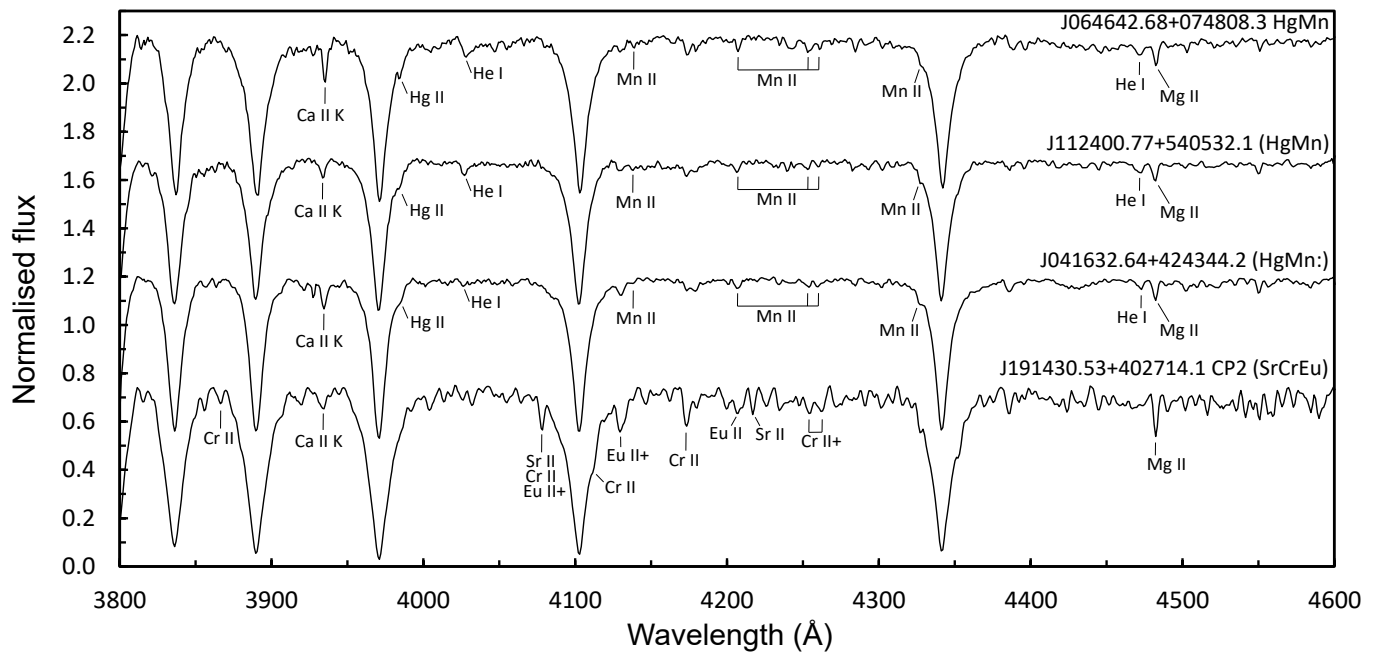}
    \caption{Blue-violet region of the LAMOST DR4 spectra of (from top to bottom) the bona fide CP3 star LAMOST J064642.68+074808.3, the good CP3 star candidate LAMOST J112400.77+540532.1, the CP3 star candidate LAMOST J041632.64+424344.2, and the CP2 star LAMOST J191430.53+402714.1. Some prominent lines of interest are identified.}
    \label{showcase2}
\end{figure*}

All stars satisfying the criteria $N$\textsubscript{det}($\lambda$3984)\,$>$\,0 \textbf{AND} $N$\textsubscript{det}(Mn\_all)\,$>$\,0 were assigned candidate status, which resulted in a candidate sample of 332 stars. As next step, the spectra of all these objects were visually inspected to sort out bona fide CP3 stars and CP3 star candidates. To this end, the spectra were overlaid with and compared to the spectrum of $\alpha$ And as shown in Figure 4.16 of \citet{gray09} and the strength of the \ion{Hg}{ii} and \ion{Mn}{ii} features was assessed. Depending on a visual estimate of the strength and clarity of the CP3 star features in the spectrum, stars were classified as either (i) bona fide CP3 stars, (ii) good CP3 star candidates, or (iii) CP3 star candidates. Figure \ref{showcase2} illustrates this process by showing the spectra of the bona fide (group i) CP3 star LAMOST J064642.68+074808.3, the good CP3 star candidate (group ii) LAMOST J112400.77+540532.1, and the CP3 star candidate (group iii) LAMOST J041632.64+424344.2. The CP3 status of stars of groups (ii) and (iii) should be confirmed by additional high-resolution spectroscopic observations.

Roughly a third of the investigated stars ($N$\,=\,125) turned out to be either bona fide CP3 stars or candidates, which highlights that the chosen approach is a viable way of identifying CP3 stars in massive spectral archives. The other two thirds are made up of stars with spectra of very low S/N (which we discarded), spectra showing artefacts at the position of the probed wavelengths (in particular null flux at 3984\,\AA), several apparently chemically normal stars, a handful of Be stars, and about 60 CP2 stars. The high number of CP2 stars is not surprising, as these objects may also show overabundances of Mn, or, more often, other strong features at the probed wavelengths. This is illustrated in the bottom panel of Figure \ref{showcase2}, which shows the spectrum of the known CP2 star LAMOST J191430.53+402714.1 (spectral type B9 IV$-$V CrEu; \citealt{huemmerich20}) that was recovered with the MKCLASS\_HgMn code. The strong \ion{Eu}{ii} 4205\,\AA\ line, in particular, can be confused with \ion{Mn}{ii} 4206\,\AA. This actually happened with a significant number of stars for which both a strong \ion{Eu}{ii} 4205\,\AA\ and a \ion{Mn}{ii} 4206\,\AA\ lines were identified by the code (cf. Table \ref{table_MKCLASS_HgMn_output}). This confusing redundancy in output will be addressed by a future update of the code. Furthermore, strong metal lines in the spectral region from 4248\,\AA\ to 4262\,\AA\ may be misidentified as \ion{Mn}{ii} 4252/9\,\AA. In the case of LAMOST J191430.53+402714.1 (Figure \ref{showcase2}), we find strong absorption features centred on 4254\,\AA\ and 4262\,\AA. We assume that these are mostly caused by strong \ion{Cr}{i} 4254.331\,\AA, \ion{Cr}{ii} 4254.56\,\AA, and \ion{Cr}{ii} 4261.92\,\AA\ lines, which goes along well with the observed significant overabundance of Cr as judged from the corresponding strong lines at 3866\,\AA, 4111\,\AA, and 4172\,\AA.

\begin{figure*}
        \includegraphics[width=\textwidth]{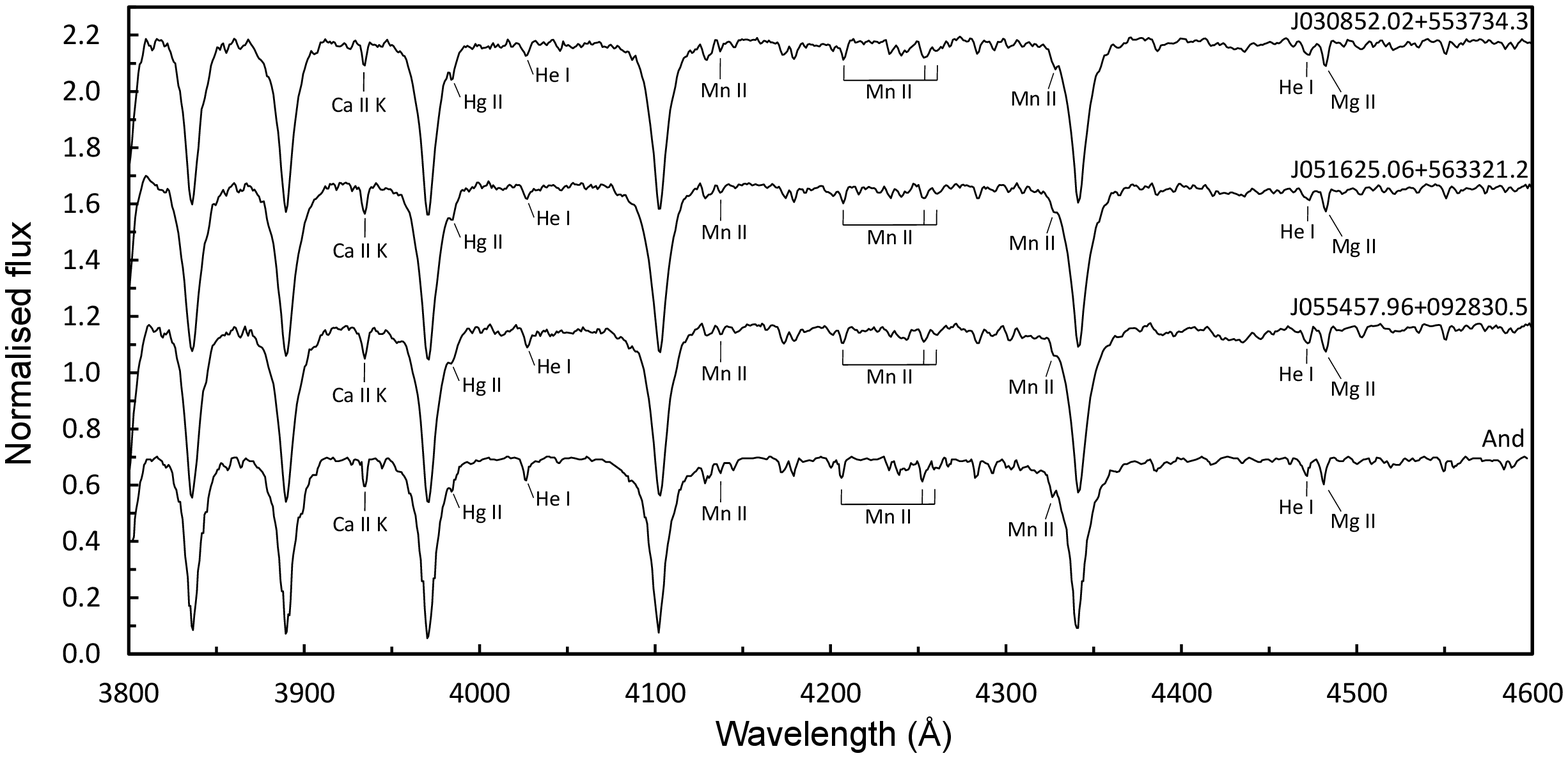}
    \caption{Showcase of three newly identified bona fide CP3 stars, illustrating the blue-violet region of the LAMOST DR4 spectra of (from top to bottom) LAMOST J030852.02+553734.3 (B8 IV HgMn), LAMOST J051625.06+563321.2 (B7 IV$-$V HgMn), and LAMOST J055457.96+092830.5 (B8 IV HgMn). The bottom spectrum is that of $\alpha$ And as adapted from Figure 4.16 of \citet{gray09}. The spectral types in parentheses correspond to the MKCLASS final types. Some prominent lines of interest are identified.}
    \label{showcase1}
\end{figure*}

\begin{table*}
\caption{Spectral classifications derived by the MKCLASS\_HgMn code. The columns denote: (1) LAMOST identifier. (2) 'raw' MKCLASS\_HgMn output using the standard star libraries \textit{libr18}, \textit{libnor36}, \textit{libsynth}, and \textit{liblamost}. (3)-(6) Number of detections $N$\textsubscript{det} of a peculiarly strong line at the specified wavelength (\AA) using all four libraries. (7) Number of detections across all investigated \ion{Mn}{ii} lines. (8) MKCLASS final type.}
\label{table_MKCLASS_HgMn_output}
\scriptsize{
\begin{center}
\begin{tabular}{|l|l|l|l|l|l|l|l|}
\hline
(1) & (2) & (3) & (4) & (5) & (6) & (7) & (8) \\
\hline
\textbf{LAMOST\_ID}	& \textbf{Output using \textit{libr18}/\textit{libnor36}/\textit{libsynth}/\textit{liblamost}}	& $N$\textsubscript{det}($\lambda$3984)	& $N$\textsubscript{det}($\lambda$4136)	& $N$\textsubscript{det}($\lambda$4206)	& $N$\textsubscript{det}($\lambda$4259)	& $N$\textsubscript{det}(Mn\_all)	& \textbf{SpT\_final} \\
\hline
J000118.65+464355.0	&	 A0 II-III  Eu4205 Hg3984 Mn4259  		&	1	&	0	&	0	&	1	&	1	&	B9.5 III HgMn	\\
	&	?	&		&		&		&		&		&		\\
	&	 B9 III  Eu4205  			&		&		&		&		&		&		\\
	&	 B9.5 III   			&		&		&		&		&		&		\\
\hline
J002635.94+562206.5	&	 A0 II-III  Eu4205 Hg3984 Mn4136 Mn4206  		&	2	&	2	&	2	&	0	&	4	&	B9 IV-V HgMn	\\
	&	 B8 V  Eu4205 Mn4206  		&		&		&		&		&		&		\\
	&	 B8 IV  Eu4205 Hg3984 Mn4136  		&		&		&		&		&		&		\\
	&	 B9 IV-V   				&		&		&		&		&		&		\\
\hline
J020922.72+533908.6	&	 B7 V  Eu4205 Mn4136 Mn4206  		&	2	&	3	&	3	&	0	&	6	&	B8 IV HgMn	\\
	&	 B8 IV  Eu4205 Hg3984 Mn4136 Mn4206  		&		&		&		&		&		&		\\
	&	 B8 IV  Eu4205 Hg3984 Mn4136 Mn4206  		&		&		&		&		&		&		\\
	&	 B9 III-IV   			&		&		&		&		&		&		\\
\hline
J030852.02+553734.3	&	 B8 III-IV  Eu4205 Hg3984 Mn4206 Mn4259  		&	3	&	0	&	4	&	4	&	8	&	B8 IV HgMn	\\
	&	 B7 III-IV  Eu4205 Hg3984 Mn4206 Mn4259  		&		&		&		&		&		&		\\
	&	 B8 III  Eu4205 Hg3984 Mn4206 Mn4259  		&		&		&		&		&		&		\\
	&	 B0 0  Mn4206 Mn4259  		&		&		&		&		&		&		\\
\hline
J031312.82+493559.8	&	 A0 II-III  Eu4205 Hg3984 Mn4136  		&	1	&	1	&	0	&	0	&	1	&	B9 III-IV HgMn	\\
	&	 ?     				&		&		&		&		&		&		\\
	&	 B9 III  Eu4205  			&		&		&		&		&		&		\\
	&	 B9 III-IV   			&		&		&		&		&		&		\\
\hline
J031943.49+561329.7	&	 B7 III  Eu4205 Mn4206  		&	1	&	0	&	3	&	0	&	3	&	B7 III HgMn	\\
	&	 B5 III  Hg3984 Mn4206  		&		&		&		&		&		&		\\
	&	 B7 III  Eu4205 Mn4206  		&		&		&		&		&		&		\\
	&	 ?     				&		&		&		&		&		&		\\
\hline
J032252.69+582806.7	&	 B6 IV  Mn4136 Mn4206 Mn4259  		&	0	&	2	&	4	&	4	&	10	&	B6 IV HgMn	\\
	&	 B7 III-IV  Mn4206 Mn4259  		&		&		&		&		&		&		\\
	&	 B8 III-IV  Eu4205 Mn4206 Mn4259  		&		&		&		&		&		&		\\
	&	 B6 IV  Mn4136 Mn4206 Mn4259  		&		&		&		&		&		&		\\
\hline
J033700.61+571139.0	&	 A0 II-III  bl4077 Eu4205 Hg3984  		&	1	&	0	&	0	&	0	&	0	&	B9 IV (HgMn)	\\
	&	 B9 III-IV  bl4077 Eu4205  		&		&		&		&		&		&		\\
	&	 B9 IV  bl4077 Eu4205  		&		&		&		&		&		&		\\
	&	 B9 IV  bl4077  			&		&		&		&		&		&		\\
\hline
J033939.69+473758.3	&	 A0 II  Eu4205 Mn4136 Mn4206  		&	2	&	3	&	3	&	0	&	6	&	B8 IV HgMn	\\
	&	 B8 IV  Eu4205 Hg3984 Mn4136 Mn4206  		&		&		&		&		&		&		\\
	&	 B9 II-III  Eu4205  		&		&		&		&		&		&		\\
	&	 B8 IV  Hg3984 Mn4136 Mn4206  		&		&		&		&		&		&		\\
\hline
J034903.76+454037.3	&	 B9 III-IV  Eu4205 Hg3984 Mn4259  		&	1	&	0	&	0	&	1	&	1	&	B9 IV HgMn	\\
	&	 B9 IV  Eu4205  			&		&		&		&		&		&		\\
	&	 B9 IV  Eu4205  			&		&		&		&		&		&		\\
	&	 B9 IV-V   				&		&		&		&		&		&		\\
\hline
\end{tabular}                                                                                                                                                          
\end{center}                                                                                                                                                                    
}                                                                                                                                                                       
\end{table*}

\section{Results} \label{results}

\subsection{Spectral classification} \label{spectral_classification}

For most stars, the spectral types derived with the four different spectral libraries are in good agreement. For the final classification, we followed the methodology of \citet{huemmerich20}. In summary, whenever a spectral type was derived more than once across the different spectral libraries, the most common spectral type was adopted as the final classification. If no common classifications existed, spectral types were favoured in the order \textit{liblamost} > \textit{libsynth} > \textit{libnor36} > \textit{libr18}. In the case of striking discrepancies, the corresponding spectra were visually inspected and the best-fitting type was adopted. For convenience, the final adopted classification is termed hereafter the 'MKCLASS final type'.

MKCLASS final types for all sample stars are presented in Table \ref{table_master1}. Bona fide CP3 stars, good CP3 star candidates, and CP3 star candidates can be identified by, respectively, the suffixes \textbf{HgMn}, \textbf{(HgMn)}, and \textbf{(HgMn:)}. Only three stars (LAMOST J052308.77+323128.3, LAMOST J055400.30+290112.2, and LAMOST J062503.02+033237.3) are common with the list of \citet{RM09}. However, none of these objects was classified as a CP3 star. They are identified and shortly commented on in Table \ref{table_master1}. Furthermore, stars contained in the sample of \citet{chojnowski20} are indicated by a corresponding footnote behind the LAMOST identifier in column two of this table. Figure \ref{showcase1} illustrates the spectra of three newly identified bona fide CP3 stars. 

\begin{figure}
        \includegraphics[width=\columnwidth]{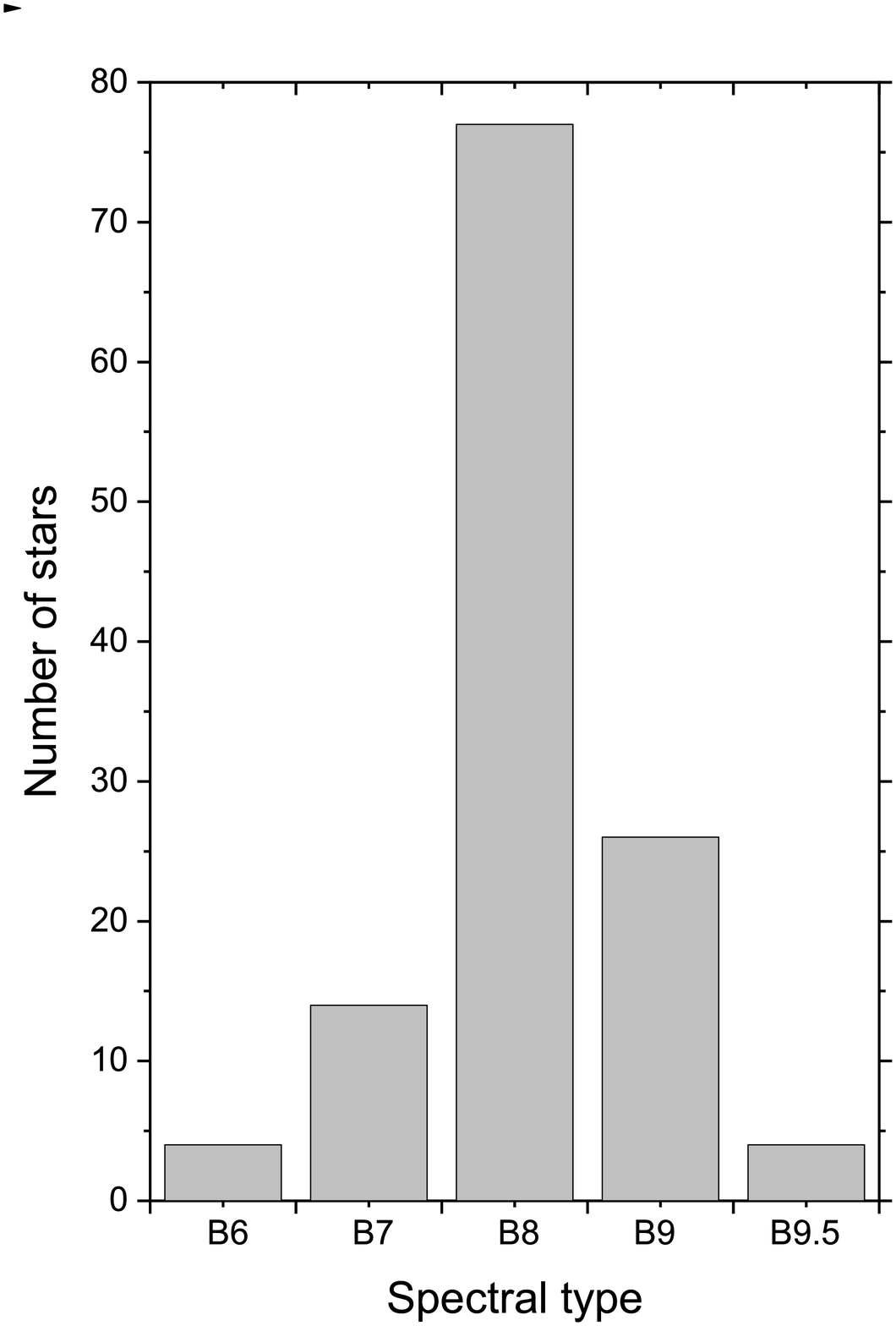}
    \caption{Histogram of the spectral temperature-type distribution of our sample stars.}
		\label{histogram_SpT}
\end{figure}

In accordance with the expectations for CP3 stars, all our sample stars are contained within the rather narrow spectral temperature-type range from B6 to B9.5 (Figure \ref{histogram_SpT}). This is in excellent agreement with the derived mass estimates (cf. Section \ref{evolutionary_status_mass_age}), which we interpret as independent proof of the applicability of our approach and the validity of the spectral classifications.

\begin{figure}
        \includegraphics[width=\columnwidth]{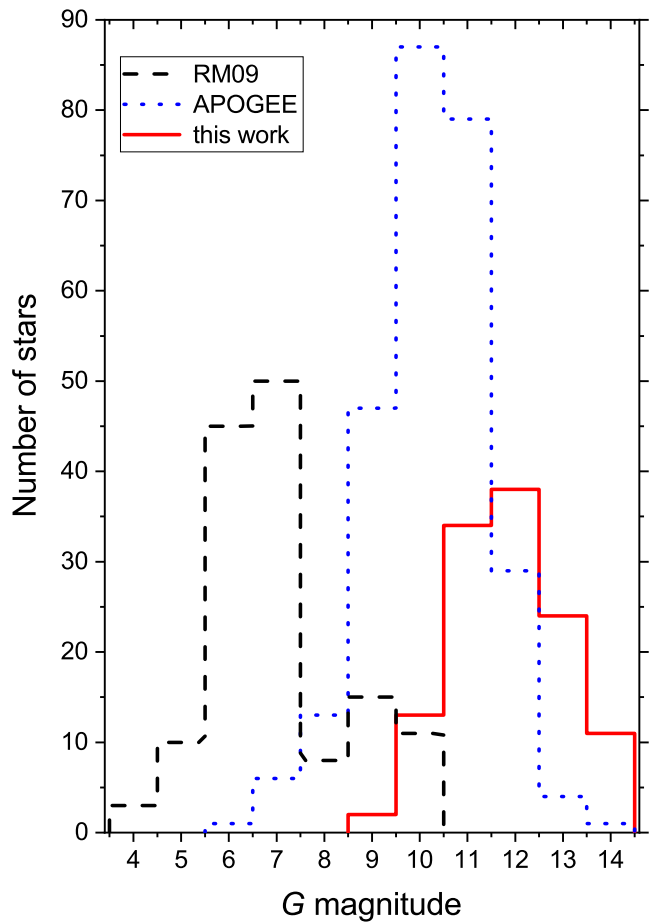}
    \caption{Histograms of the $G$ magnitudes for the CP3 stars from \citet{RM09} (black dashed line), \citet{chojnowski20} (blue dotted line), and our sample (red solid line).}
		\label{histogram_mag}
\end{figure}

\begin{figure}
        \includegraphics[width=\columnwidth]{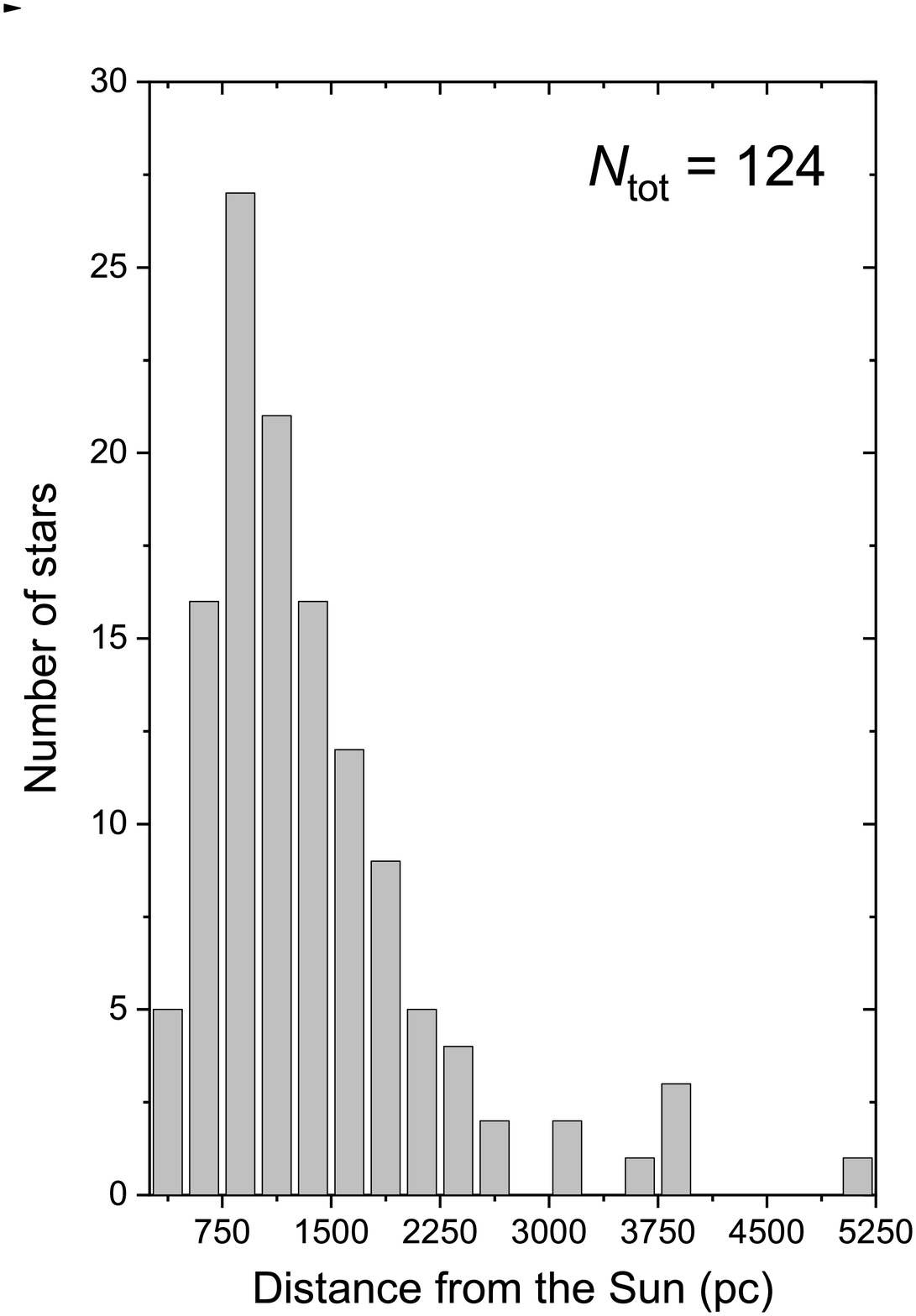}
    \caption{Histogram of the distances from the Sun. Parallax measurements from the $Gaia$ DR2 catalogue are available for all sample stars, with the single exception of LAMOST J031312.82+493559.8.}
		\label{histogram_distances}
\end{figure}

\subsection{Magnitudes, distances from the Sun, and cluster membership} \label{magnitudes_and_distances}

Histograms of the $G$ magnitudes and distances from the Sun are shown in Figures \ref{histogram_mag} and \ref{histogram_distances}. The magnitude distribution of our sample stars has a broad maximum between $G$ magnitudes 9.5 and 13.5. Our sample stars, therefore, are on average significantly fainter than the CP3 stars listed in the catalogues of \citet{RM09} and \citet{chojnowski20} (cf. Figure \ref{histogram_mag}). Most of our sample stars have distances between 500\,pc and 2.5\,kpc. There are only five objects within 500\,pc from the Sun and nine objects with distances exceeding 2.5\,kpc. In summary, our sample is a perfect extension to the CP3 stars listed in the above mentioned catalogues.

Several CP3 stars are known to be members of open clusters \citep[e.g.][]{hubrig12}, which allows a reliable age determination. LAMOST generally avoids dense regions of the sky such as star clusters. Nevertheless, we searched for possible cluster members among our sample stars within 3\,$\sigma$ of the positions, diameter, proper motions, distances and their errors of star clusters from \citet{2013A&A...558A..53K} and \citet{2018A&A...618A..93C}. Only one match was found: According to our data, LAMOST J041533.86+493944.1 belongs to the open cluster NGC 1528, an intermediate age (about 350\,Myr old) open cluster \citep{2013A&A...558A..53K} with no known CP stars. The evolutionary status of this star is further investigated in Section \ref{evolutionary_status_mass_age}.

\subsection{Evolutionary status} \label{evolutionary_status}

Here we consider the evolutionary status of our sample stars, discuss their properties in the $(BP - RP)_0$ versus $M_{\mathrm{G,0}}$ and mass versus fractional age on the main-sequence spaces, and investigate the influences of binarity and overall metallicity on our results.

\subsubsection{Colour-magnitude diagram} \label{evolutionary_status_CMD}

The colour-magnitude diagram (CMD) was constructed using the homogeneous Gaia DR2 photometry from \citet{gaia3}. As most of our sample stars are members of the Galactic disk and situated beyond 500\,pc from the Sun, interstellar reddening (absorption) needs to be taken into account. Unfortunately, Str{\"o}mgren-Crawford indices, which allow a reliable reddening estimation, are not available for most of our sample stars \citep{paunzen15}. We therefore relied on the reddening map of \citet{Green2018} and directly converted parallaxes to distances in order to interpolate within this map. Reddening values were transformed using the relations
\begin{equation}
E(B - V) = 0.76E(BP - RP) = 0.40A_G,
\end{equation}
which take into account the coefficients listed in \citet{Green2018}.

\begin{figure}
\begin{center}
\includegraphics[width=\columnwidth]{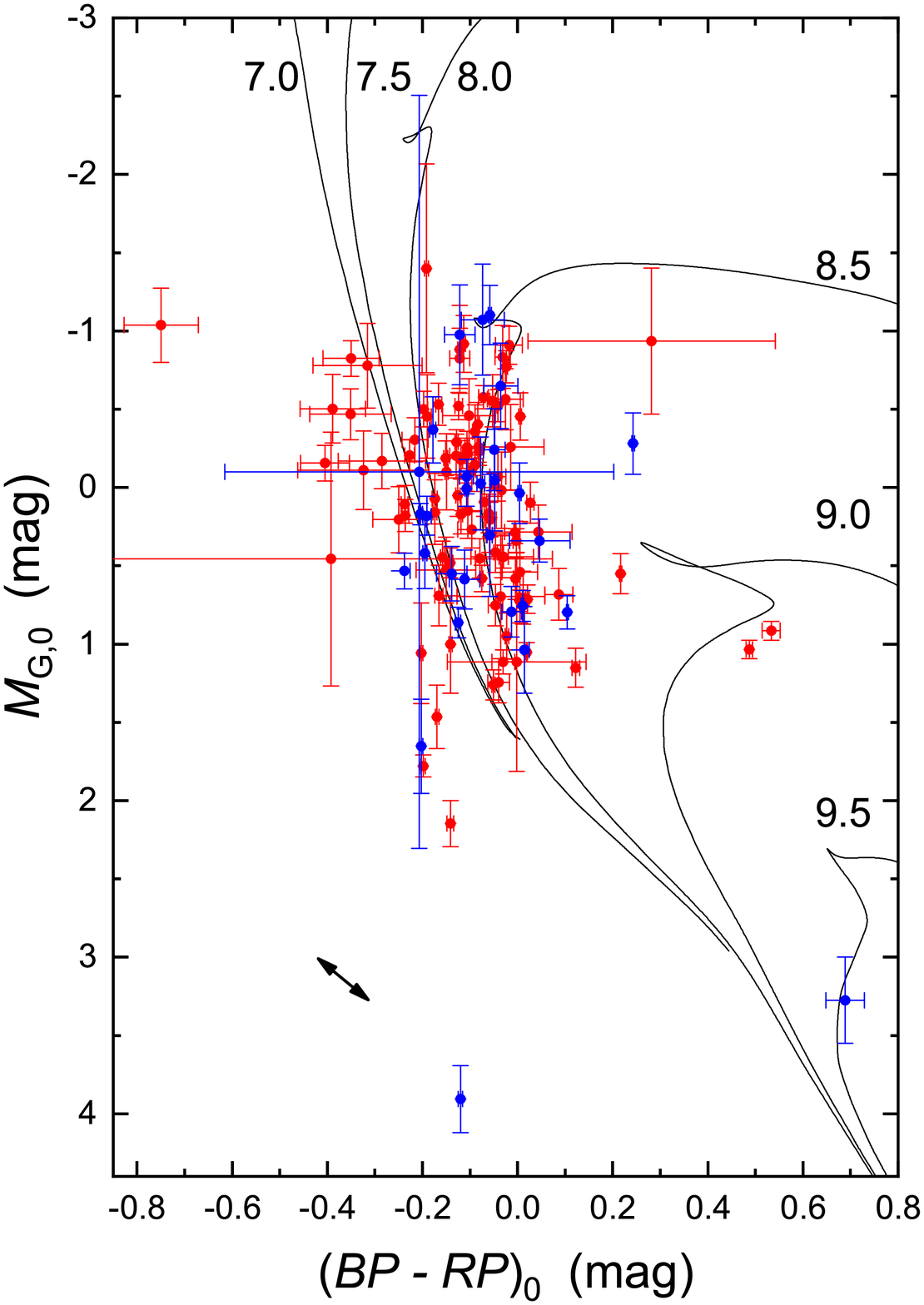}
\caption{$(BP - RP)_0$ versus $M_{\mathrm{G,0}}$ diagram of our sample stars. Also indicated are PARSEC isochrones for solar metallicity [Z]\,=\,0.02\,dex. Ages are given in logarithmic units. The arrow indicates the reddening vector according to an uncertainty of 0.1\,mag for $E(B-V)$. Red and blue symbols denote, respectively, bona fide CP3 stars and CP3 star candidates.}
\label{CMD} 
\end{center} 
\end{figure}

Figure \ref{CMD} illustrates the CMD of our sample stars, the reddening vector according to an uncertainty of 0.1\,mag for $E(B-V)$, and PARSEC isochrones \citep{bressan12} for solar metallicity [Z]\,=\,0.02\,dex.\footnote{We favour this value of solar metallicity because it has been shown to agree with recent results of helioseismology \citep{Vagnozzi2019}.} Considering the errors, eight stars are situated well below the zero-age main sequence (ZAMS). Their position cannot be explained by errors in the reddening estimation process and is likely due to inconsistent photometry. We are unable to shed more light on this matter with the available data.

The following three stars are characterised by their peculiar positions in the CMD and deserve further attention:

\textbf{LAMOST J033700.61+571139.0} (B9 IV (HgMn)): The derived age of about 3\,Gyr and the star's very red colour are not compatible with the available spectrum that clearly indicates a late B-type star and was classified accordingly by the MKCLASS\_HgMn code. The star is situated only about 260\,pc from the Sun, which rules out strong and underestimated reddening as cause of the observed aberrant colours. However, the spectral energy distribution (SED) is best fit with an effective temperature of 6000\,K. To exclude identification issues, we double checked the identifications of the LAMOST spectrum and the photometry employed for the SED fitting; no misidentifications were detected. In the literature, we also find widely different effective temperature estimates for this star: 5325\,K \citep{2006ApJ...638.1004A}, 6210\,K \citep{2019AJ....158..138S}, 6785\,K \citep{2019AJ....158...93B}, 7950\,K \citep{2019ApJS..245...34X}, and 9500\,K \citep{2018ApJS..235...16B}. The wide range of different temperature values points towards a spectroscopic binary system observed at different phases. Further spectroscopic data are needed to shed more light on the nature of this peculiar object.

\textbf{LAMOST J063246.02+030319.1} (B8 IV HgMn): The estimated visual absorption in the line-of-sight to this star amounts to $A_V$\,=\,2.42\,mag. A correction of about one magnitude would be necessary to shift its position to the ZAMS in the CMD. On close inspection, the reddening map of \citet{Green2018} reveals a significant source of absorption between 1.2 and 3.2\,kpc, with LAMOST J063246.02+030319.1 being situated just on the edge of the most dense region of the extinction source. According to \citet{2018AJ....156...58B}, the distance range for this star amounts to 1.2 to 1.4\,kpc. A direct conversion of the parallax yields a distance of 1.36\,\,kpc with an error of about 9\,\%, in agreement with the aforementioned result. We therefore assume that the line-of-sight extinction is significantly smaller than suggested by the reddening map, which is supported by the absorption value of $A_G$\,=\,1.18\,mag listed in the Gaia DR2 catalogue.  

\textbf{LAMOST J112400.77+540532.1} (B8 IV (HgMn:)): This star appears about three magnitudes too faint for its colour. It also shows a very large proper motion [$-$25.539, $-$5.957\,mas\,yr$^{-1}$] and is situated only about 350\,pc from the Sun. No further literature studies on this object are available. The star's location in the CMD would qualify it as a hot subdwarf (\citealt{2019MNRAS.482.3831P}, \citealt{2020A&A...635A.193G}). These objects are considered to be core helium-burning stars located at the blueward extension of the Horizontal Branch, the so called Extreme or Extended Horizontal Branch (EHB). To end up on the EHB, stars have to lose almost their entire hydrogen envelopes during the red-giant phase, which is most likely accomplished via binary mass transfer \citep{2016PASP..128h2001H}. Strongly peculiar hot subdwarfs are known \citep{2011MNRAS.412..363N}. LAMOST J112400.77+540532.1 shows several peculiarities, such as strong features at 4136\,\AA, 4206\,\AA\ and 4252/9\,\AA, which we interpret as being due to \ion{Mn}{ii}, but only a slight bump at 3984\,\AA. Consequently, the star was classified as a CP3 star candidate. We encourage a detailed abundance analysis using high-resolution spectroscopy to shed more light on the nature of this object.

\begin{figure}
\begin{center}
\includegraphics[width=\columnwidth]{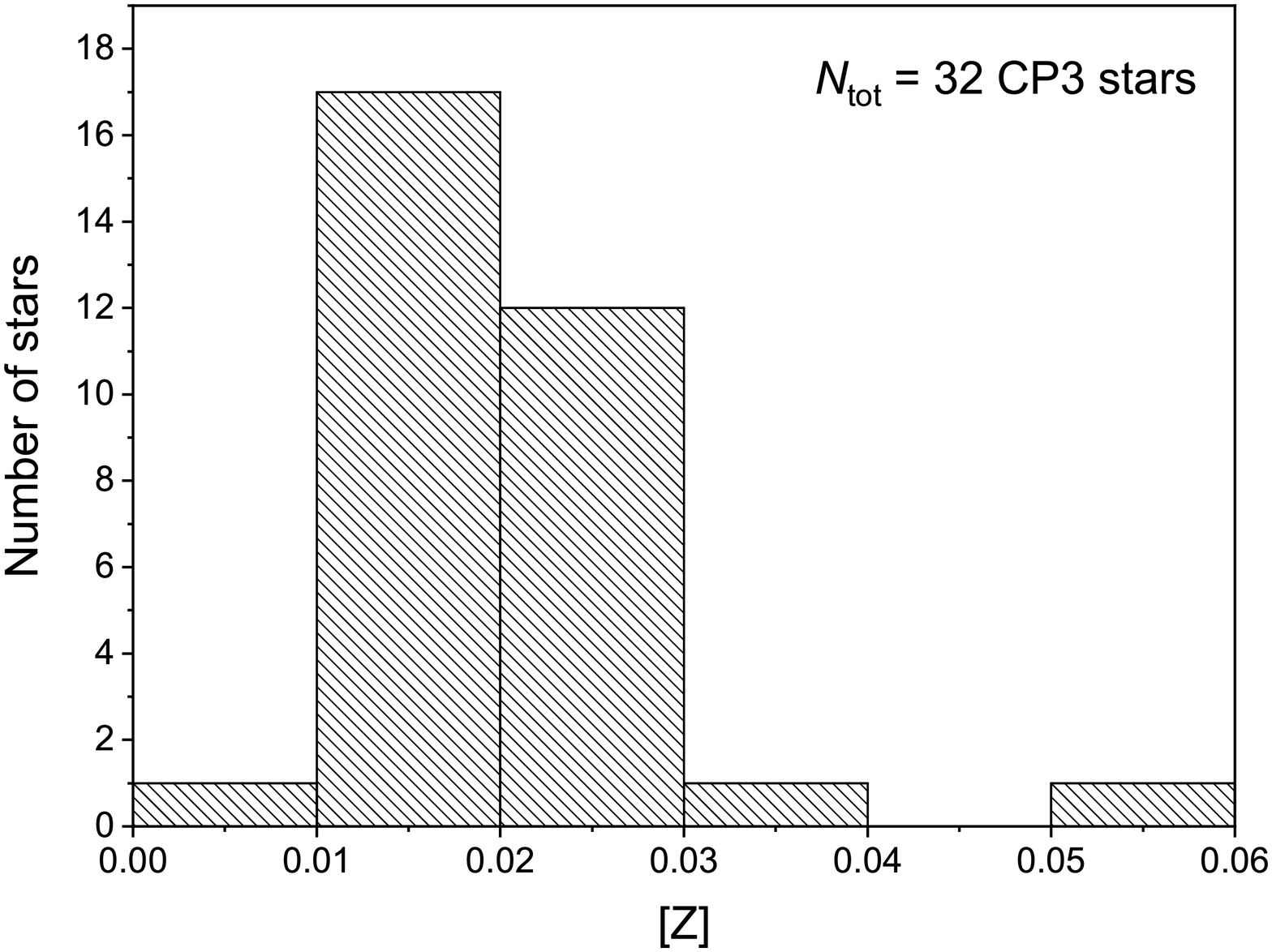}
\caption{Distribution of [Z] values for the CP3 stars from the \citet{ghazaryan18} catalogue with a least three measurements of C, N, O, and S. Almost all stars have an overall metallicity comparable to the solar value ([Z]\,=\,0.02\,dex).}
\label{plot_Z}
\end{center} 
\end{figure}

From the distribution of stars in Figure \ref{CMD}, we deduce that the majority of our sample stars is between 100\,Myr and 500\,Myr old. Several stars are obviously situated at or close to the ZAMS. To further interpret these findings, we need to consider the influences of multiplicity and the choice of the overall metallicity [Z] for the isochrones. The latter parameter has a significant influence on the derived ages (e.g. \citealt{bagnulo06}, \citealt{huemmerich20}).

\citet{schoeller10} announced that up to $\sim$90\,\% of all CP3 stars are actually part of binary systems. In classification resolution spectra, only the most extreme double-lined spectroscopic (SB2) systems can be detected. Therefore, we had to resort to other data to estimate the binarity rate among our sample stars.

As a first step, we investigated the Gaia DR2 astrometric solutions because it has been shown that a large error may indicate binarity \citep{2020AJ....159...33Z}. Furthermore, the Gaia DR2 catalogue includes a flag called 'duplicated\_source', which, if set to one, may indicate stellar multiplicity, but also observational, cross-matching, or processing problems. This flag has been set for 20 of our sample stars. We visually inspected the surroundings of these objects using the ALADIN sky atlas \citep{ALADIN1,ALADIN2} and found two stars (LAMOST J040155.24+514923.5 and LAMOST J062503.02+033237.3) with a comparably bright star in the close angular vicinity as well as four objects (LAMOST J050338.55+234349.5, LAMOST J053419.30+302407.7, LAMOST J060619.07+345118.1, and LAMOST J072016.47+141320.6) with much fainter neighbouring stars. As judged by this technique, the other fourteen stars have 'clean' surroundings.

We next investigated the parameter 'astrometric\_gof\_al' (gofAL), which denotes the goodness-of-fit statistic of the astrometric solution for the source in the along-scan direction. For the calculation of this parameter, the 'Gaussianised chi-square' statistic has been used, which should approximately follow a normal distribution with zero mean value and unit standard deviation for good fits. Values exceeding +3 indicate a bad fit to the data \citep{gaia2} and might indicate binarity. 113 sample stars show values exceeding this limit up to values of several hundreds. Interestingly, most of the objects with large gofAL values also have non-zero values for the parameters 'astrometric\_excess\_noise' and 'astrometric\_excess\_noise\_sig', which we interpret as a consistent and strong indication of binarity. In summary, using the outlined approach, only 12 objects (or 8\,\% of our sample) show no indication of binarity, which is fully in line with the binarity fraction estimate of \citet{schoeller10}. It is therefore imperative that the effects of binarity are taken into account when determining the masses and ages of our sample stars (and, for that matter, any other sample of CP3 stars).

The employed PARSEC isochrones use scaled abundances according to the abundance pattern of the Sun, that is, they were calculated assuming standard chemical composition. It is an open question if, or to what extent, this assumption can also be applied to the highly peculiar CP3 stars, which often show quite individualistic abundance patterns. If diffusion is accepted as the main mechanism behind the formation of the observed peculiar atmospheric compositions, in agreement with the current paradigm, we expect that the surface composition is not representative of the stellar interior, which should have an overall abundance close to solar. Furthermore, the current isochrone calculations assume a [Z] value for the whole star; no corresponding calculations are available that can handle different [Z] values for different stellar layers.

To evaluate the effect of metallicity on the results, we examined the [Z] distribution for CP3 stars from the \citet{ghazaryan18} catalogue having at least three measurements of the light elements C, N, O, and S, which contribute the most to the derived [Z] values. In total, 32 CP3 stars fulfil this criterion. Fig. \ref{plot_Z} illustrates the distribution of the derived [Z] values. Unfortunately, no constraints on the corresponding errors could be estimated because the errors of the individual abundance determinations are mostly not available in the reference source.  With the exception of only three stars, all objects show [Z] values between 0.01 and 0.03. Assuming a conservative error of about 10\,\%, this is in agreement with the assumption of solar abundance [Z]\,=\,0.02. Our choice of isochrones of solar abundance, therefore, seems justified and is not expected to introduce significant uncertainty.

\subsubsection{Mass versus age on the main sequence} \label{evolutionary_status_mass_age}

For theoretical considerations on the formation and evolution of CP3 stars, an accurate knowledge of their masses and ages is essential. It has already been emphasised that the majority of CP3 stars are contained within binary systems, a characteristic they share with the CP1 (Am/Fm) stars, with whom they may form a related sequence (cf. e.g. \citealt{smith74} and \citealt{wahlgren98}). Questions like whether CP3 stars are homogeneously distributed over the main sequence are important to understand the role of angular momentum transfer between the companions and the diffusion time scale in their atmospheres.

\begin{figure}
\includegraphics[width=\columnwidth]{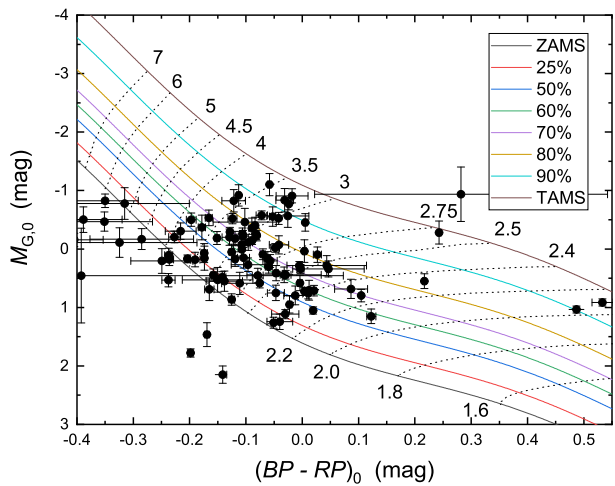}
\caption{$(BP - RP)_0$ versus $M_{\mathrm{G,0}}$ diagram including lines of constant $\tau$ and masses for solar metallicity [Z]\,=\,0.02.}
\label{plot_age_mass}
\end{figure}

In Fig. \ref{plot_age_mass}, we present the location of our sample stars in the $(BP - RP)_0$ versus $M_{\mathrm{G,0}}$ diagram, with the exception of the three outliers discussed in Section \ref{evolutionary_status_CMD}. Also included are lines of constant masses and constant $\tau$, which represents stellar age defined as the fraction of life spent on the main sequence, with the ZAMS corresponding to $\tau$\,=\,0\,\% and the terminal-age main sequence (TAMS) to $\tau$\,=\,100\,\%. Again, we used PARSEC isochrones \citep{bressan12} for solar metallicity [Z]\,=\,0.02. We stress that these isochrones were calculated for single stars and can here only serve as an approximation.

Lines of constant $\tau$ are not evenly spaced in terms of $M_{\mathrm{G,0}}$ and $(BP - RP)_0$. For instance, for a star with 3\,M$_\odot$, the same $M_{\mathrm{G,0}}$ interval of 0.45\,mag is covered in the intervals from 0\,\%\,$<$\,$\tau$\,$<$\,50\,\%, 50\,\%\,$<$\,$\tau$\,$<$\,80\,\%, and 80\,\%\,$<$\,$\tau$\,$<$\,100\,\%. The difference in colour is unevenly spaced, too, with roughly the same intervals covered from 0\,\%\,$<$\,$\tau$\,$<$\,80\,\% and 80\,\%\,$<$\,$\tau$\,$<$\,100\,\%. For lower mass stars, such as 2.4\,M$_\odot$ objects, the colour difference from the ZAMS to the TAMS already amounts to 0.5\,mag. As a consequence, depending on a star's location in the CMD, the size of the errors in both observables significantly impacts the final results.

As an overall conclusion from Fig. \ref{plot_age_mass}, most stars have masses between 2.4 and 4\,M$_\odot$, which corresponds to spectral types between B9.5\,V and B6\,V. This is in excellent agreement with the derived spectral types; without a single exception, all sample stars are found in the specified range (cf. Figure \ref{histogram_SpT}). Furthermore, our sample stars occupy the whole age range from ZAMS to TAMS, with an apparent accumulation of objects between 50\,\%\,$<$\,$\tau$\,$<$\,80\,\%.

To examine the effects of multiplicity on our results, knowledge of the mass ratio ($q$) of the components is necessary. In the literature \citep[e.g.][]{1998CoSka..27..319R,hubrig12}, mass ratios between 0.45 and 0.98 have been published for well investigated systems including at least one CP3 star component. We note that these results might be observationally biased as low-mass main-sequence companions are more difficult to detect using, for example, radial velocity studies.

\begin{figure}
\includegraphics[width=\columnwidth]{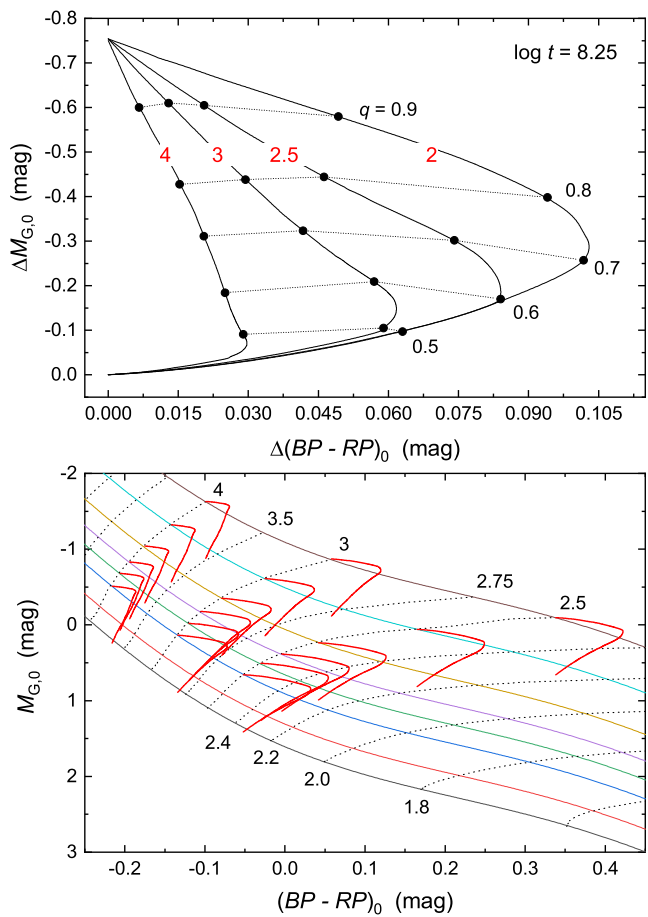}
\caption{Upper panel: Grids for different $q$ values, as calculated for primary masses of 4, 3, 2.5, and 2\,M$_\odot$ using an isochrone for $\log t$\,=\,8.25. Lower panel: Effect of binarity for different masses in the $(BP - RP)_0$ versus $M_{\mathrm{G,0}}$ diagram.}
\label{plot_effect_binarity}
\end{figure}

In the following considerations, only binary systems, that is, systems with two components, were considered. For equal masses, the maximum expected effect amounts to 0.754\,mag in terms of absolute magnitude, in the sense that the individual components are actually 0.754\,mag fainter that an assumed companionless star in the $(BP - RP)_0$ versus $M_{\mathrm{G,0}}$ diagram. However, as becomes obvious from Figure \ref{plot_effect_binarity}, colour and absolute magnitude change depending on the mass of the primary, hence the situation is complex. The upper panel illustrates the grids for different $q$ values, as calculated for primary masses of 4, 3, 2.5, and 2\,M$_\odot$ using an isochrone for $\log t$\,=\,8.25. The effect becomes noticeable in $M_{\mathrm{G,0}}$ for $q$\,$>$\,0.5 and reaches values of up to 0.1\,mag in $(BP - RP)_0$. In consequence, without the knowledge of
$q$, a correction for the effects of binarity is impossible to do on solid grounds.

The red curves in the lower panel of Figure \ref{plot_effect_binarity} show the situation for different masses and $\tau$ values. During the 'horizontal phase' of the curves (up to $q$\,$<$\,0.6 for $M$\,=\,2.5\,M$_\odot$), the positions of the stars initially shift to higher $\tau$ values and then to lower ones. As mentioned before, lines of constant $\tau$ are not evenly distributed in the $M_{\mathrm{G,0}}$ space, therefore the maximal effect for $q$\,=\,1 depends on $\tau$ itself. For instance, this can lead to decreases in the derived fractional ages from the TAMS to about $\tau$\,=\,88\,\%, from $\tau$\,=\,80\,\% to about $\tau$\,=\,55\,\%, and from $\tau$\,=\,50\,\% to the ZAMS. Derived masses generally decrease over the whole course of the curves; this effect can reach up to 15\,\% for the whole investigated mass range.

An independent test for binarity can be carried out using members of star clusters with known ages, which enables us to calculate $\tau$ for a given mass and check whether the location in the $M_{\mathrm{G,0}}$ versus $(BP - RP)_0$ diagram matches. Unfortunately, only LAMOST J041533.86+493944.1 was identified as member of the open cluster NGC 1528, which is approximately 350\,Myr old. Assuming this object to be a single star, we derive a mass of about 3.1\,M$_\odot$ and a fractional age of $\tau$\,$\approx$\,83\,\%. This is not compatible with the main-sequence lifetime of about 365\,Myr for the given stellar mass. We therefore conclude that LAMOST J041533.86+493944.1 is a binary star. Assuming $q$\,=\,0.91, we obtain $M$\,=\,2.8\,M$_\odot$ and $\tau$\,=\,73\,\%, in perfect agreement with the values derived from photometry.

Taking the above discussed aspects into account, it becomes obvious that the impact of binarity on the results is significant and cannot be ignored. However, with the available data, we are not able to investigate this issue in more detail. In consequence, because of the background of the expected high percentage of binary systems among our sample stars, no individual $\tau$ and mass values were determined. As general conclusions from our analysis, we find that the investigated CP3 stars cover the whole age range from ZAMS to TAMS and are almost homogeneously distributed at fractional ages of $\tau$\,$\le$\,80\,\%.

\subsection{Photometric variability} \label{photometric_variability}

CP3 stars were initially regarded as being among the 'quietest' (i.e. non-variable) early-type stars \citep{adelman98}. With the advent of high-precision space photometry, this picture has recently been changing. However, open questions as to the involved variability mechanisms (pulsation versus rotation) remain (cf. Section \ref{introduction}). We took advantage of the new sample of CP3 stars to identify variables using data from various photometric surveys. The following sections provide an overview over the employed data sources and shortly discuss the variable stars.

\subsubsection{Data sources} \label{photometric_variability_sub1}

For the investigation of the variability of our sample stars, photometric time series data from the following survey sources were employed.

\textbf{ASAS-3} \citep{pojmanski02}: two wide-field telescopes equipped with f/2.3 200\,mm Minolta lenses and 2048x2048 AP 10 Apogee detectors, situated at Las Campanas Observatory, Chile; $V$ band light curves available for the entire southern sky up to $\delta$\,$<$\,+28\degr; time baseline of almost 10\,yr; ASAS-3 archive\footnote{http://www.astrouw.edu.pl/asas/} contains photometry for stars in the magnitude range of 7\,$\la$\,$V$\,$\la$\,14.
	
\textbf{ASAS-SN} \citep{shappee14,kochanek17}: observations obtained at five stations, each consisting of four 14\,cm aperture Nikon telephoto lenses equipped with $V$ (two stations) or $g$ (three stations) band filters\footnote{https://asas-sn.osu.edu/}; entire visible sky is monitored every night to $V$\,$\la$\,17 mag; saturation limit 10 to 11\,mag ($V$); ongoing survey, time baseline of up to six years.
	
\textbf{$Kepler$ prime mission / K2} \citep{Kepler,Koch10}: prime mission operational from 2009 (May 2) until 2013 (May 8); after the loss of two reaction wheels, the mission (now called K2) was changed to observe the ecliptic plane \citep{Howell14}; differential photometer with 0.95\,m aperture; detectors made up of 21 modules each equipped with 2200x1024 pixel CCD detectors; single-passband (420-900\,nm) light curves taken in long-cadence (29.5\,min) and short-cadence (58.5\,s) modes; only PDC flux long-cadence data from the Mikulski Archive for Space Telescopes (MAST)\footnote{https://archive.stsci.edu/kepler/} have been employed in this study.

\textbf{SWASP} \citep{Pollacco06,Butters10}: two robotic telescopes situated at the Observatorio del Roque de los Muchachos, La Palma, and the South African Astronomical Observatory; each telescope consists of an array of eight f/1.8 200\,mm Canon lenses and 2048x2048 Andor CCD detectors; observing cadence 9 to 12\,min; observations unfiltered (prior to 2006) and obtained through a broadband filter (400-700\,nm; after 2006); single data release\footnote{https://wasp.cerit-sc.cz/form} contains light curves from 2004 to 2008 and provides good photometry in the magnitude range of 8\,$\la$\,$V$\,$\la$\,14.

\textbf{TESS} \citep{TESS3,TESS1,TESS2}: ongoing two-year all-sky survey aiming at the discovery of transiting exoplanets\footnote{https://heasarc.gsfc.nasa.gov/docs/tess/}; detectors consisting of four MIT/Lincoln Lab CCDs with 4096x4096 pixels (imaging area of 2048x2048 pixels; remaining pixels are used as a frame-store to allow rapid shutterless readout); cameras have an effective aperture size of 10\,cm and are equipped with f/1.4 lenses; single-passband (600-1000\,nm) light curves; time sampling of 2\,min.

\subsubsection{Photometric variables} \label{photometric_variability_sub2}

\begin{figure*}
\includegraphics[width=\columnwidth]{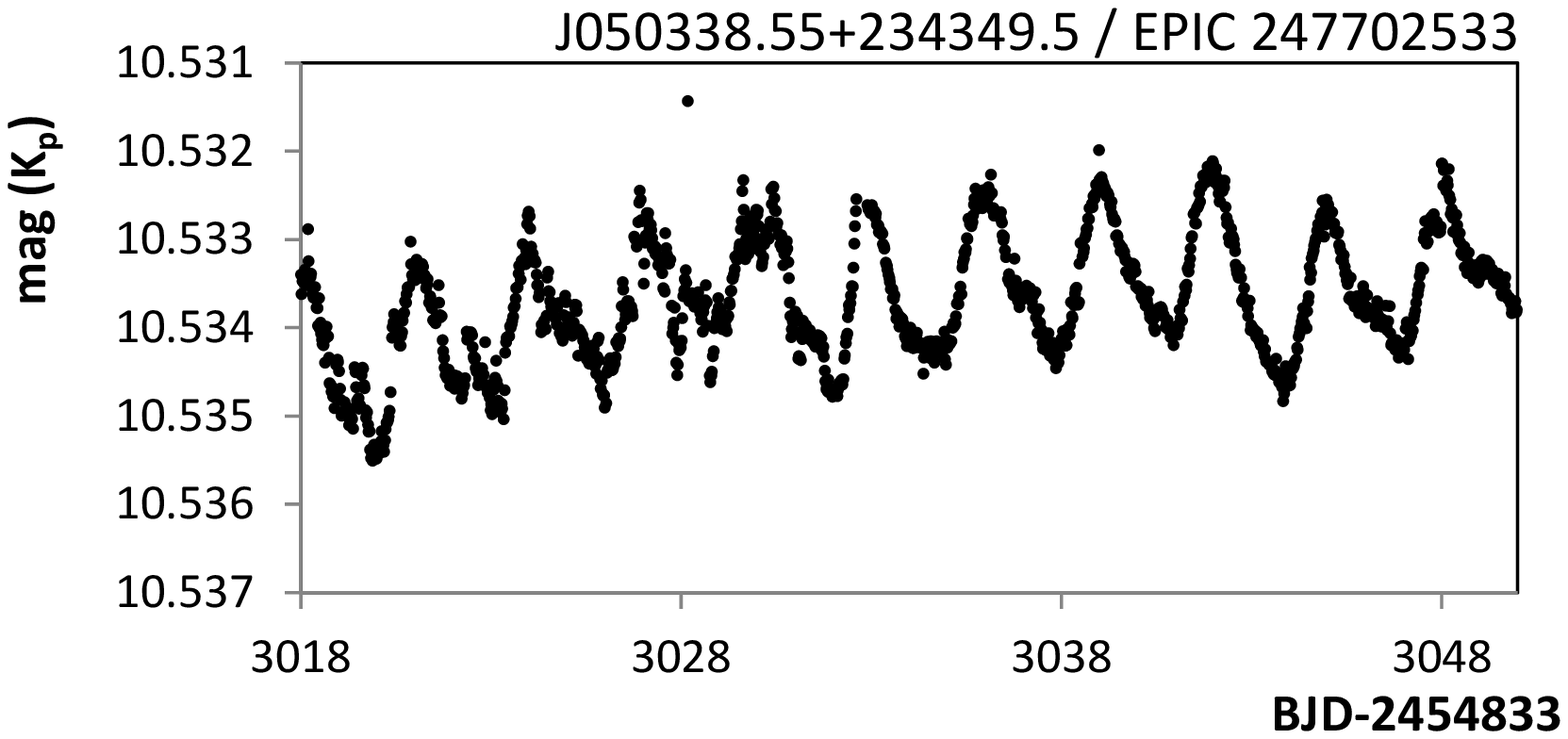}
\includegraphics[width=\columnwidth]{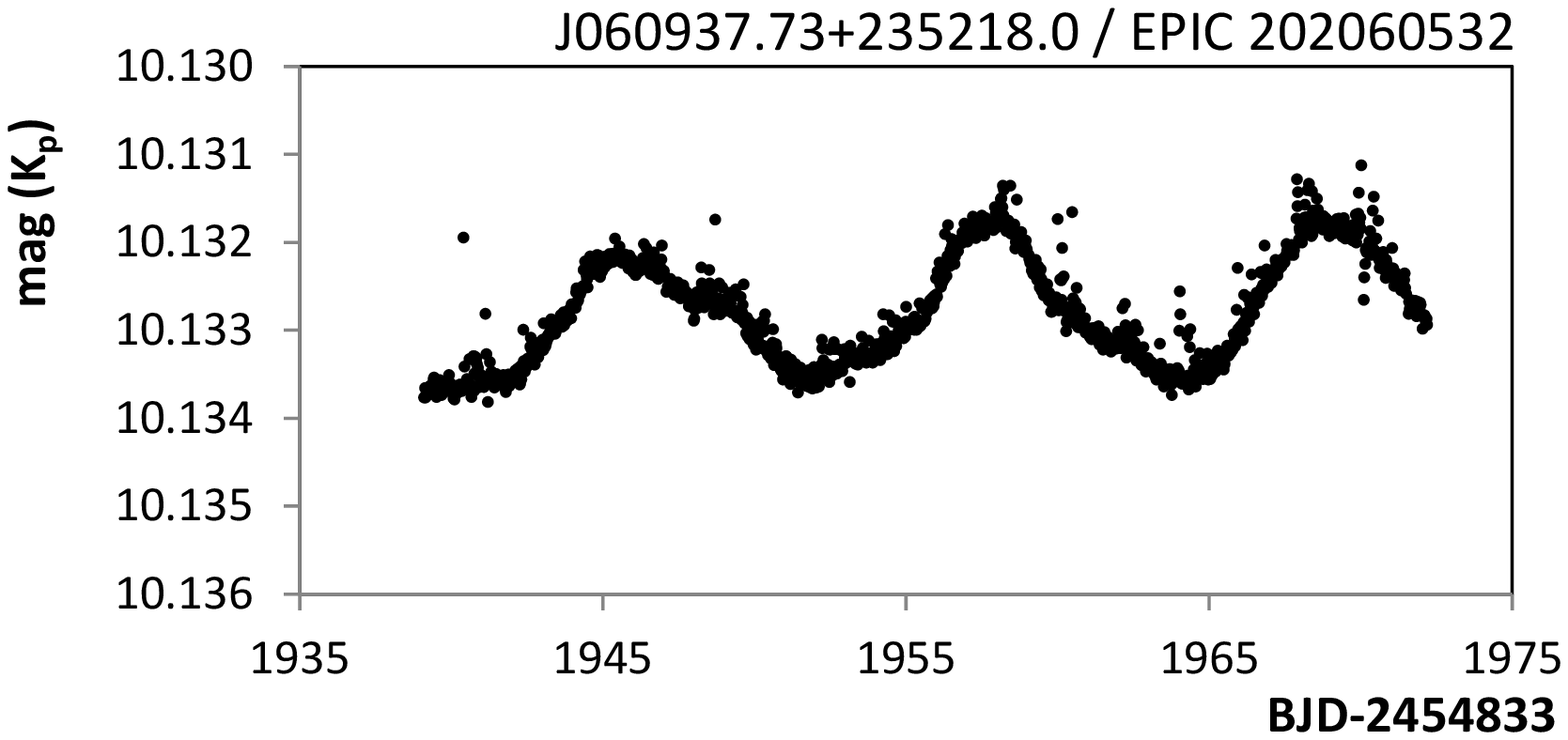}\\
\includegraphics[width=\columnwidth]{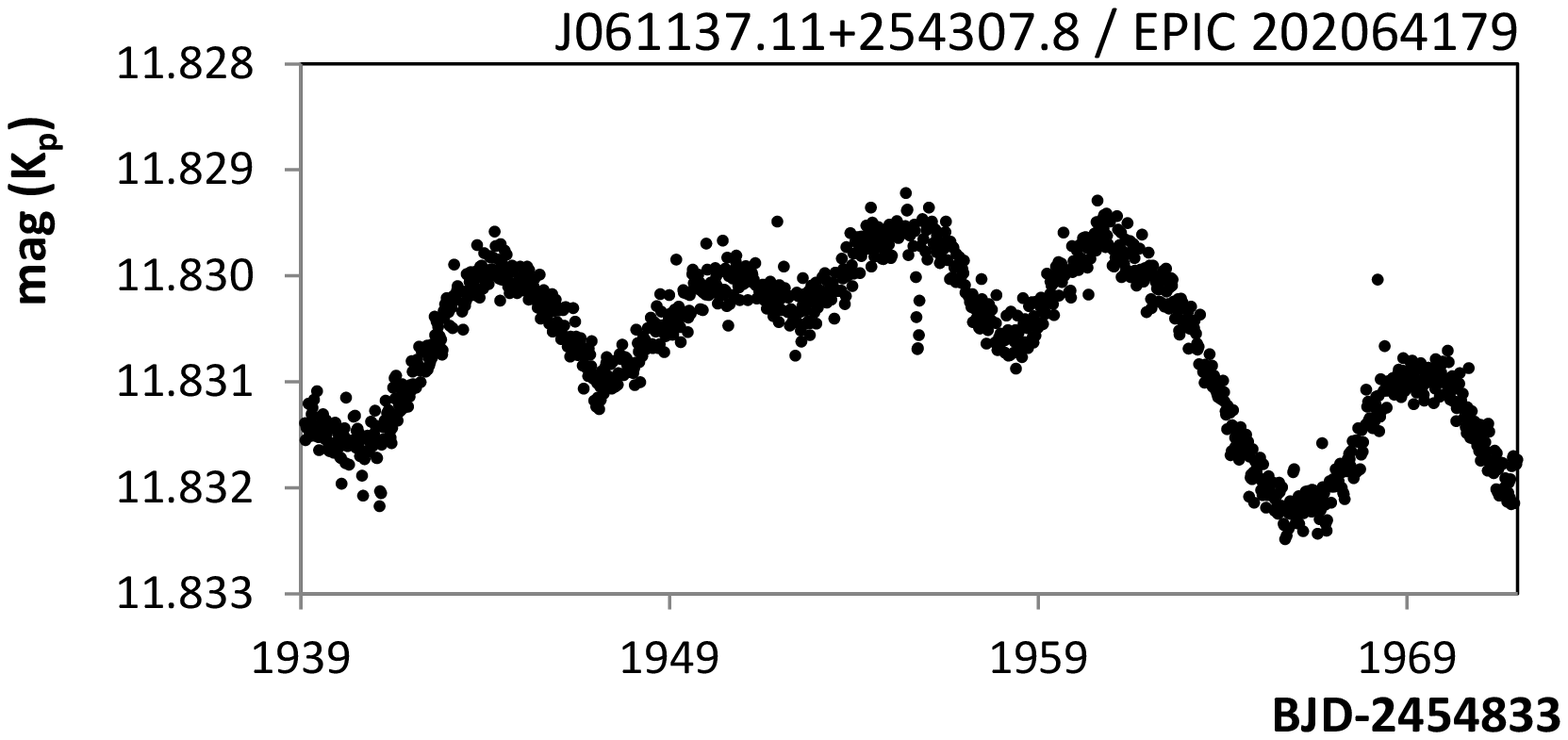}
\includegraphics[width=\columnwidth]{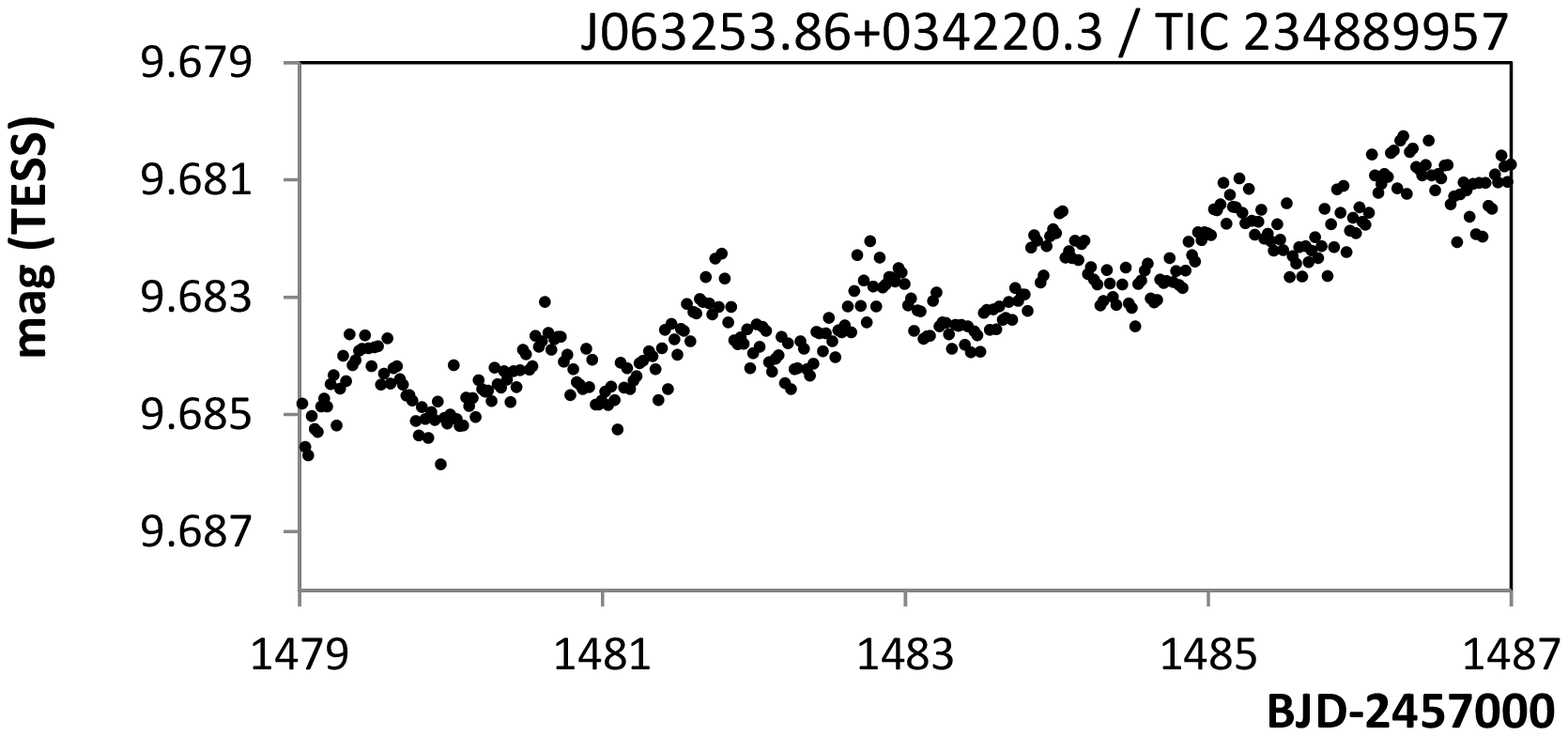}\\
\includegraphics[width=\columnwidth]{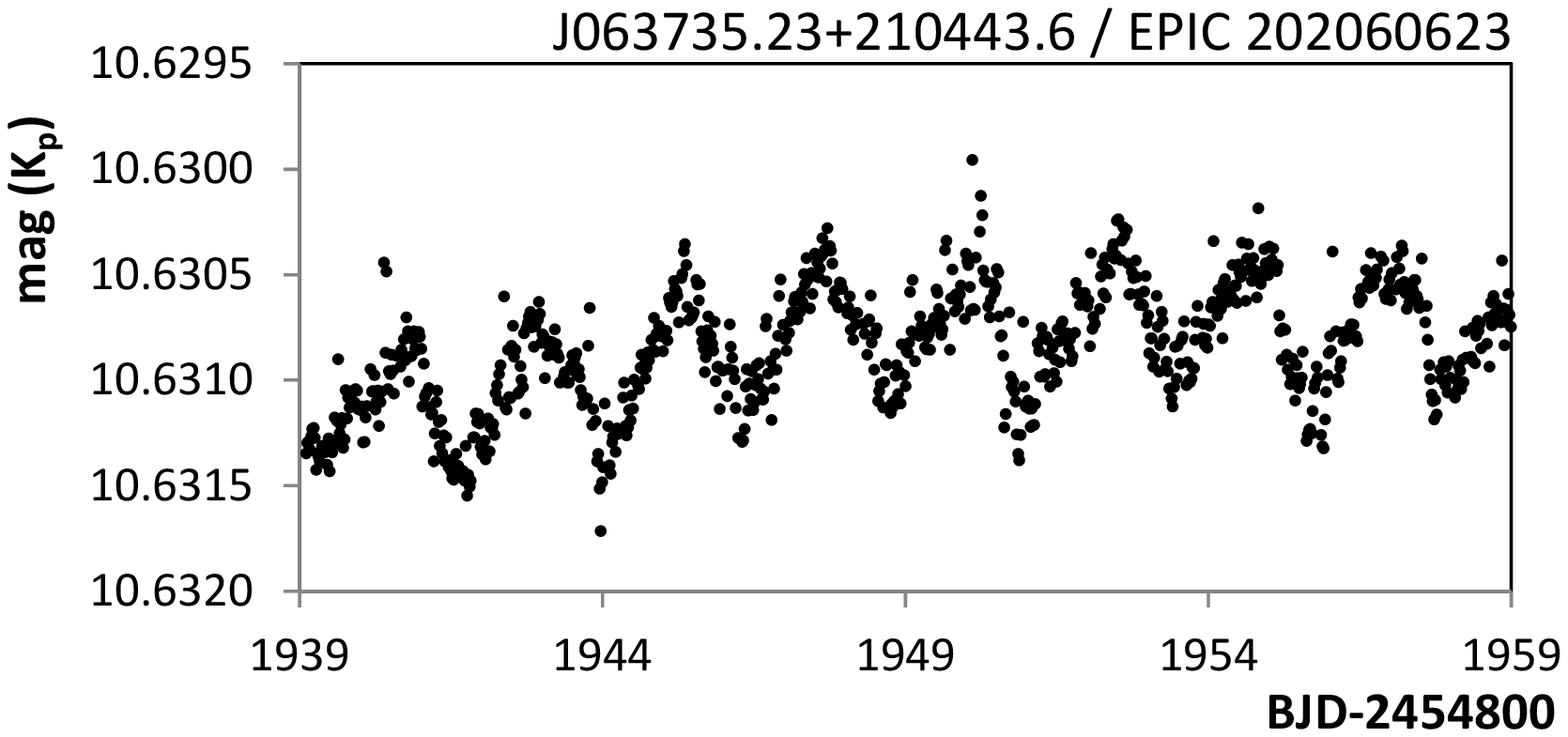}
\includegraphics[width=\columnwidth]{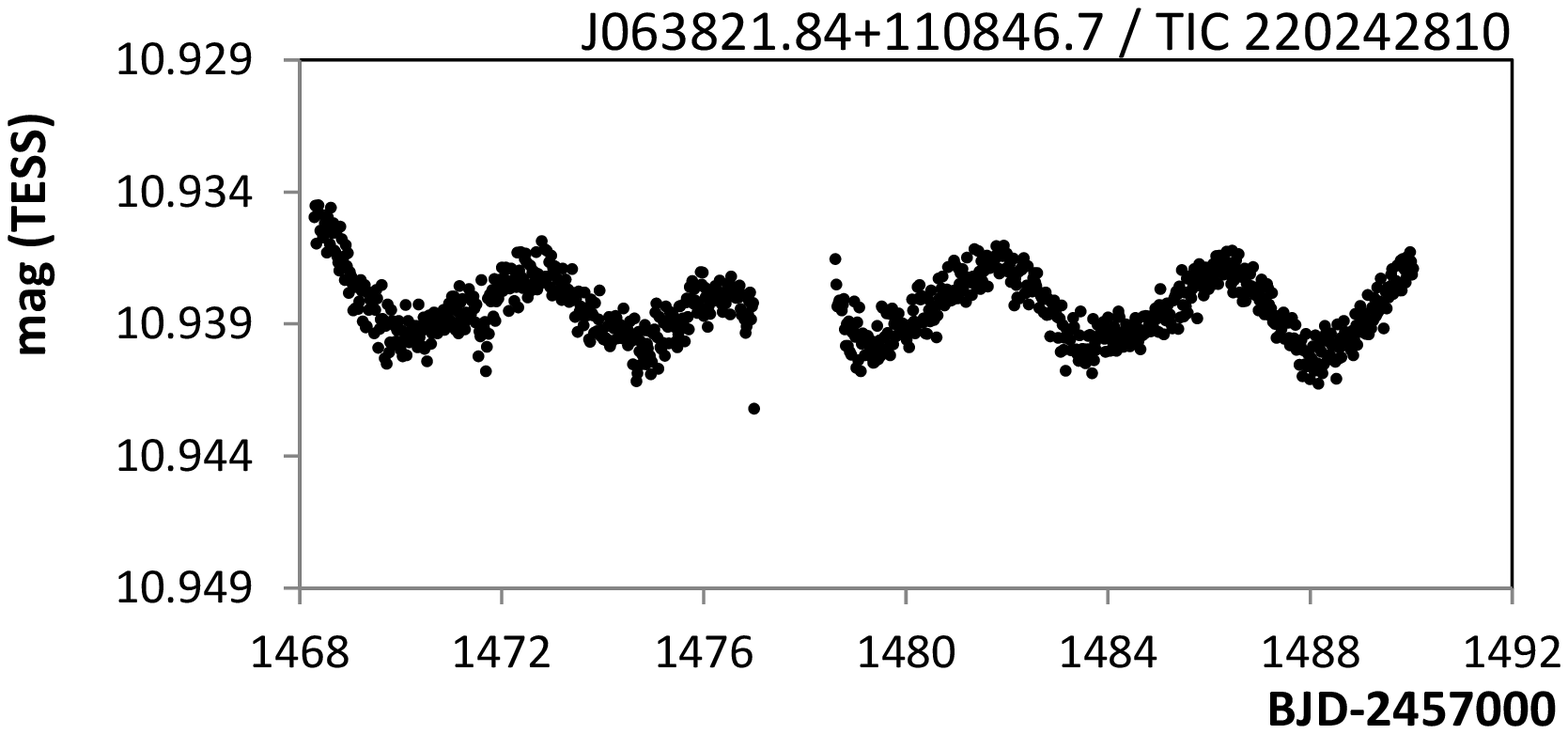}\\
\includegraphics[width=\columnwidth]{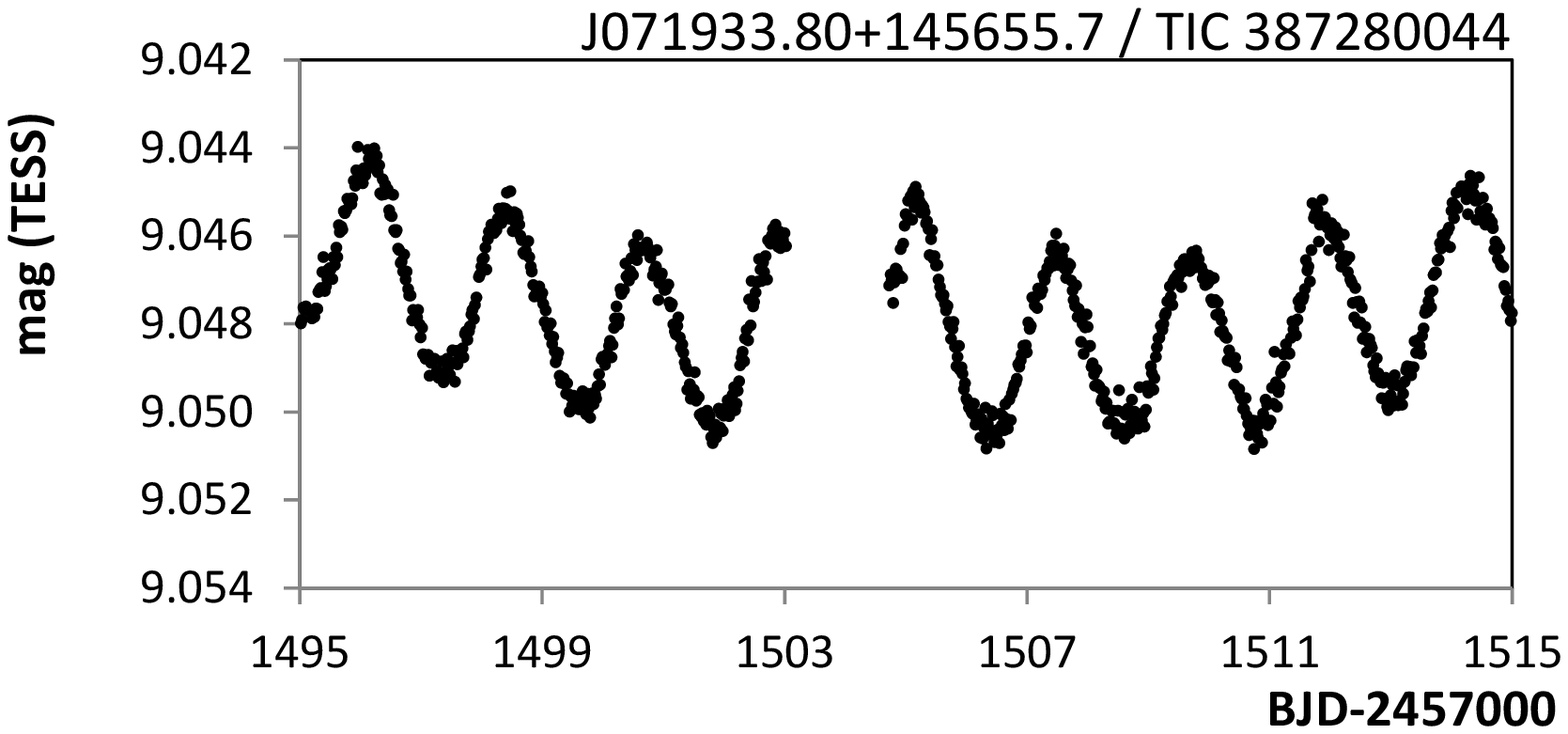}
\includegraphics[width=\columnwidth]{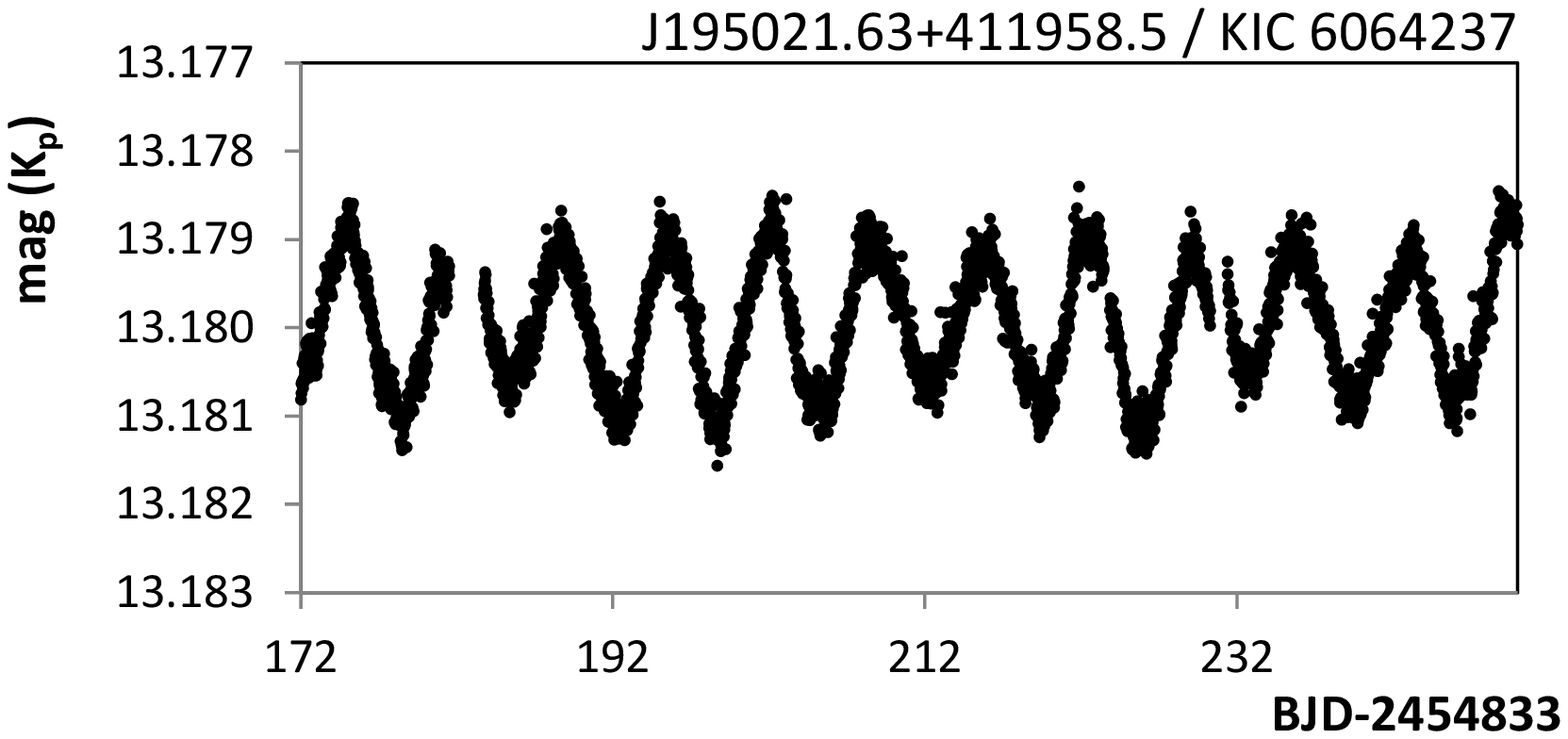}
\caption{Representative parts of the light curves of the eight CP3 stars identified as photometric variables in this study. Identifiers are indicated in the panels.}
\label{fig_LCs}
\end{figure*}

\begin{figure*}
\includegraphics[width=\columnwidth]{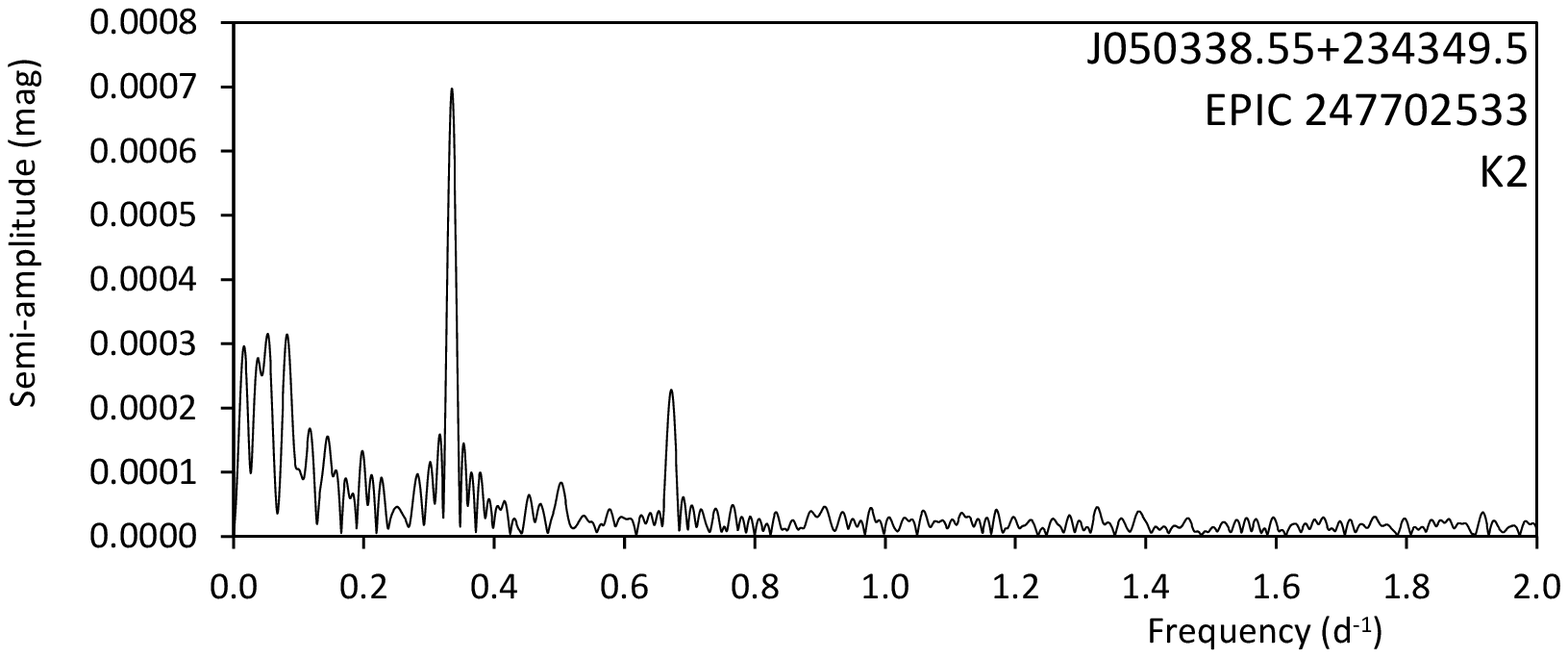}
\includegraphics[width=\columnwidth]{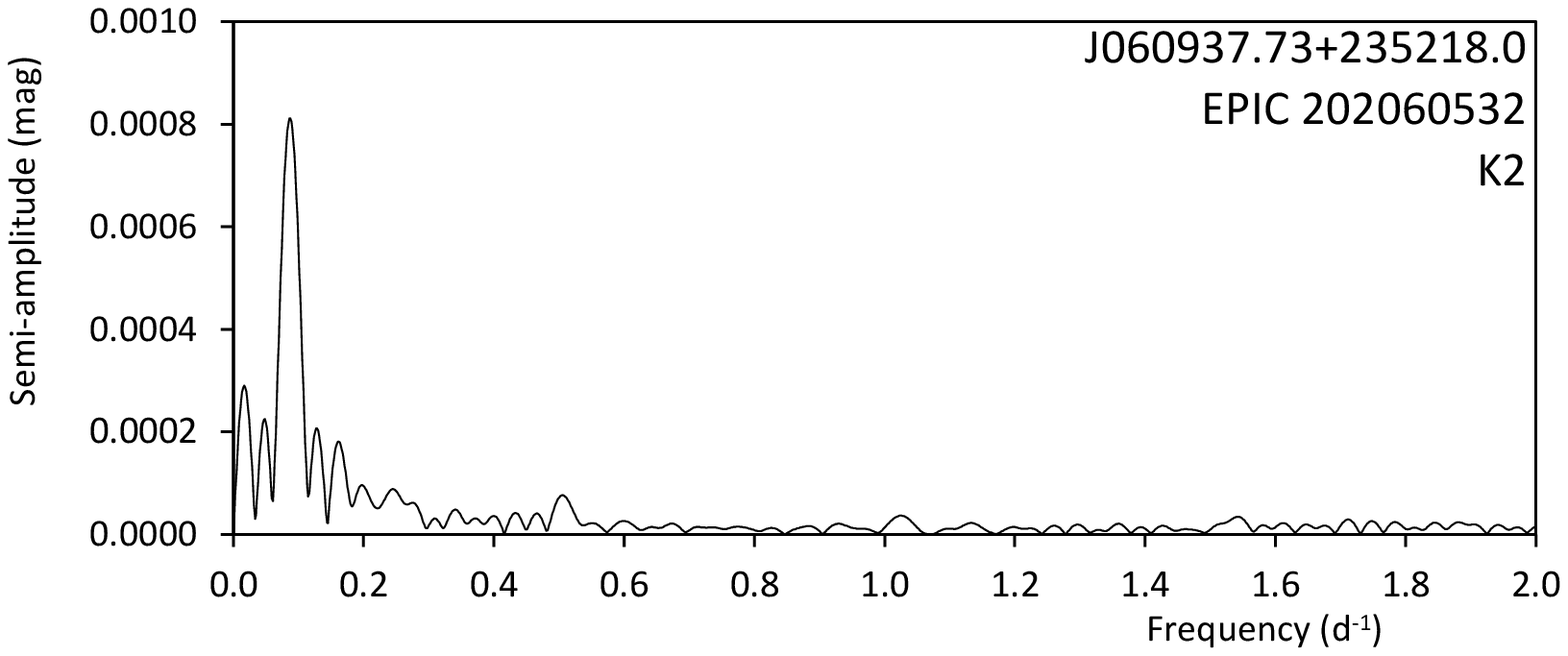}\\
\includegraphics[width=\columnwidth]{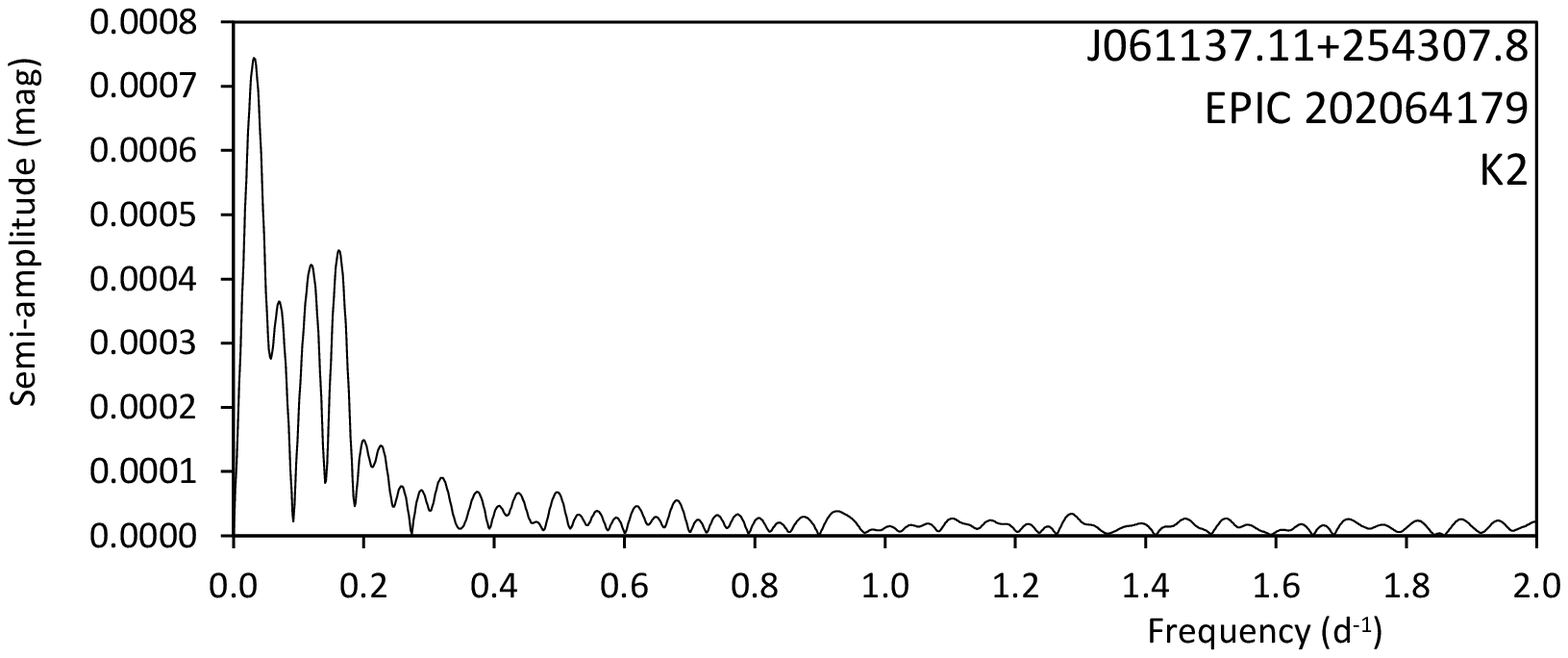}
\includegraphics[width=\columnwidth]{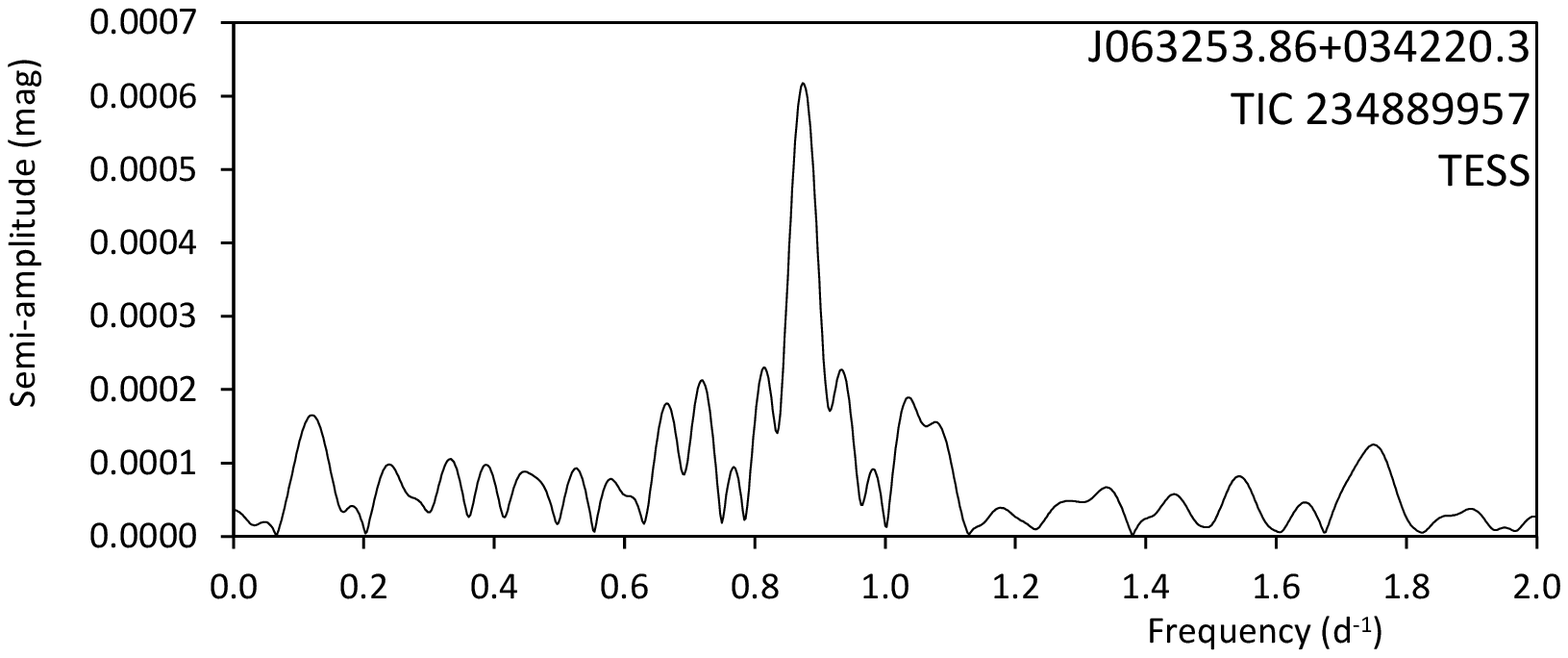}\\
\includegraphics[width=\columnwidth]{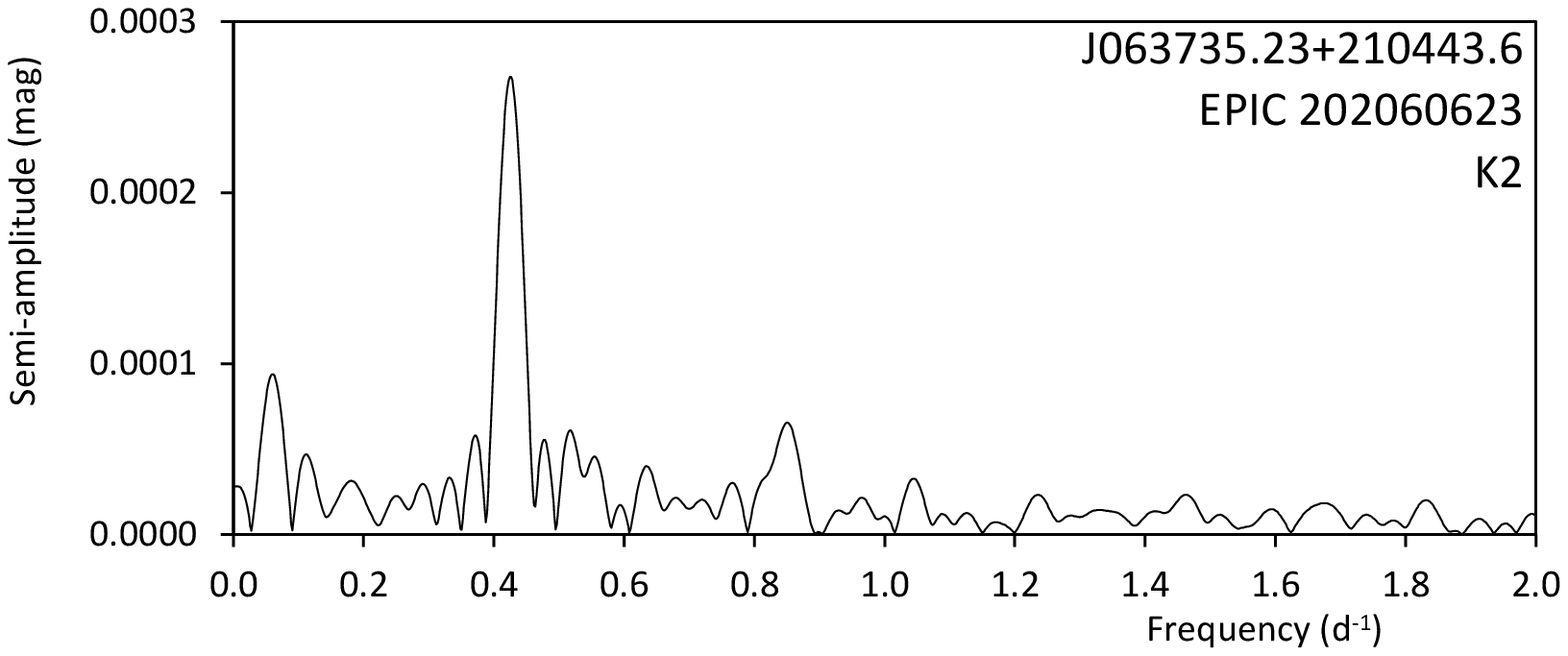}
\includegraphics[width=\columnwidth]{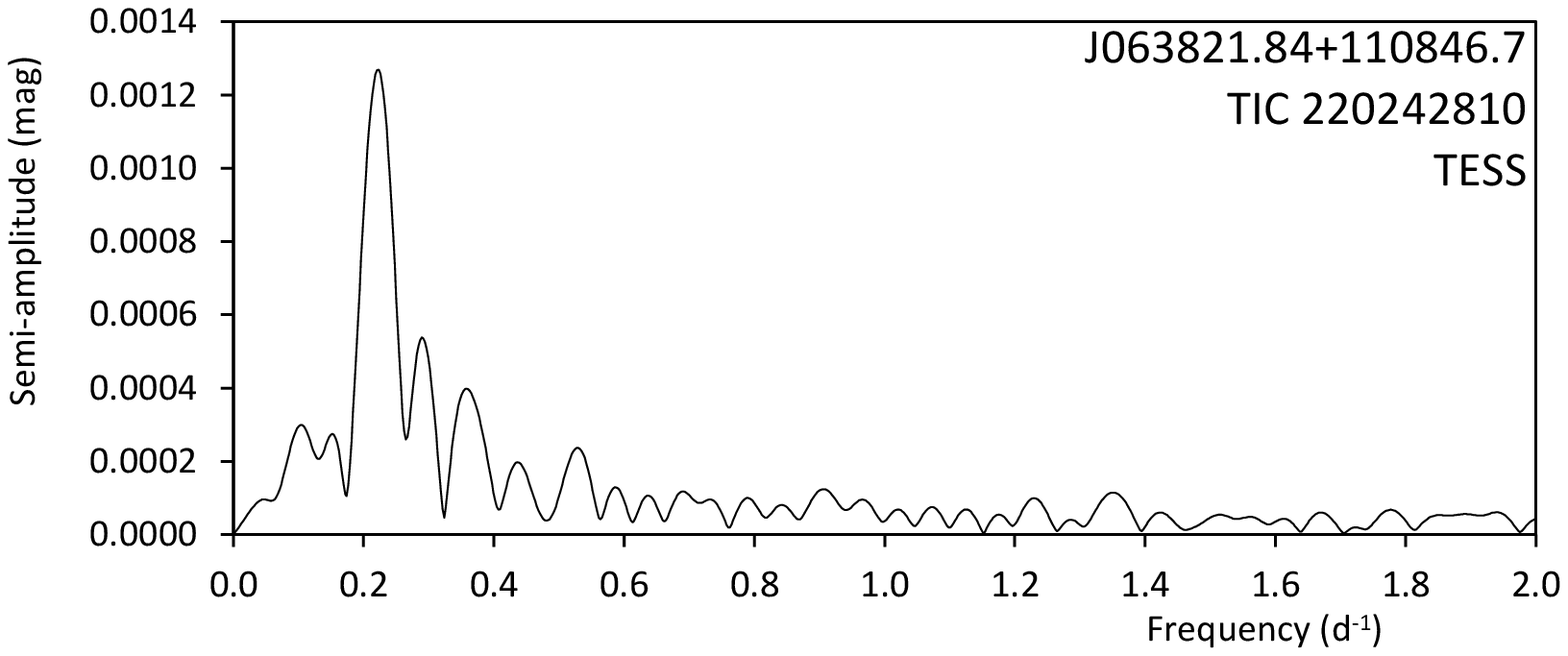}\\
\includegraphics[width=\columnwidth]{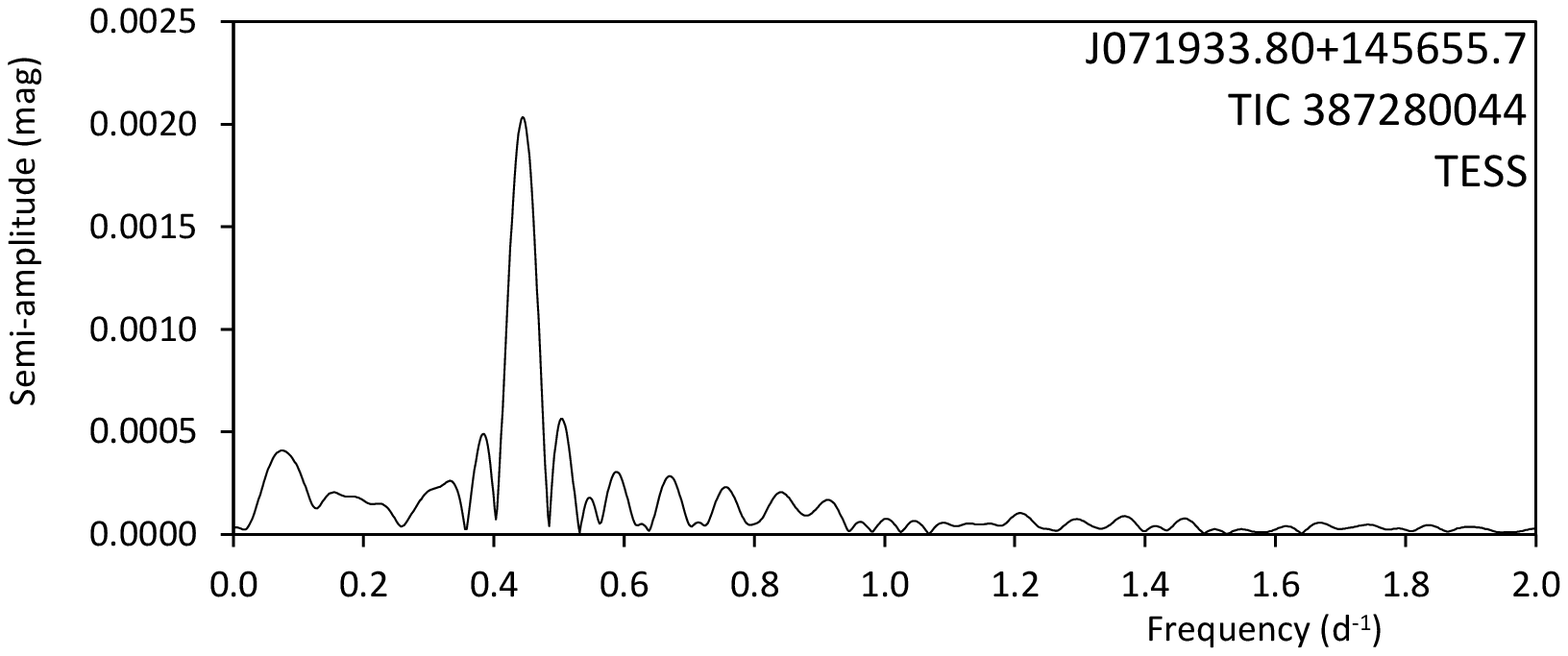}
\includegraphics[width=\columnwidth]{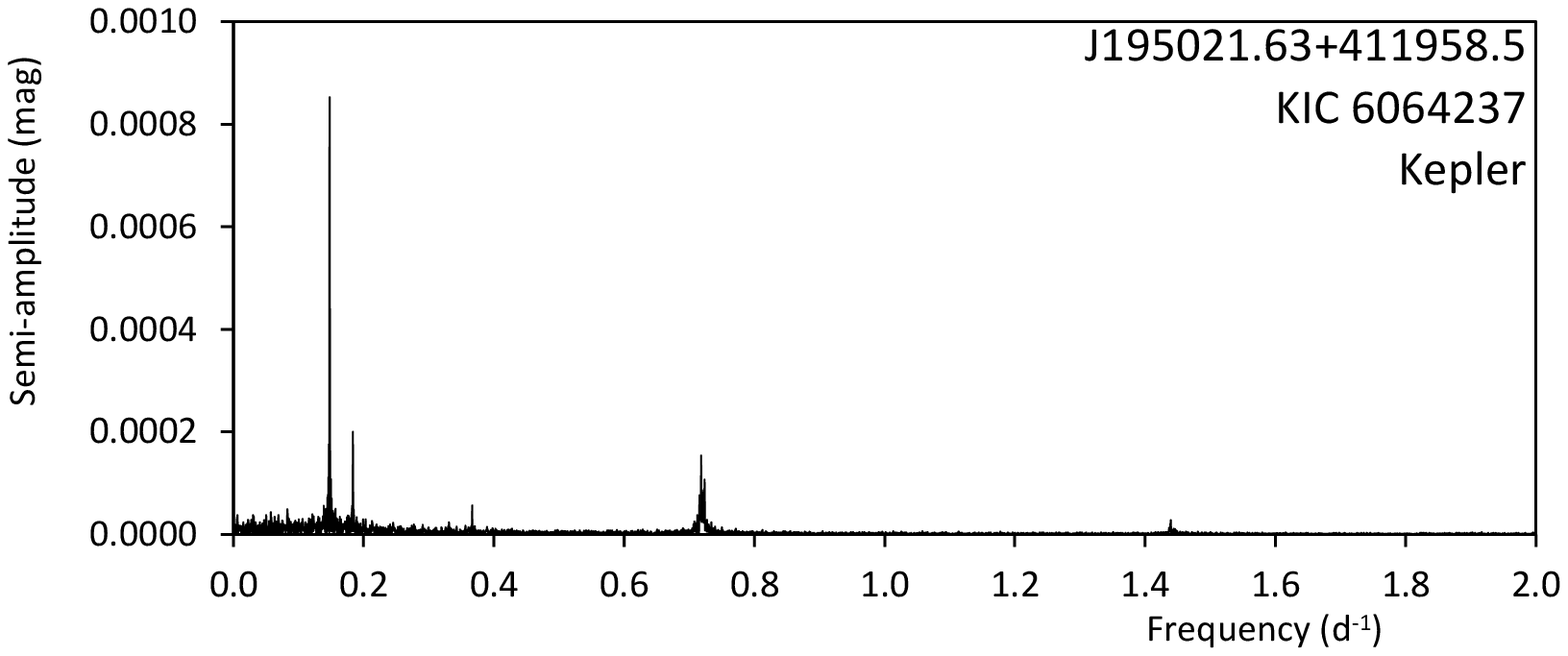}
\caption{Fourier spectra of the eight CP3 stars identified as photometric variables in this study. Identifiers and data sources are indicated in the panels.}
\label{fig_Fourier_spectra}
\end{figure*}

We identified eight photometric variable CP3 stars in the accuracy limits of the employed data sources. All variables were discovered in satellite-based data. The observed peak-to-peak amplitudes are so small that the variability would probably not have been detected in ground-based observations. Corresponding light curves and frequency spectra are shown, respectively, in Figures \ref{fig_LCs} and \ref{fig_Fourier_spectra}. Periods and uncertainties are provided in Table \ref{table_photometric_variables}. Upper limits of variability for the non-variable sources are provided in Table \ref{table_upperlimits}.

\textbf{LAMOST J050338.55+234349.5} (HD 285110; EPIC 247702533; MKCLASS final type: B8 IV	(HgMn)): This star is announced here as variable star for the first time. Based on analysis of $Kepler$ K2 data, we derive a period of $P$\,=\,2.99(2)\,d and a peak-to-peak variability amplitude of 0.0023\,mag\,($K$\textsubscript{p}). Only one significant period and corresponding harmonics are present in the Fourier spectrum, which is in line with rotational modulation. The light curve shows some irregularities, which are likely of instrumental origin.

\textbf{LAMOST J060937.73+235218.0} (HD 252428; EPIC 202060532; MKCLASS final type: B8 III$-$IV HgMn): This star was identified as a photometric variable on the basis of $Kepler$ K2 data by \citet{armstrong16}, who derived a period of $P$\,=\,9.901513\,d and a peak-to-peak variability amplitude of 0.002\,mag\,($K$\textsubscript{p}). No variability type was specified. Based on this information, the star is listed as variable star of unspecified type (type VAR) in the International Variable Star Index (VSX) of the American Association of Variable Star Observers (AAVSO) \citep{VSX}. However, the K2 data cannot be phased correctly with the period presented by \citet{armstrong16}, which is obviously wrong. From analysis of the same dataset, we derive a period of $P$\,=\,11.44(39)\,d (peak-to-peak variability amplitude of 0.0018\,mag\,($K$\textsubscript{p})), in agreement with the results of \citet{balona16}, who derived a period of 11.91(7)\,d for this star. The monoperiodic nature and the period of the variability suggest rotational variability but only three cycles have been covered, so more observations are needed.

\textbf{LAMOST J061137.11+254307.8} (TYC 1881-804-1; EPIC 202064179; MKCLASS final type: B8 IV$-$V HgMn): Based on analysis of $Kepler$ K2 data, this star was announced as a variable star of unspecified type by \citet{armstrong15}, who gave a period of $P$\,=\,6.248527\,d and derived a peak-to-peak variability amplitude of 0.007\,mag\,($K$\textsubscript{p}). \citet{balona16} investigated the same dataset and derived a period of 6.33(5)\,d. Our period solution ($P$\,=\,6.2(3)\,d; peak-to-peak variability amplitude of 0.0028\,mag\,($K$\textsubscript{p})) agrees with the above mentioned literature results. The star shows an interesting light curve and is obviously multi-periodic. We consider it a prime target for further studies aimed at deciphering the variability mechanisms at work in CP3 stars.

\textbf{LAMOST J063253.86+034220.3} (HD 46338; TIC 234889957; MKCLASS final type: B8 V (HgMn)): Based on analysis of TESS data, we here identify this star as a new variable star and derive a period of $P$\,=\,1.14(1)\,d and a peak-to-peak variability amplitude of 0.0014\,mag\,($TESS$). The monoperiodic nature of the light variability suggests rotational modulation. There is an obvious long-term trend in brightness, which may be of instrumental origin.

\textbf{LAMOST J063735.23+210443.6} (HD 260690; EPIC 202060623; MKCLASS final type: B9 III$-$IV	HgMn): This star was announced as a variable star on the basis of $Kepler$ K2 data by \citet{armstrong15}, who derived a period of $P$\,=\,2.38192\,d and a peak-to-peak variability amplitude of 0.002\,mag\,($K$\textsubscript{p}). \citet{balona16} derived a slightly but significantly different period of $P$\,=\,2.364(3)\,d, which, nevertheless, agrees with our period solution of $P$\,=\,2.34(2)\,d (peak-to-peak variability amplitude of 0.0007\,mag\,($K$\textsubscript{p})). The observed monoperiodic variability and the presence of harmonic frequencies are in agreement with rotational modulation.

\textbf{LAMOST J063821.84+110846.7} (HD 261050; TIC 220242810; MKCLASS final type: B8 III HgMn): Based on analysis of TESS data, we here identify this star as a new variable star and derive a period of $P$\,=\,4.47(14)\,d and a peak-to-peak variability amplitude of 0.0036\,mag\,($TESS$). The monoperiodic nature of the light variability suggests rotational modulation.
		
\textbf{LAMOST J071933.80+145655.7} (HD 56914; TIC 387280044; MKCLASS final type: B8 IV HgMn): Based on analysis of TESS data, we here identify this star as a new variable star and derive a period of $P$\,=\,2.25(3)\,d and a peak-to-peak variability amplitude of 0.0042\,mag\,($TESS$). The monoperiodic nature of the light variability suggests rotational modulation.

\textbf{LAMOST J195021.63+411958.5} (Gaia DR2 2076873634046444800; KIC 6064237; MKCLASS final type: B9 IV	HgMn): This star was first identified as a photometric variable on the basis of $Kepler$ Q1 data by \citet{debosscher11}. It has been proposed as a rotational variable in the literature \citep{balona13,nielsen13}. From analysis of the full $Kepler$ dataset, we derive a main period of $P$\,=\,6.7675(4)\,d and a peak-to-peak variability amplitude of 0.0024\,mag\,($K$\textsubscript{p}), in agreement with the results from former investigations ($P$\,=\,6.757\,d, \citealt{balona13}; $P$\,=\,6.768\,d, \citealt{nielsen13}). Interestingly, the star is obviously multi-periodic. We identify 11 independent and highly significant (S/N\,$>$\,7) frequencies, which are listed in Table \ref{table_KIC6064237}. The star is therefore a prime candidate for further studies as well as asteroseismological and pulsational modelling attempts.

\begin{table}
\caption{Periods of the eight photometric variable CP3 stars identified in this study. The columns denote: (1) LAMOST identifier. (2) Period derived in this study. (3) Literature period. The corresponding references are indicated in the notes.}
\label{table_photometric_variables}
\begin{tabular}{lll}
\hline
\hline
(1) & (2) & (3) \\
ID\_LAMOST & $P$\textsubscript{our}\,(d) & $P$\textsubscript{lit}\,(d) \\
\hline
J050338.55+234349.5 & 2.99(2) & $-$ \\
J060937.73+235218.0 & 11.44(39) & 9.901513$^{a}$, 11.91(7)$^{b}$ \\
J061137.11+254307.8 & 6.2(3)$^{1}$ & 6.248527$^{a}$, 6.33(5)$^{b}$ \\
J063253.86+034220.3 & 1.14(1) & $-$ \\
J063735.23+210443.6 & 2.34(2) & 2.38192$^{a}$, 2.364(3)$^{b}$ \\
J063821.84+110846.7 & 4.47(14) & $-$ \\
J071933.80+145655.7 & 2.25(3) & $-$ \\
J195021.63+411958.5 & 6.7675(4)$^{1}$ & 6.757$^{c}$, 6.768$^{d}$ \\
\hline
\multicolumn{3}{l}{References:} \\
\multicolumn{3}{l}{$^{a}$ \citet{armstrong16}, $^{b}$ \citet{balona16}, $^{c}$ \citet{balona13},} \\
\multicolumn{3}{l}{$^{d}$ \citet{nielsen13}} \\
\hline
\multicolumn{3}{l}{Note:} \\
\multicolumn{3}{l}{$^{1}$ multi-periodic} \\
\end{tabular}                                                                                                                                       
\end{table}

\begin{table}
\caption{Frequency solution for KIC\,6064237. The columns denote: (1) Frequency identification. (2) Frequency (d$^{-1}$). (3) Semiamplitude. (4) Phase, as derived
with \textsc{PERIOD04} (reference time = BJD-2454833). (5) Remark.}
\label{table_KIC6064237}
\begin{center}
\begin{tabular}{lllll}
\hline
\hline
(1) & (2) & (3) & (4) & (5) \\
No. & Frequency & Semiamplitude & Phase & Remark \\
\hline
$f_{1}$ & 0.147770	& 0.000859	& 0.383 &  \\
$f_{2}$ & 0.183330 & 0.000197 &	0.327 &  \\
$f_{3}$ & 0.718329 & 0.000171 &	0.277 &  \\
$f_{4}$ & 0.723315 & 0.000119 &	0.874 &  \\
$f_{5}$ & 0.719782	& 0.000096 & 0.248 &  \\
$f_{6}$ & 0.715784 & 0.000080 &	0.467 &  \\
$f_{7}$ & 0.722585 & 0.000082 & 0.179 &  \\
$f_{8}$ & 0.720130 & 0.000095 &	0.308 &  \\
$f_{9}$ & 0.366583 & 0.000054 &	0.044 & $2f_{2}$ \\
$f_{10}$ & 0.721365 & 0.000058 & 0.045 &  \\
$f_{11}$ & 0.030432 & 0.000034 & 0.006 & sidereal period? \\
\hline
\end{tabular}                                                                                                                                                             
\end{center}  
\end{table}

\section{Conclusions} \label{conclusion}

Using an altered version of Richard O. Gray's MKCLASS code, which was tailored to probe several \ion{Hg}{ii} and \ion{Mn}{ii} features relevant to the identification of CP3 stars, we searched for CP3 stars among the LAMOST DR4 spectra of a colour-preselected sample of early-type stars. The spectra of all selected candidates were visually inspected; non-CP3 objects were sorted out and the strength of the \ion{Hg}{ii} and \ion{Mn}{ii} features was estimated. Depending on this estimate, the resulting sample was divided into bona fide CP3 stars, good CP3 star candidates, and candidates.

The main findings of the present investigation are summarised in the following:

\begin{itemize}
	\item We present 99 bona fide CP3 stars, 19 good CP3 star candidates, and seven candidates. Most of these stars are new discoveries (only 15 stars are contained in the catalogue of \citealt{chojnowski20}). The resulting increased sample size of CP3 stars will greatly benefit in-depth statistical studies.
	\item In accordance with the expectations for CP3 stars, all our sample stars are contained within the narrow spectral temperature-type range from B6 to B9.5. This is in excellent agreement with the derived mass estimates, which denote masses between 2.4\,M$_\odot$ and 4\,M$_\odot$ for most of our sample stars, which are mostly between 100\,Myr and 500\,Myr old. 
	\item With a broad peak distribution between $G$ magnitudes 9.5 and 13.5, the sample presented here contains on average the faintest CP3 stars known, forming a perfect extension to the samples of \citet{RM09} and \citet{chojnowski20}.
	\item Our sample stars cover the whole age range from ZAMS to TAMS and are almost homogeneously distributed at fractional ages of $\tau$\,$\le$\,80\,\%, with an apparent accumulation of objects between 50\,\%\,$<$\,$\tau$\,$<$\,80\,\%.
	\item Using various indirect indicators of multiplicity, we find that only 12 objects (or 8\,\% of our sample) show no indication of binarity, in agreement with the expectation for this group of objects.
	\item We investigate the impact of binarity on the mass and age estimates and find it to be significant. We are, however, unable to investigate this issue in more detail with the available data. Because of the background of the expected high percentage of binary systems among our sample stars, no individual $\tau$ and mass values were determined. We caution that it is imperative that the effects of binarity are taken into account when determining the mass and age of any sample of CP3 stars.
	\item Using time series data from diverse photometric surveys, we investigate the photometric variability of our sample stars. Only eight photometric variables were discovered. Most of these stars show monoperiodic variability in agreement with rotational modulation. LAMOST J061137.11+254307.8 and LAMOST J195021.63+411958.5 show multi-periodic variability and are prime candidates for additional studies dealing with the origin of the photometric variability of this group of CP stars as well as asteroseismological and pulsational modelling attempts.
\end{itemize}

Future investigations will be concerned with expanding the approach presented here to other massive spectral databases, in order to further increase the sample size of Galatic CP3 stars.

\begin{acknowledgements}
We thank the referee for his/her valuable comments that helped to improve the paper. This work has been supported by the DAAD (project No. 57442043). The Guo Shou Jing Telescope (the Large Sky Area Multi-Object Fiber Spectroscopic Telescope, LAMOST) is a National Major Scientific Project built by the Chinese Academy of Sciences. Funding for the project has been provided by the National Development and Reform Commission. LAMOST is operated and managed by National Astronomical Observatories, Chinese Academy of Sciences. This work presents results from the European Space Agency (ESA) space mission Gaia. Gaia data are being processed by the Gaia Data Processing and Analysis Consortium (DPAC). Funding for the DPAC is provided by national institutions, in particular the institutions participating in the Gaia MultiLateral Agreement (MLA). The Gaia mission website is https://www.cosmos.esa.int/gaia. The Gaia archive website is https://archives.esac.esa.int/gaia. This paper makes use of data from the DR1 of the WASP data (Butters et al. 2010) as provided by the WASP consortium, and the computing and storage facilities at the CERIT Scientific Cloud, reg. no. CZ.1.05/3.2.00/08.0144 which is operated by Masaryk University, Czech Republic. This research has made use of "Aladin sky atlas" developed at CDS, Strasbourg Observatory, France.
\end{acknowledgements}

\clearpage

%
%

\bibliographystyle{aa}
\bibliography{lamost_HgMn_stars}

\appendix

\section{Essential data for our sample stars}

Table \ref{table_master1} lists essential data for our sample stars. It is organised as follows:

\begin{itemize}
\item Column 1: Internal identification number.
\item Column 2: LAMOST identifier.
\item Column 3: Alternativ identifier (HD number, TYC identifier, or GAIA DR2 number).
\item Column 4: Right ascension (J2000). Positional information was taken from GAIA DR2 \citep{gaia2,gaia3}.
\item Column 5: Declination (J2000).
\item Column 6: MKCLASS final type, as derived in this study.\footnote{We note that, as in the \citet{RM09} catalogue, the ’p’ denoting peculiarity was omitted from the spectral classifications.}
\item Column 7: Sloan $g$ band S/N of the analysed spectrum.
\item Column 8: $G$\,mag (GAIA DR2).
\item Column 9: $G$\,mag error.
\item Column 10: Parallax (GAIA DR2).
\item Column 11: Parallax error.
\item Column 12: Dereddened colour index $(BP-RP){_0}$ (GAIA DR2).
\item Column 13: Colour index error.
\item Column 14: Absorption in the $G$ band, $A_G$.
\item Column 15: Intrinsic absolute magnitude in the $G$ band, $M_{\mathrm{G,0}}$.
\item Column 16: Absolute magnitude error.
\end{itemize}

Upper limits of variability are provided in Table \ref{table_upperlimits}.

\setcounter{table}{0}  
\begin{sidewaystable*}
\caption{Essential data for our sample stars, sorted by increasing right ascension. The columns denote: (1) Internal identification number. (2) LAMOST identifier. (3) Alternativ identifier (HD number, TYC identifier or GAIA DR2 number). (4) Right ascension (J2000; GAIA DR2). (5) Declination (J2000; GAIA DR2). (6) Spectral type, as derived in this study. (7) Sloan $g$ band S/N ratio of the analysed spectrum. (8) $G$\,mag (GAIA DR2). (9) $G$\,mag error. (10) Parallax (GAIA DR2). (11) Parallax error. (12) Dereddened colour index $(BP-RP){_0}$ (GAIA DR2). (13) Colour index error. (14) Absorption in the $G$ band, $A_G$. (15) Intrinsic absolute magnitude in the $G$ band, $M_{\mathrm{G,0}}$. (16) Absolute magnitude error.}
\label{table_master1}
\begin{center}
\begin{adjustbox}{max width=\textwidth}
\begin{tabular}{lllcclcccccccccc}
\hline
\hline
(1) & (2) & (3) & (4) & (5) & (6) & (7) & (8) & (9) & (10) & (11) & (12) & (13) & (14) & (15) & (16) \\
No.	&	ID\_LAMOST	&	ID\_alt	&	RA(J2000) 	&	 Dec(J2000)    	&	SpT\_final	&	S/N\,$g$	&	$G$\,mag	&	e\_$G$\,mag	&	pi (mas)	&	e\_pi	&	$(BP-RP){_0}$	&	e\_$(BP-RP){_0}$	&	A${_G}$	&	M${_G}{_0}$	&	e\_M${_G}{_0}$	\\
\hline
1	&	J000118.65+464355.0	&	HD 224872	&	00 01 19.00	&	+46 43 55.01	&	B9.5 III HgMn	&	316	&	8.2905	&	0.0005	&	2.2517	&	0.0676	&	$-$0.089	&	0.002	&	0.197	&	$-$0.144	&	0.065	\\
2	&	J002635.94+562206.5	&	TYC 3661-39-1	&	00 26 35.91	&	+56 22 06.63	&	B9 IV-V HgMn	&	250	&	11.2054	&	0.0011	&	0.6004	&	0.0593	&	$-$0.102	&	0.056	&	0.557	&	$-$0.459	&	0.236	\\
3	&	J020922.72+533908.6	&	TYC 3685-766-1	&	02 09 22.74	&	+53 39 06.99	&	B8 IV HgMn	&	270	&	10.4932	&	0.0003	&	0.7401	&	0.0514	&	+0.005	&	0.003	&	0.294	&	$-$0.455	&	0.151	\\
4	&	J030852.02+553734.3	&	TYC 3706-249-1	&	03 08 52.03	&	+55 37 34.35	&	B8 IV HgMn	&	157	&	11.1172	&	0.0005	&	1.5127	&	0.0357	&	+0.533	&	0.019	&	1.102	&	+0.914	&	0.060	\\
5	&	J031312.82+493559.8	&	TYC 3319-890-1	&	03 13 12.72	&	+49 35 59.76	&	B9 III-IV HgMn	&	216	&	10.9302	&	0.0003	&	$-$	&	$-$	&	$-$	&	$-$	&	0.767	&	$-$	&	$-$	\\
6	&	J031943.49+561329.7	&	TYC 3706-265-1	&	03 19 43.49	&	+56 13 29.74	&	B7 III HgMn	&	215	&	11.0004	&	0.0005	&	0.8892	&	0.0632	&	+0.282	&	0.260	&	1.681	&	$-$0.935	&	0.465	\\
7	&	J032252.69+582806.7	&	TYC  3715-493-1	&	03 22 52.69	&	+58 28 06.69	&	B6 IV HgMn	&	170	&	11.5315	&	0.0005	&	1.0023	&	0.0342	&	$-$0.351	&	0.085	&	2.005	&	$-$0.469	&	0.163	\\
8	&	J033700.61+571139.0	&	TYC 3724-347-1	&	03 37 00.37	&	+57 11 39.56	&	B9 IV (HgMn)	&	164	&	10.4069	&	0.0014	&	3.9697	&	0.4858	&	+0.689	&	0.040	&	0.126	&	+3.274	&	0.275	\\
9	&	J033939.69+473758.3	&	TYC 3317-2031-1	&	03 39 39.69	&	+47 37 58.36	&	B8 IV HgMn	&	178	&	11.1676	&	0.0007	&	0.8508	&	0.0353	&	$-$0.081	&	0.003	&	1.047	&	$-$0.230	&	0.090	\\
10	&	J034903.76+454037.3	&	TYC 3326-1800-1	&	03 49 03.77	&	+45 40 37.48	&	B9 IV HgMn	&	619	&	10.1595	&	0.0011	&	2.0305	&	0.0575	&	+0.019	&	0.002	&	0.646	&	+1.051	&	0.061	\\
11	&	J035332.73+523248.9	&	Gaia DR2 251948416582201600	&	03 53 32.74	&	+52 32 48.91	&	B9 V HgMn	&	243	&	12.0045	&	0.0002	&	1.1101	&	0.0332	&	$-$0.030	&	0.024	&	1.118	&	+1.113	&	0.077	\\
12	&	J035333.90+524118.7	&	TYC 3717-313-1	&	03 53 33.91	&	+52 41 18.64	&	B9 IV HgMn	&	311	&	11.3319	&	0.0008	&	0.9816	&	0.0470	&	$-$0.096	&	0.030	&	1.022	&	+0.269	&	0.116	\\
13	&	J035805.60+452213.6	&	TYC 3327-113-1	&	03 58 05.60	&	+45 22 13.66	&	B8 IV HgMn	&	363	&	10.9552	&	0.0007	&	1.2487	&	0.0438	&	$-$0.003	&	0.006	&	1.096	&	+0.341	&	0.077	\\
14	&	J040155.24+514923.5	&	TYC 3339-14-1	&	04 01 55.23	&	+51 49 25.54	&	B8 V HgMn	&	219	&	10.4772	&	0.0005	&	1.3808	&	0.0438	&	$-$0.032	&	0.066	&	0.736	&	+0.442	&	0.135	\\
15	&	J040437.83+460141.5	&	TYC 3327-946-1	&	04 04 37.84	&	+46 01 41.51	&	B9 IV HgMn	&	252	&	11.2340	&	0.0009	&	1.1841	&	0.0603	&	$-$0.059	&	0.006	&	1.298	&	+0.303	&	0.111	\\
16	&	J041102.63+513103.7	&	TYC 3340-532-1	&	04 11 02.64	&	+51 31 03.70	&	B8 V HgMn	&	190	&	11.6903	&	0.0006	&	0.7446	&	0.0553	&	$-$0.174	&	0.005	&	0.979	&	+0.071	&	0.161	\\
17	&	J041115.00+360322.2	&	Gaia DR2 177165618554221312	&	04 11 15.01	&	+36 03 22.24	&	B9 II-III HgMn	&	209	&	12.1073	&	0.0002	&	0.3783	&	0.0490	&	$-$0.122	&	0.004	&	0.879	&	$-$0.882	&	0.281	\\
18	&	J041446.58+454448.3	&	TYC 3328-1424-1	&	04 14 46.58	&	+45 44 48.40	&	B9.5 III HgMn	&	344	&	11.2415	&	0.0010	&	0.9165	&	0.0549	&	+0.027	&	0.007	&	0.955	&	+0.097	&	0.131	\\
19	&	J041533.86+493944.1 $^{a}$	&	TYC 3336-762-1	&	04 15 34.12	&	+49 39 45.54	&	B8 IV HgMn	&	429	&	9.8972	&	0.0005	&	1.2322	&	0.0410	&	$-$0.072	&	0.010	&	0.926	&	$-$0.576	&	0.074	\\
20	&	J041632.64+424344.2	&	Gaia DR2 229032193375251328	&	04 16 32.64	&	+42 43 44.31	&	B8 V (HgMn)	&	175	&	12.1332	&	0.0003	&	0.6091	&	0.0551	&	+0.243	&	0.003	&	1.338	&	$-$0.282	&	0.196	\\
21	&	J042204.78+482811.3	&	Gaia DR2 258229381053351552	&	04 22 04.79	&	+48 28 11.42	&	B8 IV (HgMn)	&	121	&	12.2792	&	0.0002	&	0.8299	&	0.0316	&	+0.046	&	0.063	&	1.535	&	+0.339	&	0.137	\\
22	&	J042726.80+525506.2	&	TYC 3732-1016-1	&	04 27 26.82	&	+52 55 06.22	&	B8 IV-V HgMn	&	562	&	10.1456	&	0.0004	&	1.4293	&	0.0495	&	$-$0.285	&	0.092	&	1.091	&	$-$0.169	&	0.178	\\
23	&	J042829.12+511637.5	&	HD 232961	&	04 28 28.81	&	+51 16 37.58	&	B8 IV-V HgMn	&	397	&	9.8120	&	0.0005	&	1.7241	&	0.0597	&	$-$0.324	&	0.137	&	1.105	&	$-$0.110	&	0.252	\\
24	&	J042908.75+530654.3	&	Gaia DR2 272149949877398400	&	04 29 08.75	&	+53 06 54.31	&	B9 III-IV HgMn	&	174	&	12.1482	&	0.0002	&	1.0128	&	0.0335	&	$-$0.030	&	0.010	&	1.734	&	+0.441	&	0.074	\\
25	&	J043338.55+523755.1 $^{a}$	&	HD 232975	&	04 33 38.89	&	+52 37 55.13	&	B8 IV-V HgMn	&	348	&	9.5673	&	0.0002	&	1.7266	&	0.0413	&	$-$0.350	&	0.059	&	1.578	&	$-$0.825	&	0.114	\\
26	&	J043743.94+414133.8 $^{a}$	&	HD 276610	&	04 37 43.94	&	+41 41 33.81	&	B8 IV HgMn	&	247	&	10.2297	&	0.0003	&	0.9607	&	0.0451	&	$-$0.042	&	0.002	&	0.666	&	$-$0.524	&	0.102	\\
27	&	J044429.05+554643.2	&	Gaia DR2 274518164783721472	&	04 44 29.06	&	+55 46 43.19	&	B8 IV HgMn	&	113	&	13.1855	&	0.0003	&	0.3542	&	0.0249	&	$-$0.041	&	0.017	&	1.001	&	$-$0.070	&	0.155	\\
28	&	J044732.75+482124.5	&	TYC 3347-1290-1	&	04 47 32.76	&	+48 21 24.54	&	B8 IV HgMn	&	189	&	10.5235	&	0.0004	&	1.1015	&	0.0498	&	$-$0.227	&	0.010	&	0.939	&	$-$0.205	&	0.100	\\
29	&	J045516.12+581945.5	&	TYC 3746-129-1	&	04 55 16.12	&	+58 19 45.62	&	B8 IV-V HgMn	&	366	&	10.3948	&	0.0004	&	1.6412	&	0.0433	&	+0.487	&	0.007	&	0.436	&	+1.035	&	0.058	\\
30	&	J050338.55+234349.5	&	HD 285110	&	05 03 38.57	&	+23 43 49.67	&	B8 IV (HgMn)	&	222	&	10.6501	&	0.0005	&	1.5336	&	0.0694	&	+0.010	&	0.003	&	0.823	&	+0.756	&	0.098	\\
31	&	J050740.17+380705.1	&	TYC 2895-2440-1	&	05 07 40.17	&	+38 07 05.03	&	B8 IV HgMn	&	125	&	11.5741	&	0.0005	&	1.1917	&	0.0469	&	$-$0.076	&	0.003	&	1.374	&	+0.581	&	0.086	\\
32	&	J050810.73+453321.8 $^{a}$	&	TYC 3345-3314-1	&	05 08 10.81	&	+45 33 22.70	&	B8 III-IV HgMn	&	210	&	10.7819	&	0.0004	&	0.8520	&	0.0482	&	$-$0.119	&	0.003	&	0.612	&	$-$0.178	&	0.123	\\
33	&	J051309.48+551742.2	&	TYC 3739-643-1	&	05 13 09.49	&	+55 17 42.26	&	B9 III-IV HgMn	&	192	&	11.0260	&	0.0007	&	1.1686	&	0.0662	&	$-$0.061	&	0.005	&	1.197	&	+0.167	&	0.123	\\
34	&	J051503.39+323903.2	&	Gaia DR2 180704873465298432	&	05 15 03.39	&	+32 39 03.24	&	B7 IV-V (HgMn:)	&	140	&	12.1514	&	0.0004	&	0.7300	&	0.0607	&	$-$0.111	&	0.032	&	0.881	&	+0.587	&	0.189	\\
35	&	J051625.06+563321.2	&	TYC 3743-51-1	&	05 16 25.07	&	+56 33 21.30	&	B7 IV-V HgMn	&	128	&	11.0038	&	0.0007	&	1.1158	&	0.0375	&	$-$0.048	&	0.008	&	1.261	&	$-$0.019	&	0.074	\\
36	&	J051813.77+452059.7	&	TYC 3358-979-1	&	05 18 13.77	&	+45 20 59.77	&	B8 III-IV  HgMn	&	279	&	11.2963	&	0.0010	&	0.5462	&	0.0460	&	$-$0.113	&	0.003	&	0.901	&	$-$0.917	&	0.183	\\
37	&	J052056.26+085723.5	&	HD 242852	&	05 20 56.27	&	+08 57 23.53	&	B8 IV HgMn	&	241	&	10.1233	&	0.0006	&	1.6486	&	0.0734	&	$-$0.105	&	0.002	&	1.427	&	$-$0.218	&	0.097	\\
38	&	J052119.19+370619.9	&	Gaia DR2 184455827321040896	&	05 21 19.20	&	+37 06 20.04	&	B9 III-IV HgMn	&	108	&	12.8334	&	0.0005	&	0.5342	&	0.0442	&	$-$0.150	&	0.063	&	0.945	&	+0.527	&	0.210	\\
39	&	J052308.77+323128.3 $^{b}$ &	HD 242991	&	05 23 08.78	&	+32 31 28.35	&	B9 IV-V HgMn	&	302	&	10.7273	&	0.0005	&	0.9065	&	0.1014	&	$-$0.150	&	0.003	&	0.614	&	$-$0.100	&	0.243	\\
40	&	J052436.09+321553.5	&	HD 243231	&	05 24 36.10	&	+32 15 53.71	&	B7 V HgMn	&	328	&	10.6718	&	0.0004	&	1.5063	&	0.4506	&	$-$0.002	&	0.146	&	0.447	&	+1.114	&	0.699	\\
41	&	J052606.26+321808.6	&	HD 243493	&	05 26 06.27	&	+32 18 08.75	&	B9 III HgMn	&	133	&	11.6397	&	0.0004	&	0.6341	&	0.0403	&	$-$0.129	&	0.013	&	0.852	&	$-$0.202	&	0.140	\\
42	&	J052649.33+431004.9	&	TYC 2921-1297-1	&	05 26 49.33	&	+43 10 05.00	&	B9 IV HgMn	&	267	&	11.3570	&	0.0008	&	0.8990	&	0.0645	&	+0.087	&	0.029	&	0.444	&	+0.682	&	0.164	\\
43	&	J052746.31+314215.9	&	HD 243760	&	05 27 46.31	&	+31 42 15.90	&	B7 III-IV (HgMn:)	&	214	&	10.8051	&	0.0004	&	0.9702	&	0.0480	&	$-$0.108	&	0.015	&	0.810	&	$-$0.070	&	0.110	\\
44	&	J052808.91+315213.3	&	HD 243806	&	05 28 08.91	&	+31 52 13.44	&	B8 IV HgMn	&	345	&	9.7190	&	0.0005	&	1.2846	&	0.1441	&	$-$0.015	&	0.070	&	0.521	&	$-$0.258	&	0.274	\\
45	&	J053028.44+060100.9	&	Gaia DR2 3237727698424162304	&	05 30 28.44	&	+06 01 00.53	&	B9 IV-V HgMn	&	164	&	12.2845	&	0.0002	&	1.0175	&	0.0447	&	$-$0.051	&	0.003	&	1.061	&	+1.261	&	0.095	\\
46	&	J053419.30+302407.7 $^{a}$	&	HD 244826	&	05 34 19.08	&	+30 24 07.64	&	B7 III-IV HgMn	&	312	&	9.9475	&	0.0009	&	1.0973	&	0.0598	&	$-$0.018	&	0.028	&	1.058	&	$-$0.909	&	0.121	\\
47	&	J053513.37+192705.9	&	TYC 1305-1903-1	&	05 35 13.37	&	+19 27 05.96	&	B8 IV-V HgMn	&	238	&	11.7078	&	0.0004	&	0.7565	&	0.0464	&	$-$0.004	&	0.004	&	0.521	&	+0.581	&	0.133	\\
48	&	J053642.85+240205.8 $^{a}$	&	HD 245324	&	05 36 42.86	&	+24 02 05.85	&	B8 IV HgMn	&	393	&	10.6003	&	0.0006	&	0.8801	&	0.0741	&	$-$0.315	&	0.115	&	1.101	&	$-$0.778	&	0.271	\\
49	&	J053819.93+392350.0 $^{a}$	&	TYC 2914-1196-1	&	05 38 20.01	&	+39 23 49.76	&	B8 IV HgMn	&	439	&	10.7991	&	0.0005	&	0.9685	&	0.0491	&	$-$0.151	&	0.004	&	0.917	&	$-$0.187	&	0.110	\\
50	&	J053836.11+340929.7	&	TYC 2412-837-1	&	05 38 36.12	&	+34 09 29.77	&	B6 III (HgMn)	&	159	&	11.7727	&	0.0004	&	0.4500	&	0.0719	&	$-$0.073	&	0.045	&	1.111	&	$-$1.072	&	0.355	\\
\hline
\end{tabular}                                                                                                                                                                   
\end{adjustbox}
\end{center}                                                                                                                                             
\end{sidewaystable*}
\setcounter{table}{0}  
\begin{sidewaystable*}
\caption{Essential data for our sample stars, sorted by increasing right ascension. The columns denote: (1) Internal identification number. (2) LAMOST identifier. (3) Alternativ identifier (HD number, TYC identifier or GAIA DR2 number). (4) Right ascension (J2000; GAIA DR2). (5) Declination (J2000; GAIA DR2). (6) Spectral type, as derived in this study. (7) Sloan $g$ band S/N ratio of the analysed spectrum. (8) $G$\,mag (GAIA DR2). (9) $G$\,mag error. (10) Parallax (GAIA DR2). (11) Parallax error. (12) Dereddened colour index $(BP-RP){_0}$ (GAIA DR2). (13) Colour index error. (14) Absorption in the $G$ band, $A_G$. (15) Intrinsic absolute magnitude in the $G$ band, $M_{\mathrm{G,0}}$. (16) Absolute magnitude error.}
\label{table_master2}
\begin{center}
\begin{adjustbox}{max width=\textwidth}
\begin{tabular}{lllcclcccccccccc}
\hline
(1) & (2) & (3) & (4) & (5) & (6) & (7) & (8) & (9) & (10) & (11) & (12) & (13) & (14) & (15) & (16) \\
No.	&	ID\_LAMOST	&	ID\_alt	&	RA(J2000) 	&	 Dec(J2000)    	&	SpT\_final	&	S/N\,$g$	&	$G$\,mag	&	e\_$G$\,mag	&	pi (mas)	&	e\_pi	&	$(BP-RP){_0}$	&	e\_$(BP-RP){_0}$	&	A${_G}$	&	M${_G}{_0}$	&	e\_M${_G}{_0}$	\\
\hline
51	&	J054033.05+331124.6 $^{a}$	&	HD 246021	&	05 40 32.87	&	+33 11 24.47	&	B8 IV HgMn	&	249	&	10.1109	&	0.0007	&	1.5457	&	0.0742	&	$-$0.158	&	0.004	&	0.610	&	+0.446	&	0.104	\\
52	&	J054052.69+215844.5	&	HD 246183	&	05 40 52.68	&	+21 58 44.53	&	B8 V HgMn	&	407	&	9.7474	&	0.0008	&	2.2779	&	0.1308	&	$-$0.040	&	0.023	&	0.291	&	+1.244	&	0.131	\\
53	&	J054619.47+260416.7	&	HD 247308	&	05 46 19.25	&	+26 04 16.76	&	B8 IV-V (Hg)Mn	&	229	&	9.4514	&	0.0007	&	1.5906	&	0.0828	&	$-$0.107	&	0.003	&	0.453	&	+0.006	&	0.113	\\
54	&	J054927.98+302256.1	&	Gaia DR2 3444706261229672960	&	05 49 27.99	&	+30 22 56.20	&	B8 IV HgMn	&	128	&	11.9479	&	0.0003	&	0.4494	&	0.0380	&	$-$0.026	&	0.038	&	0.775	&	$-$0.564	&	0.195	\\
55	&	J055005.52+152159.5	&	TYC 1299-1938-1	&	05 50 05.53	&	+15 21 59.56	&	B8 IV (HgMn)	&	72	&	11.7326	&	0.0004	&	0.5887	&	0.0807	&	$-$0.078	&	0.008	&	0.608	&	$-$0.026	&	0.298	\\
56	&	J055030.45+244900.6	&	TYC 1866-1183-1	&	05 50 30.46	&	+24 49 00.60	&	B8 IV-V (Hg)Mn	&	174	&	11.5623	&	0.0004	&	0.9437	&	0.0487	&	$-$0.238	&	0.012	&	0.903	&	+0.533	&	0.113	\\
57	&	J055116.00+420914.5	&	Gaia DR2 192352725989945088	&	05 51 16.01	&	+42 09 14.57	&	B9 IV (HgMn)	&	154	&	12.9378	&	0.0005	&	0.3581	&	0.0320	&	+0.004	&	0.004	&	0.671	&	+0.036	&	0.194	\\
58	&	J055213.34+325807.7	&	TYC 2410-800-1	&	05 52 13.35	&	+32 58 07.76	&	B8 IV (HgMn:)	&	104	&	11.9385	&	0.0003	&	0.5058	&	0.0755	&	$-$0.048	&	0.008	&	0.506	&	$-$0.047	&	0.324	\\
59	&	J055309.48+265424.2 $^{a}$	&	HD 248667	&	05 53 09.49	&	+26 54 24.28	&	B7 III-IV HgMn	&	216	&	10.4620	&	0.0004	&	0.9185	&	0.0479	&	$-$0.197	&	0.009	&	0.778	&	$-$0.501	&	0.114	\\
60	&	J055400.30+290112.2 $^{c}$ &	TYC 1875-1748-1	&	05 54 00.30	&	+29 01 12.33	&	B7 III-IV HgMn	&	242	&	10.7807	&	0.0006	&	0.6908	&	0.0529	&	$-$0.053	&	0.005	&	0.531	&	$-$0.553	&	0.166	\\
61	&	J055457.96+092830.5	&	HD 249170	&	05 54 57.97	&	+09 28 30.50	&	B8 IV HgMn	&	189	&	11.0284	&	0.0006	&	1.3536	&	0.0754	&	+0.044	&	0.071	&	1.402	&	+0.284	&	0.172	\\
62	&	J055505.90+212949.6	&	Gaia DR2 3423922746981082240	&	05 55 05.90	&	+21 29 49.70	&	B9 IV (HgMn:)	&	69	&	13.1423	&	0.0004	&	0.4814	&	0.0365	&	$-$0.138	&	0.032	&	1.005	&	+0.550	&	0.174	\\
63	&	J055744.95+255028.1 $^{a}$	&	HD 249589	&	05 57 44.95	&	+25 50 28.13	&	B8 IV HgMn	&	280	&	10.2700	&	0.0006	&	1.0163	&	0.0620	&	$-$0.216	&	0.027	&	0.610	&	$-$0.305	&	0.141	\\
64	&	J055811.62+303219.5	&	TYC 2406-2337-1	&	05 58 11.62	&	+30 32 19.56	&	B9 III-IV HgMn	&	114	&	12.0241	&	0.0004	&	0.8139	&	0.0551	&	$-$0.023	&	0.004	&	0.628	&	+0.949	&	0.147	\\
65	&	J055854.37+292654.8	&	TYC 1875-2341-1	&	05 58 54.37	&	+29 26 54.87	&	B8 IV HgMn	&	191	&	11.7884	&	0.0003	&	0.3839	&	0.0352	&	$-$0.031	&	0.016	&	0.544	&	$-$0.834	&	0.201	\\
66	&	J060122.46+301948.7	&	TYC 2419-83-1	&	06 01 22.46	&	+30 19 48.83	&	B8 IV (HgMn)	&	65	&	12.6351	&	0.0004	&	0.2966	&	0.0363	&	$-$0.035	&	0.036	&	0.645	&	$-$0.649	&	0.273	\\
67	&	J060238.73+295105.4 $^{a}$	&	HD 250569	&	06 02 38.74	&	+29 51 05.46	&	B8 IV-V HgMn	&	130	&	11.0837	&	0.0006	&	0.9253	&	0.0434	&	$-$0.236	&	0.007	&	0.736	&	+0.179	&	0.102	\\
68	&	J060405.77+290624.1	&	Gaia DR2 3437444781498443904	&	06 04 05.77	&	+29 06 24.21	&	B7 IV (HgMn:)	&	69	&	12.9727	&	0.0005	&	0.4879	&	0.0620	&	+0.014	&	0.010	&	0.378	&	+1.037	&	0.276	\\
69	&	J060605.43+290405.2	&	HD 251407	&	06 06 05.44	&	+29 04 04.94	&	B8 IV-V HgMn	&	205	&	10.5333	&	0.0006	&	1.0777	&	0.0597	&	$-$0.237	&	0.007	&	0.589	&	+0.106	&	0.121	\\
70	&	J060619.07+345118.1	&	TYC 2427-1141-1	&	06 06 19.07	&	+34 51 18.25	&	B8 V HgMn	&	206	&	10.5629	&	0.0005	&	0.9602	&	0.0808	&	$-$0.389	&	0.069	&	0.978	&	$-$0.503	&	0.219	\\
71	&	J060709.47+390944.1	&	TYC 2925-320-1	&	06 07 09.47	&	+39 09 44.07	&	B7 III HgMn	&	247	&	9.6155	&	0.0006	&	1.1503	&	0.0708	&	$-$0.166	&	0.006	&	0.452	&	$-$0.532	&	0.134	\\
72	&	J060712.43+243850.5	&	HD 251752	&	06 07 12.44	&	+24 38 48.86	&	B8 III-IV HgMn	&	442	&	10.6328	&	0.0003	&	1.0120	&	0.0429	&	$-$0.070	&	0.005	&	0.568	&	+0.091	&	0.092	\\
73	&	J060732.64+215421.7	&	Gaia DR2 3423513793078031104	&	06 07 32.65	&	+21 54 21.76	&	B8 V (HgMn)	&	113	&	12.8900	&	0.0013	&	0.4509	&	0.4775	&	$-$0.207	&	0.409	&	1.260	&	$-$0.100	&	2.405	\\
74	&	J060738.70+210326.1	&	Gaia DR2 3375294676281816320	&	06 07 38.70	&	+21 03 26.21	&	B8 IV-V HgMn	&	109	&	13.3276	&	0.0004	&	0.5674	&	0.0311	&	$-$0.392	&	0.466	&	1.641	&	+0.456	&	0.810	\\
75	&	J060937.73+235218.0	&	HD 252428	&	06 09 37.73	&	+23 52 18.03	&	B8 III-IV HgMn	&	559	&	10.0629	&	0.0008	&	1.1282	&	0.0528	&	$-$0.106	&	0.005	&	0.582	&	$-$0.257	&	0.102	\\
76	&	J061137.11+254307.8	&	TYC 1881-804-1	&	06 11 37.10	&	+25 43 07.87	&	B8 IV-V HgMn	&	231	&	11.9100	&	0.0003	&	0.5742	&	0.0642	&	$-$0.034	&	0.011	&	0.687	&	+0.018	&	0.243	\\
77	&	J061144.23+245921.9	&	Gaia DR2 3426691381682039040	&	06 11 44.24	&	+24 59 21.95	&	B8 IV (HgMn)	&	108	&	13.3828	&	0.0004	&	0.2225	&	0.0322	&	$-$0.121	&	0.032	&	1.095	&	$-$0.976	&	0.319	\\
78	&	J061237.92+331656.7	&	Gaia DR2 3452046261553987072	&	06 12 37.93	&	+33 16 56.72	&	B8 III-IV HgMn	&	107	&	13.1882	&	0.0004	&	0.4392	&	0.0257	&	+0.217	&	0.003	&	0.851	&	+0.550	&	0.127	\\
79	&	J061436.72+252945.2	&	Gaia DR2 3426827003867129344	&	06 14 36.73	&	+25 29 45.30	&	B9 IV-V HgMn	&	170	&	12.3439	&	0.0003	&	0.7204	&	0.0560	&	+0.012	&	0.009	&	0.928	&	+0.704	&	0.169	\\
80	&	J062219.81+224023.5	&	HD 255857	&	06 22 19.82	&	+22 40 23.54	&	B8 IV-V HgMn	&	428	&	10.6444	&	0.0006	&	1.1178	&	0.0636	&	$-$0.079	&	0.010	&	0.434	&	+0.453	&	0.125	\\
81	&	J062425.14+190413.7	&	Gaia DR2 3372370495047722880	&	06 24 25.16	&	+19 04 13.74	&	B8 IV-V HgMn	&	118	&	14.6247	&	0.0005	&	0.2290	&	0.0336	&	+0.005	&	0.038	&	0.883	&	+0.541	&	0.325	\\
82	&	J062503.02+033237.3	$^{d}$ &	HD 256951	&	06 25 03.02	&	+03 32 37.33	&	B7 IV HgMn	&	345	&	10.5086	&	0.0005	&	1.0687	&	0.0847	&	$-$0.089	&	0.005	&	1.010	&	$-$0.357	&	0.172	\\
83	&	J062524.46+224236.2	&	TYC 1879-333-1	&	06 25 24.47	&	+22 42 36.25	&	B9.5 IV HgMn	&	170	&	12.4303	&	0.0003	&	0.5753	&	0.0497	&	$-$0.249	&	0.055	&	1.025	&	+0.205	&	0.211	\\
84	&	J062719.70+095803.7	&	Gaia DR2 3327481867692763776	&	06 27 19.71	&	+09 58 03.76	&	B9 III-IV (HgMn)	&	113	&	14.0369	&	0.0004	&	0.2443	&	0.0291	&	$-$0.049	&	0.023	&	1.218	&	$-$0.242	&	0.261	\\
85	&	J063246.02+030319.1	&	TYC 150-604-1	&	06 32 46.03	&	+03 03 19.13	&	B8 IV HgMn	&	180	&	11.6210	&	0.0003	&	0.7336	&	0.0666	&	$-$0.749	&	0.078	&	1.985	&	$-$1.037	&	0.238	\\
86	&	J063253.86+034220.3	&	HD 46338	&	06 32 53.87	&	+03 42 20.34	&	B8 V (HgMn)	&	188	&	9.6598	&	0.0005	&	1.4824	&	0.0465	&	$-$0.204	&	0.004	&	0.346	&	+0.169	&	0.068	\\
87	&	J063735.23+210443.6	&	HD 260690	&	06 37 35.31	&	+21 04 43.76	&	B9 III-IV HgMn	&	565	&	10.5966	&	0.0006	&	0.7065	&	0.0866	&	$-$0.189	&	0.008	&	0.295	&	$-$0.453	&	0.266	\\
88	&	J063821.84+110846.7	&	HD 261050	&	06 38 21.85	&	+11 08 46.76	&	B8 III HgMn	&	136	&	10.7161	&	0.0004	&	1.0286	&	0.0485	&	$-$0.024	&	0.003	&	1.550	&	$-$0.772	&	0.102	\\
89	&	J063910.31+522527.3	&	HD 46980	&	06 39 10.32	&	+52 25 30.14	&	B8 IV HgMn	&	572	&	9.8145	&	0.0005	&	1.2655	&	0.0726	&	$-$0.119	&	0.004	&	0.155	&	+0.171	&	0.125	\\
90	&	J063950.80+345537.7 $^{a}$	&	TYC 2443-805-1	&	06 39 50.80	&	+34 55 37.79	&	B7 III-IV HgMn	&	188	&	10.1067	&	0.0007	&	0.9902	&	0.0443	&	$-$0.082	&	0.003	&	0.344	&	$-$0.259	&	0.097	\\
91	&	J064224.70+130935.7	&	Gaia DR2 3352682021366285568	&	06 42 24.70	&	+13 09 35.66	&	B8 IV HgMn	&	162	&	14.2209	&	0.0004	&	0.2732	&	0.0459	&	$-$0.035	&	0.092	&	0.708	&	+0.695	&	0.399	\\
92	&	J064424.23+280442.5	&	TYC 1901-773-1	&	06 44 24.23	&	+28 04 42.49	&	B8 IV HgMn	&	83	&	12.7250	&	0.0005	&	0.8249	&	0.0563	&	$-$0.141	&	0.007	&	0.160	&	+2.147	&	0.148	\\
93	&	J064459.62+202512.7	&	TYC 1338-533-1	&	06 44 59.63	&	+20 25 12.72	&	B9 IV-V   (HgMn)	&	260	&	12.1338	&	0.0005	&	0.4987	&	0.0898	&	$-$0.059	&	0.006	&	0.318	&	+0.305	&	0.391	\\
94	&	J064509.27+285032.7	&	HD 262952	&	06 45 09.27	&	+28 50 32.71	&	B6 III (HgMn)	&	132	&	11.1320	&	0.0009	&	0.5622	&	0.0550	&	$-$0.178	&	0.006	&	0.250	&	$-$0.368	&	0.212	\\
95	&	J064604.00+024628.2	&	TYC 152-650-1	&	06 46 04.00	&	+02 46 28.27	&	B8 IV HgMn	&	181	&	11.7663	&	0.0003	&	0.7475	&	0.0444	&	$-$0.047	&	0.015	&	0.383	&	+0.752	&	0.131	\\
96	&	J064642.68+074808.3	&	Gaia DR2 3133992491496045184	&	06 46 42.69	&	+07 48 08.34	&	B9 III HgMn	&	106	&	12.4021	&	0.0003	&	0.7542	&	0.0428	&	+0.122	&	0.008	&	0.638	&	+1.151	&	0.124	\\
97	&	J065006.33+023041.0	&	Gaia DR2 3126488805672228352	&	06 50 06.34	&	+02 30 41.09	&	B8 IV (HgMn:)	&	128	&	13.0658	&	0.0006	&	0.4873	&	0.0354	&	$-$0.012	&	0.020	&	0.710	&	+0.794	&	0.161	\\
98	&	J065011.54+014917.3 $^{a}$	&	TYC 148-118-1	&	06 50 11.54	&	+01 49 17.43	&	B8 IV HgMn	&	183	&	11.8240	&	0.0003	&	0.6449	&	0.0384	&	$-$0.046	&	0.004	&	0.457	&	+0.414	&	0.129	\\
99	&	J065237.92+323912.9	&	TYC 2440-341-1	&	06 52 37.92	&	+32 39 13.01	&	B8 IV (HgMn)	&	176	&	12.1561	&	0.0004	&	0.5164	&	0.0541	&	$-$0.195	&	0.004	&	0.303	&	+0.418	&	0.227	\\
100	&	J065253.84-005714.3	&	TYC 4801-1648-1	&	06 52 53.84	&	-00 57 14.38	&	B8 V HgMn	&	196	&	11.4605	&	0.0003	&	0.6814	&	0.0552	&	$-$0.104	&	0.011	&	0.479	&	+0.149	&	0.177	\\
\hline
\end{tabular}                                                                                                                                                                   
\end{adjustbox}
\end{center}                                                                                                                                             
\end{sidewaystable*}
\setcounter{table}{0}  
\begin{sidewaystable*}
\caption{Essential data for our sample stars, sorted by increasing right ascension. The columns denote: (1) Internal identification number. (2) LAMOST identifier. (3) Alternativ identifier (HD number, TYC identifier or GAIA DR2 number). (4) Right ascension (J2000; GAIA DR2). (5) Declination (J2000; GAIA DR2). (6) Spectral type, as derived in this study. (7) Sloan $g$ band S/N ratio of the analysed spectrum. (8) $G$\,mag (GAIA DR2). (9) $G$\,mag error. (10) Parallax (GAIA DR2). (11) Parallax error. (12) Dereddened colour index $(BP-RP){_0}$ (GAIA DR2). (13) Colour index error. (14) Absorption in the $G$ band, $A_G$. (15) Intrinsic absolute magnitude in the $G$ band, $M_{\mathrm{G,0}}$. (16) Absolute magnitude error.}
\label{table_master3}
\begin{center}
\begin{adjustbox}{max width=\textwidth}
\begin{tabular}{lllcclcccccccccc}
\hline
(1) & (2) & (3) & (4) & (5) & (6) & (7) & (8) & (9) & (10) & (11) & (12) & (13) & (14) & (15) & (16) \\
No.	&	ID\_LAMOST	&	ID\_alt	&	RA(J2000) 	&	 Dec(J2000)    	&	SpT\_final	&	S/N\,$g$	&	$G$\,mag	&	e\_$G$\,mag	&	pi (mas)	&	e\_pi	&	$(BP-RP){_0}$	&	e\_$(BP-RP){_0}$	&	A${_G}$	&	M${_G}{_0}$	&	e\_M${_G}{_0}$	\\
\hline
101	&	J065545.39+093941.0	&	HD 266502	&	06 55 45.30	&	+09 39 41.06	&	B8 III-IV HgMn	&	236	&	10.6756	&	0.0006	&	0.8030	&	0.0507	&	$-$0.096	&	0.008	&	0.312	&	$-$0.113	&	0.138	\\
102	&	J065942.88+003951.0	&	HD 289522	&	06 59 42.89	&	+00 39 51.03	&	B9.5 IV-V (HgMn)	&	332	&	11.2633	&	0.0006	&	0.9924	&	0.0442	&	$-$0.125	&	0.005	&	0.384	&	+0.863	&	0.097	\\
103	&	J071933.80+145655.7	&	HD 56914	&	07 19 33.80	&	+14 56 55.76	&	B8 IV HgMn	&	348	&	9.0445	&	0.0006	&	1.3497	&	0.0874	&	$-$0.084	&	0.004	&	0.099	&	$-$0.404	&	0.141	\\
104	&	J072016.47+141320.6 $^{a}$	&	HD 57109	&	07 20 16.48	&	+14 13 20.58	&	B8 IV HgMn	&	289	&	9.0677	&	0.0004	&	1.4537	&	0.0540	&	$-$0.129	&	0.003	&	0.170	&	$-$0.290	&	0.081	\\
105	&	J072147.31+245451.4	&	HD 57338	&	07 21 47.32	&	+24 54 51.50	&	B9 IV HgMn	&	593	&	9.5273	&	0.0013	&	0.9135	&	0.0804	&	$-$0.122	&	0.021	&	0.156	&	$-$0.825	&	0.191	\\
106	&	J072836.11+095219.2	&	TYC 768-1132-1	&	07 28 36.11	&	+09 52 19.20	&	B8 IV HgMn	&	144	&	11.8213	&	0.0005	&	0.6217	&	0.0553	&	$-$0.165	&	0.008	&	0.099	&	+0.690	&	0.193	\\
107	&	J074457.09+541600.8	&	TYC 3782-390-1	&	07 44 57.10	&	+54 16 00.82	&	B8 III-IV HgMn	&	247	&	10.9709	&	0.0007	&	0.7474	&	0.0709	&	$-$0.174	&	0.004	&	0.181	&	+0.158	&	0.206	\\
108	&	J091838.94+195023.7	&	Gaia DR2 635840844129715712	&	09 18 38.95	&	+19 50 23.76	&	B9 IV-V HgMn	&	130	&	13.1622	&	0.0003	&	0.3847	&	0.0551	&	$-$0.141	&	0.003	&	0.086	&	+1.002	&	0.311	\\
109	&	J112400.77+540532.1	&	TYC 3825-801-1	&	11 24 00.21	&	+54 05 30.76	&	B8 IV (HgMn:)	&	107	&	11.7136	&	0.0005	&	2.8385	&	0.2800	&	$-$0.120	&	0.004	&	0.073	&	+3.906	&	0.214	\\
110	&	J130109.88-031648.1	&	Gaia DR2 3685094588647559040	&	13 01 09.88	&	-03 16 48.15	&	B8 IV (HgMn)	&	108	&	13.0938	&	0.0005	&	0.5376	&	0.0748	&	$-$0.203	&	0.004	&	0.093	&	+1.653	&	0.302	\\
111	&	J133147.11-023941.1	&	Gaia DR2 3637524870906723328	&	13 31 47.14	&	-02 39 41.98	&	B8 IV HgMn	&	195	&	13.4212	&	0.0005	&	0.4261	&	0.0398	&	$-$0.169	&	0.004	&	0.105	&	+1.464	&	0.203	\\
112	&	J154534.12+160842.8	&	Gaia DR2 1196145299559569792	&	15 45 34.13	&	+16 08 42.80	&	B8 IV HgMn	&	188	&	12.7487	&	0.0003	&	0.4955	&	0.0734	&	$-$0.202	&	0.004	&	0.166	&	+1.058	&	0.321	\\
113	&	J184142.65+482514.8	&	TYC 3531-659-1	&	18 41 42.64	&	+48 25 13.73	&	B8 IV HgMn	&	137	&	11.2487	&	0.0007	&	1.3830	&	0.0451	&	$-$0.198	&	0.005	&	0.174	&	+1.778	&	0.071	\\
114	&	J192837.06+520029.0	&	TYC 3555-1845-1	&	19 28 37.06	&	+52 00 29.17	&	B8 IV (HgMn)	&	133	&	12.3400	&	0.0003	&	0.2319	&	0.0202	&	$-$0.058	&	0.003	&	0.269	&	$-$1.102	&	0.189	\\
115	&	J194139.54+403833.9 $^{a}$	&	HD 186254	&	19 41 39.28	&	+40 38 33.72	&	B8 IV HgMn	&	184	&	8.5859	&	0.0005	&	1.6616	&	0.0662	&	$-$0.124	&	0.008	&	0.209	&	$-$0.520	&	0.088	\\
116	&	J195021.63+411958.5	&	Gaia DR2 2076873634046444800	&	19 50 21.63	&	+41 19 58.60	&	B9 IV HgMn	&	147	&	13.2381	&	0.0003	&	0.3797	&	0.0189	&	$-$0.057	&	0.013	&	0.931	&	+0.204	&	0.110	\\
117	&	J195652.72+270157.7	&	Gaia DR2 2027500408007197824	&	19 56 52.72	&	+27 01 57.81	&	B8 V HgMn	&	184	&	12.1590	&	0.0001	&	0.8913	&	0.0277	&	+0.004	&	0.003	&	1.186	&	+0.723	&	0.068	\\
118	&	J202548.00+363352.4	&	Gaia DR2 2057692374515650816	&	20 25 48.01	&	+36 33 52.47	&	B7 IV-V HgMn	&	140	&	12.4641	&	0.0002	&	0.7678	&	0.0261	&	$-$0.405	&	0.051	&	2.046	&	$-$0.156	&	0.114	\\
119	&	J205110.47+355929.8	&	TYC 2700-379-1	&	20 51 10.47	&	+35 59 29.81	&	B7 IV (Hg)Mn	&	253	&	11.3640	&	0.0009	&	0.7526	&	0.0425	&	$-$0.190	&	0.008	&	0.566	&	+0.180	&	0.123	\\
120	&	J213215.68+425837.2	&	Gaia DR2 1967805344746109696	&	21 32 15.69	&	+42 58 37.34	&	B8 IV-V (HgMn)	&	119	&	12.8857	&	0.0003	&	0.5965	&	0.0291	&	+0.105	&	0.002	&	0.967	&	+0.797	&	0.106	\\
121	&	J213851.08+164506.8	&	Gaia DR2 1773423160802357376	&	21 38 51.10	&	+16 45 07.28	&	B6 III HgMn	&	230	&	13.4933	&	0.0005	&	0.1242	&	0.0382	&	$-$0.191	&	0.004	&	0.363	&	$-$1.399	&	0.667	\\
122	&	J230239.47+555203.1	&	TYC 3989-742-1	&	23 02 39.47	&	+55 52 03.14	&	B9 IV-V HgMn	&	193	&	11.8371	&	0.0003	&	0.7309	&	0.0304	&	+0.021	&	0.004	&	0.443	&	+0.713	&	0.090	\\
123	&	J231115.48+543738.0	&	TYC 4002-1726-1	&	23 11 15.49	&	+54 37 38.01	&	B8 IV HgMn	&	174	&	10.9405	&	0.0007	&	1.0549	&	0.0453	&	$-$0.142	&	0.006	&	0.576	&	+0.481	&	0.094	\\
124	&	J231124.89+570726.6	&	TYC 4006-596-1	&	23 11 24.89	&	+57 07 26.63	&	B8 IV-V HgMn	&	156	&	11.3446	&	0.0005	&	0.9658	&	0.0319	&	$-$0.005	&	0.002	&	0.984	&	+0.285	&	0.072	\\
125	&	J232346.41+551605.3	&	TYC 4003-471-1	&	23 23 46.41	&	+55 16 05.32	&	B8 IV-V HgMn	&	195	&	11.8367	&	0.0003	&	0.5395	&	0.0330	&	$-$0.125	&	0.005	&	0.446	&	+0.051	&	0.133	\\
\hline
\multicolumn{16}{l}{Notes:} \\
\multicolumn{16}{l}{$^{a}$ Contained in the sample of APOGEE CP3 stars of \citet{chojnowski20}.} \\
\multicolumn{16}{l}{$^{b}$ \citet{RM09}: Renson 8947; spectral type: B9 Si. LAMOST spectrum shows typical signature of a CP3 star; not a classical CP2 star.} \\
\multicolumn{16}{l}{$^{c}$ \citet{RM09}: Renson 10597; spectral type: B8 Si; CP star of doubtful nature. LAMOST spectrum shows typical signature of a CP3 star; not a classical CP2 star.} \\
\multicolumn{16}{l}{$^{d}$ \citet{RM09}: Renson 11944; spectral type: A0; CP star of doubtful nature. LAMOST spectrum shows typical signature of a CP3 star.} \\
\hline
\hline
\end{tabular}                                                                                                                                                                   
\end{adjustbox}
\end{center}                                                                                                                                             
\end{sidewaystable*}

\setcounter{table}{1}
\begin{table*}
\caption{Upper limits of variability for the stars identified as non-variable in the accuracy limits of the employed data sources. The columns denote: (1) Internal identification number. (2) LAMOST identifier. (3) Upper limit of variability as derived from SWASP data. (4) Upper limit of variability as derived from ASAS-SN data. (5) Upper limit of variability as derived from ASAS-3 data. (6) Upper limit of variability as derived from TESS data. All values denote semiamplitudes given in millimagnitudes.}
\label{table_upperlimits}
\begin{center}
\begin{adjustbox}{max width=0.99\textwidth}
\begin{tabular}{|l|l|c|c|c|c|l|l|c|c|c|c|}
\hline
(1) & (2) & (3) & (4) & (5) & (6) & (1) & (2) & (3) & (4) & (5) & (6) \\
No.	&	ID\_LAMOST	&	UL\textsubscript{SW}	&	UL\textsubscript{AS}	&	UL\textsubscript{AS3}	&	UL\textsubscript{TE}	&	No.	&	ID\_LAMOST	&	UL\textsubscript{SW}	&	UL\textsubscript{AS}	&	UL\textsubscript{AS3}	&	UL\textsubscript{TE}	\\
\hline
1	&	J000118.65+464355.0	&	4.7	&		&		&		&	61	&	J055457.96+092830.5	&		&	3.8	&	10.3	&	0.5	\\
2	&	J002635.94+562206.5	&		&	5.7	&		&		&	62	&	J055505.90+212949.6	&	4.5	&	4.1	&		&		\\
3	&	J020922.72+533908.6	&	5.7	&	2.8	&		&		&	63	&	J055744.95+255028.1	&	9.7	&		&		&		\\
4	&	J030852.02+553734.3	&	2.4	&	3.2	&		&		&	64	&	J055811.62+303219.5	&	25.1	&	3.5	&		&		\\
5	&	J031312.82+493559.8	&	4.0	&	3.5	&		&		&	65	&	J055854.37+292654.8	&	3.0	&	3.0	&		&		\\
6	&	J031943.49+561329.7	&	2.7	&	7.5	&		&		&	66	&	J060122.46+301948.7	&	4.7	&	4.0	&		&		\\
7	&	J032252.69+582806.7	&	10.0	&	7.6	&		&		&	67	&	J060238.73+295105.4	&	2.5	&	3.2	&		&		\\
8	&	J033700.61+571139.0	&	3.0	&	3.2	&		&		&	68	&	J060405.77+290624.1	&	3.4	&	6.2	&		&		\\
9	&	J033939.69+473758.3	&	12.1	&	2.0	&		&		&	69	&	J060605.43+290405.2	&	2.7	&	3.3	&		&		\\
10	&	J034903.76+454037.3	&	4.9	&		&		&		&	70	&	J060619.07+345118.1	&	2.8	&	4.4	&		&		\\
11	&	J035332.73+523248.9	&	19.2	&	3.4	&		&		&	71	&	J060709.47+390944.1	&	2.8	&		&		&		\\
12	&	J035333.90+524118.7	&	8.4	&	3.8	&		&		&	72	&	J060712.43+243850.5	&	2.5	&	5.4	&		&		\\
13	&	J035805.60+452213.6	&	4.6	&	2.8	&		&		&	73	&	J060732.64+215421.7	&	44.2	&	4.8	&		&		\\
14	&	J040155.24+514923.5	&	3.6	&	5.0	&		&		&	74	&	J060738.70+210326.1	&	4.5	&	7.2	&		&		\\
15	&	J040437.83+460141.5	&	3.5	&	2.6	&		&		&	77	&	J061144.23+245921.9	&	9.6	&	10.8	&		&		\\
16	&	J041102.63+513103.7	&	4.7	&	5.9	&		&		&	78	&	J061237.92+331656.7	&	2.5	&	5.5	&		&		\\
17	&	J041115.00+360322.2	&	2.3	&	7.0	&		&		&	79	&	J061436.72+252945.2	&	3.4	&	6.2	&		&		\\
18	&	J041446.58+454448.3	&	8.0	&	4.0	&		&		&	80	&	J062219.81+224023.5	&	5.7	&	3.2	&		&		\\
19	&	J041533.86+493944.1	&	6.7	&		&		&		&	81	&	J062425.14+190413.7	&	6.1	&		&		&		\\
20	&	J041632.64+424344.2	&	2.7	&	5.0	&		&		&	82	&	J062503.02+033237.3	&		&	5.0	&	10.0	&	0.4	\\
21	&	J042204.78+482811.3	&	5.7	&	6.8	&		&		&	83	&	J062524.46+224236.2	&		&	4.9	&		&		\\
22	&	J042726.80+525506.2	&		&		&		&		&	84	&	J062719.70+095803.7	&		&		&		&	0.7	\\
23	&	J042829.12+511637.5	&	2.1	&		&		&		&	85	&	J063246.02+030319.1	&		&	9.9	&	11.0	&	0.5	\\
24	&	J042908.75+530654.3	&	2.1	&	3.5	&		&		&	89	&	J063910.31+522527.3	&	4.8	&		&		&		\\
25	&	J043338.55+523755.1	&	6.5	&		&		&		&	90	&	J063950.80+345537.7	&	3.2	&		&		&		\\
26	&	J043743.94+414133.8	&	5.0	&	5.7	&		&		&	91	&	J064224.70+130935.7	&		&		&	20.2	&		\\
27	&	J044429.05+554643.2	&	15.4	&	11.0	&		&		&	92	&	J064424.23+280442.5	&	1.8	&	5.4	&		&		\\
28	&	J044732.75+482124.5	&	5.8	&	5.3	&		&		&	93	&	J064459.62+202512.7	&		&	4.6	&		&		\\
29	&	J045516.12+581945.5	&	6.5	&	5.8	&		&		&	94	&	J064509.27+285032.7	&	2.0	&	3.2	&		&		\\
31	&	J050740.17+380705.1	&	2.0	&	3.0	&		&		&	95	&	J064604.00+024628.2	&		&	4.6	&	11.8	&	0.4	\\
32	&	J050810.73+453321.8	&	2.8	&	2.4	&		&		&	96	&	J064642.68+074808.3	&		&	6.9	&	20.0	&	0.4	\\
33	&	J051309.48+551742.2	&	2.5	&	3.7	&		&		&	97	&	J065006.33+023041.0	&		&	7.0	&		&	0.5	\\
34	&	J051503.39+323903.2	&	20.0	&	10.9	&		&		&	98	&	J065011.54+014917.3	&		&	3.0	&	11.1	&	0.4	\\
35	&	J051625.06+563321.2	&	3.4	&	4.8	&		&		&	99	&	J065237.92+323912.9	&	3.6	&	4.5	&		&		\\
36	&	J051813.77+452059.7	&	1.3	&	3.0	&		&		&	100	&	J065253.84-005714.3	&		&	6.2	&	10.5	&	0.4	\\
37	&	J052056.26+085723.5	&		&	3.5	&	4.2	&		&	101	&	J065545.39+093941.0	&		&	7.1	&	10.8	&	0.4	\\
38	&	J052119.19+370619.9	&	2.6	&	5.6	&		&		&	102	&	J065942.88+003951.0	&		&	6.0	&	10.0	&		\\
39	&	J052308.77+323128.3	&	1.7	&	4.5	&		&		&	104	&	J072016.47+141320.6	&		&		&	7.0	&	0.4	\\
40	&	J052436.09+321553.5	&	5.8	&	5.1	&		&		&	105	&	J072147.31+245451.4	&	2.8	&		&		&		\\
41	&	J052606.26+321808.6	&	3.8	&	4.5	&		&		&	106	&	J072836.11+095219.2	&		&	5.7	&	12.3	&	0.4	\\
42	&	J052649.33+431004.9	&	3.2	&	5.7	&		&		&	107	&	J074457.09+541600.8	&	11.6	&	6.0	&		&		\\
43	&	J052746.31+314215.9	&	1.5	&	4.0	&		&		&	108	&	J091838.94+195023.7	&		&	9.2	&		&		\\
44	&	J052808.91+315213.3	&	1.8	&		&		&		&	109	&	J112400.77+540532.1	&		&	6.0	&		&		\\
45	&	J053028.44+060100.9	&		&	4.5	&	19.8	&	0.5	&	110	&	J130109.88-031648.1	&		&	1.8	&		&		\\
46	&	J053419.30+302407.7	&	5.2	&		&		&		&	111	&	J133147.11-023941.1	&		&	7.3	&		&		\\
47	&	J053513.37+192705.9	&		&	4.2	&	25.1	&		&	112	&	J154534.12+160842.8	&		&	5.6	&		&		\\
48	&	J053642.85+240205.8	&	2.4	&	6.1	&		&		&	113	&	J184142.65+482514.8	&	8.4	&	33.0	&		&	0.4	\\
49	&	J053819.93+392350.0	&	5.4	&	2.7	&		&		&	114	&	J192837.06+520029.0	&	5.4	&	5.9	&		&	0.4	\\
50	&	J053836.11+340929.7	&	2.3	&	3.3	&		&		&	115	&	J194139.54+403833.9	&		&		&		&	0.4	\\
51	&	J054033.05+331124.6	&	3.2	&		&		&		&	117	&	J195652.72+270157.7	&		&	8.2	&		&	0.5	\\
52	&	J054052.69+215844.5	&	2.8	&		&		&		&	118	&	J202548.00+363352.4	&	4.3	&	4.9	&		&	0.5	\\
53	&	J054619.47+260416.7	&	2.3	&		&		&		&	119	&	J205110.47+355929.8	&		&	9.3	&		&	0.4	\\
54	&	J054927.98+302256.1	&	2.6	&	2.6	&		&		&	120	&	J213215.68+425837.2	&	4.0	&	11.0	&		&	0.5	\\
55	&	J055005.52+152159.5	&		&	6.5	&	14.0	&	0.3	&	121	&	J213851.08+164506.8	&		&	10.0	&		&		\\
56	&	J055030.45+244900.6	&	3.1	&	4.7	&		&		&	122	&	J230239.47+555203.1	&	8.1	&	5.4	&		&	0.4	\\
57	&	J055116.00+420914.5	&	5.6	&	10.3	&		&		&	123	&	J231115.48+543738.0	&	6.2	&	8.4	&		&		\\
58	&	J055213.34+325807.7	&	2.8	&	8.3	&		&		&	124	&	J231124.89+570726.6	&	13.3	&	4.7	&		&		\\
59	&	J055309.48+265424.2	&	2.3	&	3.1	&		&		&	125	&	J232346.41+551605.3	&	5.8	&	17.3	&		&		\\
60	&	J055400.30+290112.2	&	3.3	&	2.3	&		&		&		&		&		&		&		&		\\
\hline
\end{tabular}                                                                                                                                                                  
\end{adjustbox}
\end{center}                                                                                                                                             
\end{table*}

\end{document}